\documentclass[12pt]{article}
\usepackage{amsfonts,amssymb,amsmath,amsthm}
\usepackage{latexsym}
\usepackage{verbatim}
\usepackage{epsfig}
\usepackage{comment}
\usepackage{soul}
\usepackage{color}
\usepackage{multirow}
\usepackage{chemarr}
\usepackage{mathrsfs}  
\usepackage{makecell}
\usepackage{subcaption}
\usepackage{graphicx}
\usepackage{amsbsy}
\usepackage{amscd}
\usepackage{bm}
\usepackage[nonamebreak,square,numbers, sort]{natbib}

\def\spacingset#1{\renewcommand{\baselinestretch}%
{#1}\small\normalsize} \spacingset{1}

\providecommand{\M}[1]{\mathbf#1}
\providecommand{\mc}[1]{\mathcal#1}
\providecommand{\mc}[1]{\mathcal#1}

\newcommand{\R}{{\mathbb R}}

\def\x{\mathbf{x}}

\DeclareMathOperator{\E}{\mathbf{E}}
\DeclareMathOperator{\p}{\mathbf{P}}

\DeclareMathOperator{\cov}{Cov}

\providecommand{\T}{\top} 

\providecommand{\wt}[1]{\widetilde{#1}}
\providecommand{\wh}[1]{\widehat{#1}}

\providecommand{\nnorm}[1]{ \lVert#1 \rVert}
\newcommand{\scp}[2]{\left\langle#1, #2\right\rangle}
\newcommand{\nscp}[2]{\langle#1, #2\rangle}

\newcommand{\abs}[1]{\left | #1 \right |}

\newcommand{\dev}[2]{\Big\lvert _{{#1}={#2}}}

\newcommand{\blanco}[1]{  }
\newcommand{\deriv}[3]{%
\ifthenelse{#1 = 1}{\frac{d\,#2}{d\,#3}}{\frac{d^{{#1}} #2}{d{#3}^{{#1}}}}
} 
\newcommand{\partials}[3]{%
\ifthenelse{#1 = 1}{\frac{\partial\,#2}{\partial\,#3}}{\frac{\partial^{#1}
    #2}{\partial#3^{#1}}}
} 
\def\su{\sum_{i=1}^n}

\def \coloneq{\mathrel{\mathop:}=}
\def \invcoloneq{=\mathrel{\mathop:}}

\def \lec{\preceq}

\newtheorem{theo}{Theorem}
\newtheorem{propo}{Theorem}

\newtheorem{theoD}{Theorem}
\newtheorem{lemmachenA}{Theorem}
\newtheorem{lemmachenB}{Theorem}

\newtheorem{theoremD}[theoD]{Theorem D.}
\newtheorem{prop}[propo]{Proposition}

\newtheorem{lemmaA}[lemmachenA]{Lemma A.}
\newtheorem{lemmaB}[lemmachenB]{Lemma B.}

\usepackage{algorithmic}
\usepackage{algorithm}

\usepackage{tikz}
\usepackage{bbm}
\usepackage{colortbl}
\usepackage{enumitem}
\definecolor{gray}{rgb}{.75, .75, .75}
\newcommand\LL[1]{\multicolumn{1}{|l}{#1}}
\newcommand\RR[1]{\multicolumn{1}{l|}{#1}}

\newcommand{\sss}[2]{\mbox{{\normalsize #1} \;({\footnotesize #2})}}

\usepackage{bbm}
\usepackage{tikz}

\newcommand{\symfootnote}[1]{%
\let\oldthefootnote=\thefootnote%
\stepcounter{mpfootnote}%
\addtocounter{footnote}{-1}%
\renewcommand{\thefootnote}{\fnsymbol{mpfootnote}}%
\footnote{#1}%
\let\thefootnote=\oldthefootnote%
}

\begin{document}

\title{\bf {\Large \begin{tabular}{l}Estimation in exponential family Regression based on \\linked data contaminated by mismatch error
\end{tabular}}}

\author{Zhenbang Wang$^{\star}$ $\qquad$ Emanuel Ben-David$^{\dagger}$ $\qquad$ Martin Slawski$^{\star}$ \thanks{
    The first and the last author are partially supported by NSF award CRII: CIF: 1849876}\hspace{.2cm}\\[2ex]
    $^{\star}${\normalsize Department of Statistics, George Mason University}
    $\qquad$ $\dagger${\normalsize U.S.~Census, CSRM}}
\date{}
\maketitle

\bigskip
\bigskip
\bigskip

\begin{abstract}
Identification of matching records in multiple files can be a challenging and error-prone task. Linkage error can considerably  affect subsequent statistical analysis based on the resulting linked file.  Several recent papers have studied post-linkage linear regression analysis with the response variable in one file and the covariates in a second file from the perspective of the "Broken Sample Problem" and ``Permuted Data". In this paper, we present an  extension of this line of research to exponential family response given the assumption of a small to moderate number of mismatches. A method based on observation-specific offsets to account for potential mismatches and $\ell_1$-penalization is proposed, and its
statistical properties are discussed. We also present sufficient conditions for the recovery of the correct correspondence
between covariates and responses if the regression parameter is known. The proposed approach is compared to established baselines, namely the methods by Lahiri-Larsen and Chambers, both theoretically and empirically based on synthetic and real data. The results indicate that substantial improvements over those methods can be achieved even if only limited information about the linkage process is available. 
\end{abstract}
\noindent%
{\it Keywords:} Record linkage, Broken Sample Problem, Generalized Linear models, Penalized Estimation, Permutation 

\newpage

\spacingset{1.4}

\section{Introduction}
A tacit assumption in regression is that response-predictor pairs correspond to the same statistical unit. In practice, this assumption is often violated at least in part when different subsets of variables are collected in an asynchronous fashion and are subsequently combined into a single data set. Roughly speaking, the latter amounts to merging multiple data sets given agreement on a set of matching variables shared across those data sets; Figure \ref{fig:record_linkage_illustration} serves as an illustration. 

This setting is well-studied in the field of \emph{record linkage},  e.g., \cite{Newcombe, Fellegi69, Christen2012, Herzog2007}. A principal reason for its importance is considerable potential in reducing efforts for data collection in the situation that a research question of interest can be answered simply by combining existing databases. In a nutshell, probabilistic record linkage is concerned with the identification of matching records, i.e., pieces of information contained in different data sets belonging to the same statistical unit, given only approximate identifiers. The uncertainty associated with those introduces two types of linkage errors, (i) \emph{missed matches} and (ii) \emph{mismatches}. The former refers to two matching records not being linked, while (ii) refers to two records erroneously linked in the sense that those records belong to different statistical units. The present work is concerned with the consequences of mismatches on subsequent regression analysis. Since the work of Neter \cite{Neter65} in 1965, it is well known that mismatches can negatively affect model fit and parameter estimation, specifically bearing a strong potential for attenuation bias. Following up on \cite{Neter65}, a variety of papers discuss strategies for bias correction in linear
regression with mismatches in the response variable given specific information about the linkage process, including work by Scheuren and Winkler \cite{Scheuren93, Scheuren97}, Lahiri and Larsen \cite{Lahiri05} and Chambers \cite{Chambers2009}. Generalizations of this line of research beyond one-to-one matching and linear models are discussed in \cite{Han2019} and \cite{KimChambers2012} via estimating equation-based approaches.  

A somewhat more direct paradigm for dealing with mismatches in the response variable originates in the ``Broken Sample Problem", a term used in a series of papers by De Groot and collaborators, e.g., \cite{DeGroot1980, DeGroot1976}. In brief, mismatches are modeled via an unknown index permutation; the latter is typically regarded as a nuisance parameter, but inference for it might be of interest for the purpose of pinpointing and correcting errors in the linkage process. While this paradigm was widely regarded as infeasible in the record literature due to the combinatorial nature and the associated computational challenges, it has recently experienced a surge of interest, fueled by advances in computing and an array
of engineering and machine learning problems that can be cast as linear regression with unknown permutation \cite{Unnikrishnan2015, wainwright_2019, Abid2017, Abid2018, SlawskiRahmaniLi2018, ZhangSlawskiLi2019, Tsakiris2018, Tsakiris2018b}. 

\begin{figure}[t]
\hspace*{2ex}
  \begin{minipage}{0.48\textwidth}
   \vspace*{1ex}
 {\small {\bfseries File A}}\\[1ex]
{\footnotesize \begin{tabular}{llclll}
\texttt{ID} & \texttt{Age} & \texttt{Sex} & \texttt{ZIP} & \texttt{Edu} & \texttt{Salary(\$)} \\
                 \hline \\[-2.15ex]
     1  & \cellcolor{gray}45  & \cellcolor{gray}F & \cellcolor{gray}47134 & Master &  6,030   \\[-0.5ex]
     2  & 36 & M & 31526 & Doctorate & 8,427 \\[-0.5ex]
     \cline{2-4}
     3  & \LL{25} & M & \RR{63312} & Bachelor & 5,616 \\ 
     \cline{2-4} \\[-4ex]
     4 & 30 & M & 17621 & High School & 3,408 \\[-0.5ex]
     5  & \cellcolor{gray}45 & \cellcolor{gray}F & \cellcolor{gray}47134  & Doctorate & 7,799 \\
     \\[-3.5ex]\cline{2-4}
     6 & \LL{25} & M &  \RR{63312} & Master & 6,500 \\
     \cline{2-4} \\[-4ex]
     7 & 55 & F & 17621 & High School & 3,266\\[-0.5ex]
     8 & 34 & F & 17621 & Bachelor & 4,084     
               \end{tabular}}
           \end{minipage}
           \begin{minipage}{0.48\textwidth}
             \vspace*{0.5ex}
           {\small {\bfseries  File B}}\\[1ex]
{\footnotesize \begin{tabular}{llccc}
\texttt{ID} & \texttt{Age} & \texttt{Sex} & \texttt{ZIP} & \texttt{weeks\_unemployed}\\
                 \hline\\[-2.15ex]
                 5 & \cellcolor{gray}45 & \cellcolor{gray}F & \cellcolor{gray}47134 & 7 \\[-0.5ex]
                 7 & 55 & F & 17621  & 21 \\[-0.5ex]
                 \\[-4ex]
                 \cline{2-4}
                 3 & \LL{25} & M & \RR{63312} & 13 \\
                 \cline{2-4}\\[-4ex]
                 2 & 36 & M & 31526  & 5 \\[-0.5ex]
                 1 & \cellcolor{gray}45 & \cellcolor{gray}F & \cellcolor{gray}47134 &  11 \\[-0.5ex]
                 4 & 30 & M & 17621 & 19 \\    
                  \\[-4ex]
                 \cline{2-4}
                 6 & \LL{25} & M & \RR{63312} & 9\\    
                 \cline{2-4}     
                 8 & 34 & F & 17621 & 15\\ 
\end{tabular}}
\end{minipage}
\vspace*{-.3ex}
\caption{Schematic illustration of the basic setting studied in this paper. Two files A and B are linked to study how a response variable contained in one file (here given by duration of unemployment in weeks) depends on covariates (here given by education level and previous monthly salary) contained in another file. File linkage based on quasi-identifiers, here given by the triple (\texttt{Age}, \texttt{Gender}, \texttt{ZIP}), can be error-prone due to ambiguities (highlighted by boxes and grey color, respectively), and bear a potential for mismatch error affecting post-linkage regression.}\label{fig:record_linkage_illustration}
\end{figure}

In this paper, we adopt the ``Broken Sample" formulation for generalized linear regression models with exponential family response \cite{Mcc1989} as an alternative to the methods in Lahiri-Larsen \cite{Lahiri05,Han2019} and Chambers \cite{Chambers2009, KimChambers2012} to account for potential mismatches in the response variable. The approach taken herein arises as a natural generalization of work by Slawski and Ben-David \cite{SlawskiBenDavid2017} for Gaussian response. An appealing property of the approach is that no information about the linkage process is required, in contrast to the methods put forth by Lahiri \& Larsen and Chambers. This can be an important advantage if file linkage has been performed by a third party, and the data analyst is only provided the linked file, a situation that is not uncommon in practice given that file linkage is often based on sensitive personal information such as names or addresses. 

In return, the number of mismatches that can be tolerated by the proposed approach is limited to at most a small linear fraction of the number of samples; while such stringent limit on the fraction of mismatches may be potentially improved upon, parameter estimation in the setting of arbitrary mismatch contamination becomes in general infeasible from both computational and statistical viewpoints \cite{Hsu2017}.     
\vskip2ex
\noindent {\bfseries Contributions}. In this paper, we study the "Broken Sample" problem, also known as "Regression with Unknown Permutation" or "Regression with Shuffled Data", for generalized linear models. While the work herein is in the same spirit as prior work on this subject, the mechanism generating mismatches is not required to be a permutation; instead, we work with the sample-to-register linkage paradigm in \cite{KimChambers2012, Chambers2019improved} in which the number of responses (observed sample) is allowed to be smaller than the number of predictors (contained in the register). As outlined above, primary interest concerns the regime of "sparsely mismatched" data in which the fraction of mismatches is subject to specific limits as elaborated in the sequel. The main technical contributions herein concern (1) restoration of the correctly matching records ("permutation recovery") for known regression parameter, (2) estimation of the regression parameter via computationally tractable schemes. In combination, (1) and (2) pave the way for "plug-in" estimation of the (generalized) permutation, thereby sidestepping the computational barriers that are associated with joint estimation. Specifically, (1) is shown to be reducible to sorting, and recovery results are derived under certain separability conditions naturally extending those for linear regression \cite{Pananjady2017, SlawskiBenDavid2017}; regarding (2), we follow the route taken in \cite{SlawskiBenDavid2017} in which sparse mismatch contamination is captured by observation-specific dummy variables and $\ell_1$-penalization whose statistical analysis is based on techniques in \cite{negahban2012unified}. The proposed approach is compared to the Lahiri-Larsen-type method in \cite{Han2019} theoretically as well as empirically by means of bike sharing data from \cite{fanaee2014event}.  
\vskip2ex
\noindent {\bfseries Related work}. There is a rapidly growing body of literature on regression with unknown permutation, starting from \cite{Unnikrishnan2015, Pananjady2016}. The paper \cite{Pananjady2016} presents necessary and sufficient conditions for permutation recovery for linear models with Gaussian design. Extensions to multivariate linear models are considered in \cite{Pananjady2017, ZhangSlawskiLi2019}. The papers \cite{Abid2017, Hsu2017} show that consistent estimation of the regression parameter is impossible without substantial additional assumptions. Tsakiris and collaborators \cite{Tsakiris2018, Tsakiris2018b} have studied important theoretical aspects such as well-posedness from an algebraic perspective, and have also put forth practical computational schemes such as a branch-and-bound algorithm (cf.~also \cite{emiya2014compressed}) and concave maximization \cite{Peng2020}. An approximate EM scheme with a Markov-Chain-Monte-Carlo (MCMC) approximation of the E-step is discussed in \cite{Abid2018, Wu1998}. The latter work in turn bears a relationship with the Bayesian approach in \cite{Gutman13} and its implementation via Gibbs sampling.  Approaches to linear and multivariate linear regression with sparsely mismatched data are studied in \cite{SlawskiBenDavid2017, SlawskiRahmaniLi2018, SlawskiBenDavidLi2019, SlawskiDiaoBenDavid2019}. 

In comparison, relatively few papers have considered regression with permuted data outside the standard linear model: examples include spherical regression \cite{Shi2018}, univariate isotonic regression and statistical seriation \cite{Carpentier2016, Weed2018, Flammarion16, Ma2020, Balabdoui2020}, and binary regression \cite{wang2018signal}. 

The method for regression parameter estimation considered herein has been studied in prior work to deal with generic contaminations in linear regression \cite{She2012}, logistic regression \cite{tibshirani2013robust} and other generalized linear models \cite{yang2013onrobust}. Unlike the present paper, none of these works contain a rigorous statistical analysis.   

Lastly, as indicated at the beginning of the introduction, there is a separate line of research focusing on parameter estimation under mismatch error in the response given at least a fair amount of knowledge about the linkage process. 
We refer to the surveys \cite{Han2019, ZhangBook}. Recovery of the underlying permutation is not considered in those works.

\subsection{Problem Statement}\label{sec:PS}
We consider a regression setup in which the response and the predictor variables are contained in two
files $F_{\M{x}} = \{ \M{x}_j \}_{j = 1}^N \subset \R^d$ and $F_y = \{ y_i \}_{i = 1}^n \subset \R$, respectively, $n \leq N$. Record linkage yields a merged file $F_{\M{x} \Join y} = \{ (\M{x}_{\ell_i}, y_i) \}_{i = 1}^n$ with $\ell_i \neq \ell_j$ for $i \neq j$, i.e., resulting from complete and one-to-one linkage of $F_{\M{x}}$ and $F_y$. Linkage is typically based on additional contextual information (matching variables); however, we generally assume herein that only the merged file $F_{\M{x} \Join y}$ is given and no
further information about the linkage process is available. The case $N = n$ applies to ``sample-to-sample" linkage, with two pieces of information pertaining to the same set of entities collected via two separate
samples; the case $N > n$ applies to ``sample-to-register" linkage \cite{Chambers2019improved}, which occurs, e.g., when linking population surveys conducted on a sample of individuals to an administrative database (see, e.g., \cite{Abowd2019} for an example of contemporary interest). Following \cite{Chambers2009, KimChambers2012, Han2019, Chambers2019improved}, we assume that each $\M{x}_j$ is associated with a corresponding latent response variable $y_j^*$, $1 \leq j \leq N$, while $y_i = y_{\pi^*(i)}^*$, $1 \leq i \leq n$, for a map $\pi^*: \{1,\ldots, n\} \rightarrow \{1,\ldots,N\}$. For simplicity, we refer to $\pi^*$ as ``permutation" even if $N > n$. The linked pair $(\M{x}_{\ell_i}, y_i)$ is called a \emph{mismatch} if $\pi^*(i) \neq \ell_i$, $1 \leq i \leq n$. Without loss of generality, we assume that $\ell_i = i$, $1 \leq i \leq n$, so that $F_{\M{x} \Join y} = \{ (\M{x}_{i}, y_i) \}_{i = 1}^n$. 

In this paper, we assume that the distribution of $y_j^*|\M{x}_j$, $1 \leq j \leq N$, follows a generalized linear model (GLM) \cite{Mcc1989}, i.e., the corresponding conditional density is given by
\begin{equation}\label{eq:expfamily}
f_j(y;\vartheta_j) = \exp\left \{\frac{y\vartheta_j - \psi(\vartheta_j)}{a(\phi)} + c(y,\phi)\right \},  
\end{equation}
where $\vartheta_j$ and $\phi$ are referred to as natural parameter and scale parameter, respectively, and 
$a(\cdot)$, $\psi(\cdot)$, and $c(\cdot)$ are all known functions referred to as scale function, cumulant, and partition function, respectively; unless stated otherwise, we assume GLMs with canonical link or canonical parameterization, i.e., 
$\vartheta_j = \eta_j \coloneq \M{x}_j^{\T} \beta^*$, $1 \leq j \leq N$. The $\{ \eta_j \}_{j = 1}^N$ are referred to as
linear predictors based on a regression parameter $\beta^*$ of interest. To simplify notation, the intercept is typically absorbed into the $\{ \M{x}_j \}_{j = 1}^N$ even though occasionally we spell out its presence by writing $\eta_j = \beta_0^* + \M{x}_j^{\T} \beta^*$, $1 \leq j \leq N$. If $\{ (\M{x}_j, y_j^*) \}_{j = 1}^N$ were given, an estimate for 
$\beta^*$ could be obtained by minimizing the following negative log-likelihood corresponding to \eqref{eq:expfamily}: 
\begin{equation}\label{eq:nloglik_oracle}
\min_{\beta \in \R^d} \ell^*(\beta), \qquad \ell^*(\beta) \coloneq -\sum_{j = 1}^N \{ y_j^* \M{x}_j^{\T} \beta  - \psi(\M{x}_j^{\T} \beta) \}. 
\end{equation}
However, inference for $\beta^*$ based on the merged file $\{ (\M{x}_i, y_i) \}_{i= 1}^n$ is in general far from straightforward due to the presence of mismatches. It is well known that the na\"ive approach that amounts to 
substitution of $\{ (\M{x}_j, y_j^*) \}_{j = 1}^N$ in \eqref{eq:nloglik_oracle} by $\{ (\M{x}_i, y_i) \}_{i = 1}^n$ can exhibit massive bias even if the fraction of mismatched pairs is small. A natural alternative is to consider the 
joint negative log-likelihood of both $\beta^*$ and the unknown permutation $\pi^*$:
\begin{equation}\label{eq:nloglik_joint}
\min_{\beta \in \R^d, \, \pi  \in \mc{P}(n,N)} \ell(\beta, \pi), \qquad \ell(\beta, \pi) \coloneq -\su \{ y_i \M{x}_{\pi(i)}^{\T} \beta  + \psi(\M{x}_{\pi(i)}^{\T} \beta) \},
\end{equation}
where $\mc{P}(n,N) = \{\pi: \; \{1, \ldots, N\} \rightarrow \{1,\ldots,n\}, \; \pi \, \text{is injective}  \}$; we shall use
$\mc{P}(n)$ as a shorthand for all permutations of $\{1,\ldots,n \}$. 

Formulation \eqref{eq:nloglik_joint} involves several obstacles. First, if $n < N$, $F_{\M{x}}$ would need to be given in order to evaluate $\su \psi(\M{x}_{\pi(i)}^{\T} \beta)$; by contrast, if $n = N$, the sum does not depend on $\pi$ since the latter becomes a proper permutation.  Recent results on the case $n = N$ and the linear model, which corresponds to $\psi(z) = z^2/2$, imply that the optimization problem \eqref{eq:nloglik_joint} is intractable \cite{Pananjady2016, Hsu2017}. Moreover, putting computational intractability aside, the minimizer of \eqref{eq:nloglik_joint} fails to yield consistent estimators of $\beta^*$ or $\pi^*$ without suitable lower bounds on $\nnorm{\beta^*}_2^2/\phi^2$ \cite{Pananjady2016, Hsu2017, SlawskiBenDavid2017, Abid2017}. 

An alternative viewpoint is to think of $\pi^*$ as a random 
variable depending on contextual information used for linking $F_{\M{x}}$ and $F_y$, and to focus on inference 
for $\beta^*$. At a high level, this is the strategy adopted in \cite{KimChambers2012, Lahiri05, Han2019}. The success
of this line of work shows that it is well possible to obtain accurate estimators of $\beta^*$ if, loosely speaking, the
distribution of $\pi^*$ is concentrated on a subset of $\mc{P}(n,N)$ of manageable size. As elaborated below, the effectiveness of this approach can be particularly well understood in the situation that $\pi^*$ is known to ``block-structured" into a good number of blocks, where the blocks arise from contextual information (cf.~Figure \ref{fig:record_linkage_illustration}). In the absence of the latter, a similar reduction can be achieved under the assumption that mismatches occur sparsely in $F_{\M{x} \Join y}$ in the sense that $\pi^*(i) \neq i$ is the exception rather than the rule, that is the fraction $k/n$ is ``small", where $k = |\{i:\,\pi^*(i) \neq i \}|$ denotes the number of mismatches. This assumption is often justifiable given that record linkage tends to provide largely accurate albeit not perfect matchings, particularly if rich contextual information is available when linking $F_{\M{x}}$ and $F_y$ even though such information may not be 
available to the analyst of $F_{\M{x} \Join y}$.

While past work on the subject has predominantly focused on estimation of the regression parameter, there is a clear
path towards estimating the permutation $\pi^*$ from $F_{\M{x} \Join y}$ in the case $n = N$. In fact, for any fixed $\beta$, the optimization problem in $\pi$ only, i.e., 
\begin{equation}\label{eq:nloglik_pi}
\min_{\pi  \in \mc{P}(n)} -\su \{ y_i \M{x}_{\pi(i)}^{\T} \beta  + \psi(\M{x}_{\pi(i)}^{\T} \beta) \} = \min_{\pi \in \mc{P}(n)} -\su y_i \M{x}_{\pi(i)}^{\T} \beta + c,
\end{equation}
where $c = \su \psi(\M{x}_i^{\T} \beta)$, is a specifically simple linear assignment problem \cite{Burkard2009} whose solution reduces to sorting \cite{DeGroot1980, Pananjady2017}. This observation suggests the estimation of $\pi^*$ based 
on \eqref{eq:nloglik_pi} with $\beta$ replaced by an estimator of $\beta^*$. The accuracy of this scheme has not been studied for generalized linear models with the exception of the Gaussian linear model \cite{Pananjady2017, SlawskiBenDavid2017}. Below, we present some first insights into this questions for other selected generalized 
linear models. 

\vskip2ex
\noindent {\bfseries Outline}. In \S\ref{sec:Est_reg}, we present our approach for estimation of the regression parameter, and investigate its properties theoretically as well as empirically via simulations. Section \S\ref{sec:PR} is devoted to permutation recovery for known regression parameter, i.e., \eqref{eq:nloglik_pi} above with $\beta = \beta^*$. A comparison of the proposed approach and the methods by Lahiri-Larsen and Chambers is provided in \S\ref{sec:Com_LL} and \S\ref{sec:realdata}, which also contains a case study on real data. We conclude with a summary and an overview on potential directions of future research in \S \ref{sec:Con}.

\subsection{Notation}
For the convenience of the reader, we here collect essential notations used in this paper. For a positive integer $\ell$, $\M{1}_{\ell}$ and $I_{\ell}$ denote the vector of ones and the identity matrix, respectively, of dimension $\ell$. The $n$-by-$d$ design matrix associated with covariates $\{ \M{x}_i \}_{i=1}^n$ contained in the merged file $F_{\M{x} \Join y}$ is denoted by $\M{X}$; unless noted otherwise, we assume that $\M{X}$ includes the column for the intercept (i.e., $\M{X} = [\M{1}_{n} \; \M{X}_0]$). Likewise, the values for the response in $F_{\M{x} \Join y}$ are collected in a vector $\M{y}= (y_i)_{i = 1}^n$. The function $\mathbb{I}(\cdot)$ represents the indicator function with value one if its argument is
true and zero else. With some abuse of notation, if $f$ is a function of a single argument and $\M{v}$ is a vector of dimension $\ell$, we write $f(\M{v})$ for $(f(v_1),\ldots,f(v_{\ell}))^{\T}$. We let $a \vee b = \max\{a,b\}$ and $a \wedge b = \min\{a,b\}$. We make use of the usual Landau notation in terms of $O$, $o$, $\Omega$ and $\Theta$. We often use $a \lesssim b$, $b \gtrsim a$, and $a \asymp b$ as shortcuts for $a = O(b)$, $b = \Omega(a)$ and $a = \Theta(b)$, respectively. Numerical constants are denoted by $C, C', C_1, c, c_1$ etc.~whose values may change from line to line.
We use the symbols $\eta$ and $\mu$ (with varying subscripts) to refer to the linear and conditional expectation of the response $y$ given covariates $\M{x}$. The associated mappings (the link function and its inverse) are denoted by $g$ and $h$, respectively, as depicted in the diagram below.  
\begin{center}
\fbox{\begin{tikzpicture}
  \draw (6,-.1) node(mu2) {\textcolor{white}{{ $\mu = \E[y|\M{x}]$}}};
   \draw (6,.1) node(mu) {$\mu = \E[y|\M{x}]$};
   \draw (0,-.1) node(LP2) {\textcolor{white}{$\eta = \M{x}^{\T} \beta$}};
   \draw (0,.1) node(LP) {$\eta = \M{x}^{\T} \beta$};
  \draw[->,very thick] (LP) -- (mu) node[midway, above] {{\scriptsize $h(\cdot) = \psi'(\cdot)$}};
  \draw[<-,very thick] (LP2) -- (mu2) node[midway, below] {{\scriptsize $g(\cdot) = h^{-1}(\cdot)$}};
\end{tikzpicture}}
\end{center}
\section{Estimation of the regression parameter}\label{sec:Est_reg}
In this section, we formulate our approach for estimating the regression parameter in GLMs in the presence of mismatch error, i.e., in the setting outlined in $\S$\ref{sec:PS}. A bound on the $\ell_2$-estimation error of the proposed approach is presented subsequently, which is complemented by numerical results based
on simulated data. 
\subsection{Approach}
Defining $o_i^*  = (\M{x}_{\pi^*(i)} - \M{x}_i)^{\T} \beta^*$, we have that 
$y_i | \M{x}_i, o_i^*$ follows a GLM with linear predictor $\eta_{\pi^*(i)} = \M{x}_i^{\T} \beta^* + o_i^*$, $1 \leq i \leq n$. Clearly, $\pi^*(i) = i$ implies that $o_i^* = 0$ and in turn $\eta_{\pi^*(i)} = \eta_i = \M{x}_i^{\T} \beta^*$, $1 \leq i \leq n$. Accordingly, the underlying idea is to a fit a generalized linear
model in which each linear predictor is augmented by an individual ``dummy variable" or ``offset" in order to account for potential mismatches. Without additional constraints or regularization, such approach is not meaningful since it is overparameterized and trivially achieves perfect fit. However, in a sparse mismatch regime with 
$\pi^*(i) = i$ holding for all except for $k$ indices, the use of sparsity-promoting penalties like the $\ell_1$-penalty becomes a natural choice. This gives rise to the following formulation: for $\theta \in \R^{d + n}$, we consider the partitioning $\theta = [\beta^{\T} \; \xi^{\T}]^{\T}$ with 
$\beta \in \R^d$ and $\xi \in \R^n$. We then consider estimation based on minimizing the penalized negative log-likelihood given by 
\begin{equation}\label{eq:objective_theory}
\ell_{\text{pen}}(\theta) = \ell(\theta) + \lambda \nnorm{\xi}_1, \qquad \ell(\theta) \coloneq \frac{1}{n} \left\{ -\nscp{\M{X} \beta + \sqrt{n} \xi}{\M{y}} + \su \psi(\M{x}_i^{\T} \beta + \sqrt{n} \xi_i) \right \},    
\end{equation}
where $\M{X}$ is the usual design matrix with rows $\{ \M{x}_i^{\T} \}_{i = 1}^n$, $\M{y} = (y_i)_{i = 1}^n$, 
and $\lambda \geq 0$ is a tuning parameter whose choice will be discussed below. In \eqref{eq:objective_theory}, the dummy variables $\xi = (\xi_i)_{i = 1}^n$ are in correspondence to the $\{ o_i^* \}_{i = 1}^n$ above; re-scaling by $n^{-1/2}$ is done exclusively for technical reasons, since this choice turns out to be convenient for the theoretical analysis of \eqref{eq:objective_theory} presented below. 

Formulations of the form \eqref{eq:objective_theory} or similar have been considered in prior work in different contexts. The use of dummy variables to deal with data contamination in linear models has been discussed in She and Owen \cite{She2012}, Laska et al.~\cite{Laska2009}, Nguyen \& Tran \cite{Nguyen2013}, and more recently in Bhatia et al.~\cite{Bhatia2017}. Extensions to generalized linear models have been proposed in \cite{tibshirani2013robust, yang2013onrobust, liu2019minimizing}. Slawski \& Ben-David \cite{SlawskiBenDavid2017} study and analyze this approach in detail for mismatch contamination, and the present paper arises as a direct extension of their work. It is worth emphasizing that despite prior work on the formulation \eqref{eq:objective_theory}, the latter has not been studied specifically for mismatch contamination. Moreover, none of the earlier works on \cite{tibshirani2013robust, yang2013onrobust, liu2019minimizing} contain a complete theoretical analysis as 
provided herein.

We note in passing that \eqref{eq:objective_theory} can be applied broadly in situations beyond linkage of $F_{\M{x}}$ and $F_y$. For example, rather common situations are (i) a subset of the covariates is contained in the same file in the response, (ii) file linkage involves more than two files, each containing different subsets of the covariates and/or the response. Both (i) and (ii) can be addressed via \eqref{eq:objective_theory} based on suitable choices of the variables $\{ o_i^* \}_{i = 1}^n$. 
\subsection{Computation}
There are various ways of solving the convex optimization problem \eqref{eq:objective_theory}. A particularly suitable approach that exploits structure specific to \eqref{eq:objective_theory} is block coordinate descent with blocks formed by $\beta$ and $\xi$, respectively. The key observation is that for any fixed value 
of $\beta$, minimization over $\xi$ can be performed in closed form via a soft thresholding-type update \cite{DonohoJohnstone1994}. On the other hand, note that for any fixed $\xi$ minimization with respect to $\beta$ amounts to fitting the underlying GLM with offset $\sqrt{n}\xi_i$ for observation $i$, $1 \leq i \leq n$. Since the proposed algorithm is already iterative, alternating between updates of $\beta$ and $\xi$, we only perform a single iteration of weighted least squares (aka Fisher Scoring) when updating $\beta$; this is equivalent to minimizing the quadratic Taylor approximation of the objective (with $\xi$ treated as fixed) around the current iterate $\wh{\beta}^{(t)}$. A schematic description of the algorithm is provided below.  
\begin{algorithm}[tbh]
\caption{Block coordinate descent algorithm for \eqref{eq:objective_theory}}
Initialize $\wh{\xi}^{(0)} = 0$ and $\wh{\beta}^{(0)}$ as the ordinary GLM estimate based on $(\M{X}, \M{y})$.\par 
\textbf{1. Update for  $\xi$:}
\begin{align*}
    & \wh{\xi}^{(t+1)}_{i} \leftarrow \mathbb{I} \left \{\frac{|y_i - \wh{\mu}_i^{(t)}|}{\sqrt{n}} - \lambda > 0 \right \}  \cdot \frac{\left( (\psi^{'})^{-1}(y_i - s_i \lambda ) - \wh{\eta}_{i}^{(t)} \right)}{\sqrt{n}}, \; s_i = \text{sign}(y_i - \wh{\mu}_i^{(t)})\sqrt{n}, \quad 1 \leq i \leq n.
\end{align*}
\textbf{2. Update for $\beta$ :}
\begin{equation*}
    \wh{\beta}^{(t+1)} \leftarrow \wh{\beta}^{(t)} + (\M{X}^{\T} W^{(t)} \M{X})^{-1} \M{X}^{\T} (\M{y} - \psi^{'}(\M{X}\wh{\beta}^{(t)} + \sqrt{n}\wh{\xi}^{(t+1)})) 
\end{equation*}
where $ \wh{\eta}_i^{(t)} = \x^{\T}_{i}\wh{\beta}^{(t)}$, $\wh{\mu}_i^{(t)} = \psi^{'}(\wh{\eta}_{i}^{(t)})$, $V_i^{(t)} = \psi^{''}(\wh{\eta}_{i}^{(t)} + \sqrt{n}\wh{\xi}_{i}^{(t+1)})$, $1 \leq i \leq n$, and $W^{(t)} = \text{diag}\{ V_i^{(t)}\}_{i = 1}^n$.
\end{algorithm}

Given extensive numerical experience, Algorithm 1 converges in practice after few iterations. In order to establish convergence theoretically, the two updates above would need to be combined with a suitable mechanism for step size selection \cite{Bertsekas1999}. Since the latter is standard in the optimization literature, we refrain from discussing this aspect in detail to avoid digressions.   

\subsection{Incorporating blocking variables}\label{subsec:blocking}
Recall that $o_i^* = (\M{x}_{\pi^*(i)} - \M{x}_i)^{\T} \beta^*$, $1 \leq i \leq n$. Note that if $N = n$ so that 
$\pi^*$ is a permutation, it is easy to see that $\su o_i^*  = 0$. As a result, the additional constraint 
$\su \xi_i = 0$ may be added to the optimization problem \eqref{eq:objective_theory}. This simple observation 
can be put to much more use if $\pi^*$ is known to be ``block-structured" in the sense that the data set
can be partitioned into disjoint groups $G_1, \ldots, G_K \subset \{1,\ldots,n\}$ such that 
$i \in G_j$ for some $j \in \{1,\ldots, K\}$ implies that $\pi^*(i) \in G_j$, $1 \leq i \leq n$; in other words,
the permutation only moves indices within, but not across groups. With the same reasoning as above, we then 
have $\sum_{i \in G_j} o_i^* = 0$, which accordingly yields the constraints 
$\sum_{i \in G_j} \xi_i = 0$, $1 \leq j \leq K$, to be added to \eqref{eq:objective_theory}. Specifically, this yields the following optimization problem 
\begin{equation}\label{eq:objective_constrained}
\min_{\theta} \ell(\theta) + \lambda \nnorm{\xi}_1 \; \; \text{subject to} \; \M{C}\xi = \M{0},
\end{equation}
where $\M{C} \in \R^{K \times n}$ has entries $C_{ji} = 1$ if $i \in G_j$ and zero otherwise, $1 \leq j \leq K, \, 1 \leq i \leq n$. 
In particular, in the case of singleton groups with $G_j = \{ i \}$ for $i \in \{1,\ldots,n\}$, it immediately follows that $\xi_i = 0$ and the corresponding variable can be eliminated in \eqref{eq:objective_constrained}. If $K$ is large, 
this yields a substantial number of extra constraints whose integration in \eqref{eq:objective_constrained} can considerably boost performance relative to the unconstrained minimizer not taking any advantage of the block
structure of $\pi^*$. The latter arises when additional knowledge about the linkage process is available. More specifically, it is common that matching records are known to agree on certain combinations of variables present for the records in both $F_{\M{x}}$ and $F_y$ (e.g., demographic variables such as gender, age, and/or race, approximate geographical location based on ZIP code, approximate time stamps, etc.). Those variables are typically referred to as blocking variables, and the corresponding groups $\{ G_j \}_{j = 1}^K$ are given by subsets of observations
sharing the same values for all blocking variables. 

In summary, while the approach \eqref{eq:objective_theory} works without any knowledge about the linkage process and the existence of blocking variables, it is possible to achieve enhancements if such information is available. This aspect is investigated in detail in $\S$\ref{sec:realdata}.

\subsection{Analysis}
In the sequel, we derive a non-asymptotic upper bound on the $\ell_2$-error $\nnorm{\wh{\theta} - \theta^*}_2$, where
$\theta^* = [\beta^{*\T} \; \xi^{*\T}]^{\T}$ with $\xi_i^* = \frac{1}{\sqrt{n}} (\M{x}_{\pi^*(i)} - \M{x}_i)^{\T} \beta^*$ and 
$\wh{\theta} = [\wh{\beta}^{\T} \; \wh{\xi}^{\T}]^{\T}$ denotes a minimizer of $\ell_{\text{pen}}$ in \eqref{eq:objective_theory}. Before stating the final result in Theorem \ref{theo:estimation_error}, 
we present and discuss the assumptions underlying that result. 
\vskip2ex
\noindent {\bfseries Assumptions and Conditions}. 
\begin{itemize}
\item[({\bfseries A})] The rows $\{ \M{x}_i \}_{i = 1}^n$ of $\M{X}$ are i.i.d.~copies of a random vector $\M{x}$ with the 
            following properties: 1) there exists a positive definite matrix $\Sigma$ with uniformly bounded eigenvalues, i.e., $\sigma_{\min} I_d \lec \Sigma \lec \sigma_{\max} I_d$, such that $\scp{v}{\Sigma^{-1/2} \M{x}}$ is
            a sub-Gaussian random variable with sub-Gaussian norm at most $K$ for all $v \in \R^d$\symfootnote{cf., e.g.,\cite[$\S$2.5]{Vershynin2018} for a concise discussion of sub-Gaussian random variables and their properties.}, 2) there exists a constant $r > 0$ such that $\p(\nnorm{\M{x}}_2 \leq r) = 1$.
\end{itemize}            
We note that ({\bfseries A}) allows for the inclusion of an intercept by requiring that the first component of $\M{x}$ equals
one with probability one. In addition, the entries of $\M{x}$ neither need to be mean zero nor uncorrelated; with $\Sigma = \E[\M{x} \M{x}^{\T}]$ chosen as the population second moment matrix, only uniform upper and lower bounds
for its eigenvalues are required. The condition that the essential support of $\M{x}$ is contained in an $\ell_2$-ball of bounded radius 
is imposed to ensure bounded linear predictors, as further elaborated below. Note that if $\M{x}$ is sub-Gaussian with unbounded support (e.g., $\M{x} \sim N(0,I_d)$), the truncation $T(\M{x}) \coloneq \M{x} \mathbb{I}(\nnorm{\M{x}}_2 \leq r) + \frac{r \M{x}}{\nnorm{\M{x}}_2} \mathbb{I}(\nnorm{\M{x}}_2 > r)$ conforms with $(\M{A})$.    
\begin{itemize}           
\item[({\bfseries C1})] There exist sequences $\nu_n, \epsilon_n = o(1)$ as $n \rightarrow \infty$ such that 
                        $\nnorm{\nabla \ell(\theta^*)}_{\infty} \leq \nu_n$ with probability at least $1 - \epsilon_n$. 
\item[({\bfseries C2})] There exists a sequence $\epsilon_n' = o(1)$ as $n \rightarrow \infty$ and constants $R > 0$, $0 < \lambda_R \leq \Lambda_{R} < \infty$ such that with probability at least $1 - \epsilon_n'$
                        \begin{align*}
                        &\min_{1 \leq i \leq n} \min_{u: \nnorm{u}_2 \leq R} \psi_i''(\theta^* + u) \geq \lambda_{R}, \qquad
                        \frac{32}{\sigma_{\min} \, \lambda_{R} } (\lambda + \nu_n)\sqrt{d + k} \leq R, (\#)\\
                        &\max_{1 \leq i \leq n} \max_{u: \nnorm{u}_2 \leq R} \psi_i''(\theta^* + u) \leq \Lambda_{R}. 
                        \end{align*}
where $\theta \mapsto \psi_i(\theta) \coloneq \psi(\M{x}_i^{\T} \beta + \sqrt{n} \xi_i)$ and $\psi_i''(\theta) \coloneq \frac{d^2}{dz^2} \psi(z) \dev{z}{\M{x}_i^{\T} \beta + \sqrt{n} \xi_i}$, $1 \leq i \leq n$. 
\end{itemize}
Condition ({\bfseries C1}) holds under assumption ({\bfseries A}) with $\nu_n = C \sqrt{\frac{\log (d + n)}{n}}$ and  
 $\epsilon_n = c/n$ if additionally  at least one of the following two properties holds \cite{negahban2012unified}:
\begin{itemize}
\item[(i)] $\psi''$ is uniformly bounded,
\item[(ii)] $\E[\max_{|u| \leq 1} \psi''(\M{x}^{\T} \beta^* + u)^\alpha] \leq B$ for some $\alpha \geq 2$, $1 \leq i \leq n$. 
\end{itemize} 
Property (i) is satisfied for logistic regression, while property (ii) is satisfied for Poisson regression if $\nnorm{\beta^*}_2$ is 
uniformly bounded, which together with the boundedness assumption $\p(\nnorm{\M{x}}_2 \leq r) = 1$ in ({\bfseries A}) implies that the linear predictor is uniformly bounded. Boundedness of the $\{ \M{x}_i \}_{i = 1}^n$ and 
of $\beta^*$ is also needed for ({\bfseries C2}) to hold since $\psi''$ can generally only be lower and upper bounded on a compact 
interval; as a result, similar boundedness assumptions are commonly imposed in the literature (e.g., \cite[Definition 8.1]{Koltchinskii2011}, \cite[Example 1]{VandeGeer2008}). 

Note that in ({\bfseries C2}), a valid radius $R$ needs to satisfy the 
condition (\#) for the error bound in Theorem \ref{theo:estimation_error} below not to be vacuous. With the choice $\lambda \asymp \nu_n$ as indicated by that theorem, and the scaling $\nu_n \lesssim \sqrt{\log(d + n)/n}$ as discussed above, (\#) is of the form 
\begin{equation}\label{eq:lattenzaun}
\frac{C'}{\sigma_{\min} \lambda_R} \sqrt{\frac{\log\{d + n\} (d+k)}{n}} \leq R,
\end{equation}
which translates to a condition on the sample size of the form $n \geq C_R \, \sigma_{\min}^2 \, \log\{d+n\} \, (d+k)$, where $C_R$ is a constant depending on $R$. In turn, the latter condition restricts the number of mismatches $k$ to be at most of the order 
$n / \log n$. The left hand side of \eqref{eq:lattenzaun} coincides with the bound on the $\ell_2$-estimation error stated in the theorem below, which asserts consistent estimation given $\frac{\log \{ d + n\} \, (d+k)}{n} \rightarrow 0$ and $\nu_n$, $\lambda$ scaling as discussed above. 
\begin{theo}\label{theo:estimation_error} Suppose that assumption \emph{({\bfseries A})} and conditions \emph{({\bfseries C1})}, \emph{({\bfseries C2})} hold. Consider any minimizer of $\wh{\theta}$ of $\ell_{\text{\emph{pen}}}$ in with $\lambda = \lambda_n$ chosen such that $2\nu_n \leq \lambda \leq C \nu_n$ for some $C > 0$. Then there exists constants $C' > 0$, $c \in (0,1)$, so that if $n \geq C'\,\frac{\sigma_{\max}}{\sigma_{\min}} \cdot \left(\frac{\Lambda_R}{\lambda_R} \right)^2 \big\{ (d+k) \log\left(\textstyle\frac{n}{d+k} \right) \vee \log n \big\}$, 
it holds that 
\begin{equation*}
\nnorm{\wh{\theta} - \theta^*}_2  \leq \inf \left \{0 < r \leq R: \; r > \frac{32}{\sigma_{\min} \, \lambda_{R}} (\lambda + \nu_n)\sqrt{d + k}  \right \} 
\end{equation*}
with probability at least $1 - \epsilon_n - \epsilon_n' - 2/n$.
\end{theo}
\noindent In addition to the condition on the sample size implied by \eqref{eq:lattenzaun}, the above statement involves a second
condition on $n$ which is slightly less stringent in terms of the required ratio $n/(d+k)$, but additionally involves
a dependency on the condition number of $\Sigma$ and the Hessian of $\psi$ in a neighborhood around the true parameter, none of which, however, entails any additional requirements beyond ({\bfseries A}), ({\bfseries C1}), and ({\bfseries C2}).


\subsection{Simulations}
We here present selected simulation results that are intended to corroborate and complement the analysis of the preceding section. 
Data is generated according to model \eqref{eq:expfamily} with $n = 1000$, $d = 50$, and the entries of $\beta^*$ are drawn i.i.d.~from the $N(0,1)$-distribution, and subsequently normalized so that the $\ell_2$-norm equals a specific value (see below). The entries of the design matrix are drawn from the uniform distribution on the interval $[-\sqrt{3},\sqrt{3}]$\footnote{For space reasons, we here only present results for this specific class of random designs. Additional simulation results contained in a supplementary file show that the outcome is similar for a wide range of random designs.}. The following distributions for the response are considered. 
\begin{itemize}
\item \emph{Poisson}:  intercept $\beta_{0}^* = 2$, $\nnorm{\beta^*}_2 \in \{0.5, 1, 2\}$.
\item \emph{Binomial}: the number of Bernoulli trials per observation is fixed as $m = 25$, $\beta_{0}^* = 2$, and $\nnorm{\beta^*}_{2} \in \{0.5,1,2\}$; the case of binary response ($m = 1$) is discussed in a dedicated paragraph. 
\item \emph{Gamma}: the shape parameter is fixed as $\nu = 50$ (or equivalently, the dispersion parameter $\phi$ is set to $1/\nu$), $\beta_{0}^* = \{2, 4, 8\}$ and $\nnorm{\beta^*}_{2} \in \{1, 2, 4\}$. 
\end{itemize}
The map $\pi^*$ is drawn uniformly at random from the set of permutations on $\{1,\ldots,n\}$ that move exactly $k$ indices, where $k/n \in \{0.05, 0.1, \ldots, 0.4\}$\footnote{This can be achieved by first selecting a random subset of size $k$, and then generating a random permutation of that subset (if the resulting permutation happens to have a fixed point, it is rejected and drawn again).}. 

\begin{figure}
\begin{center}
\begin{tabular}{ccc}
    Poisson & Binomial & Gamma \\ 
     \hspace*{-3.5ex} \includegraphics[width = 0.33\textwidth]{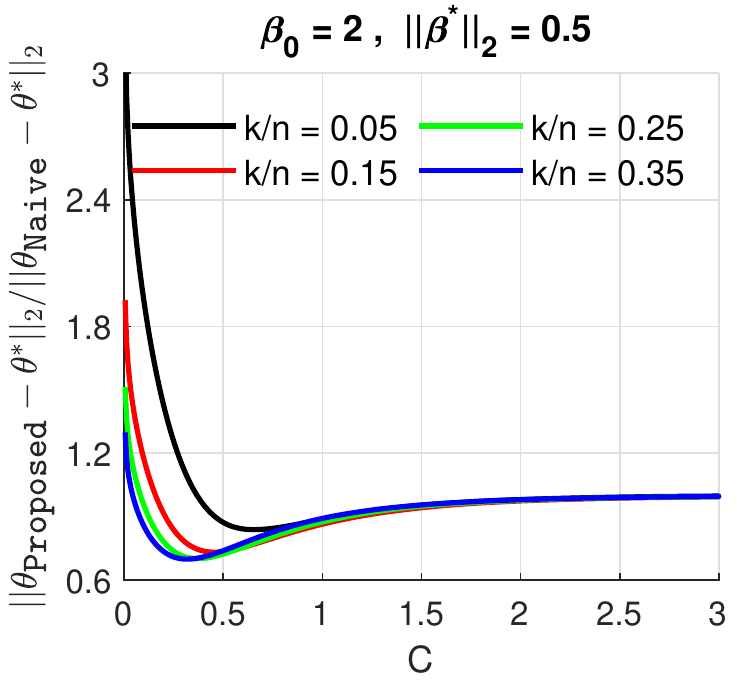}
    & \hspace*{-3.5ex} \includegraphics[width = 0.33\textwidth]{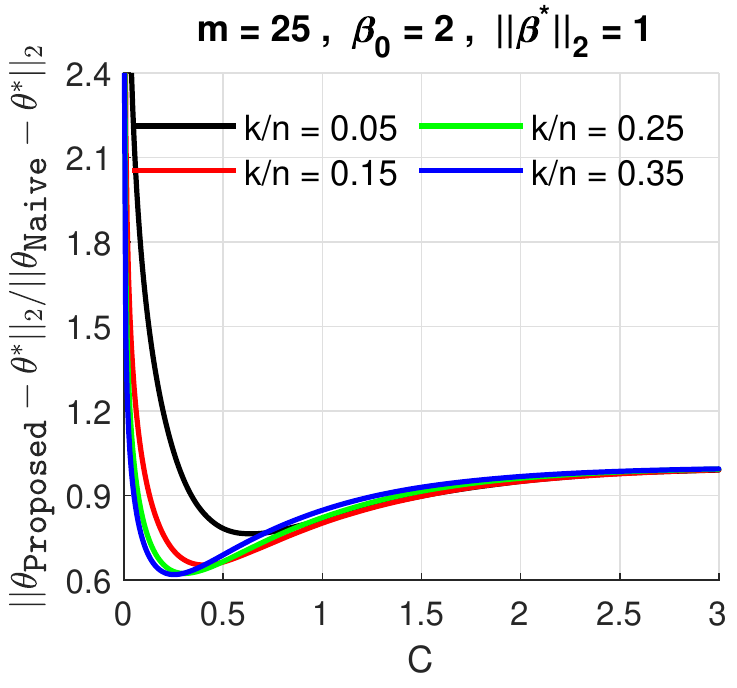} 
   & \hspace*{-3.5ex} \includegraphics[width = 0.33\textwidth]{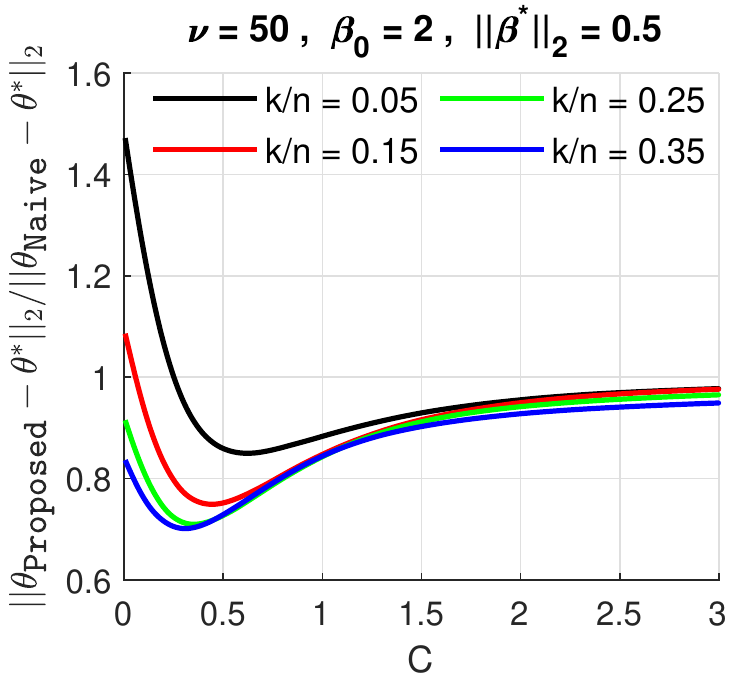}\\
    \hspace*{-3.5ex} \includegraphics[width = 0.33\textwidth]{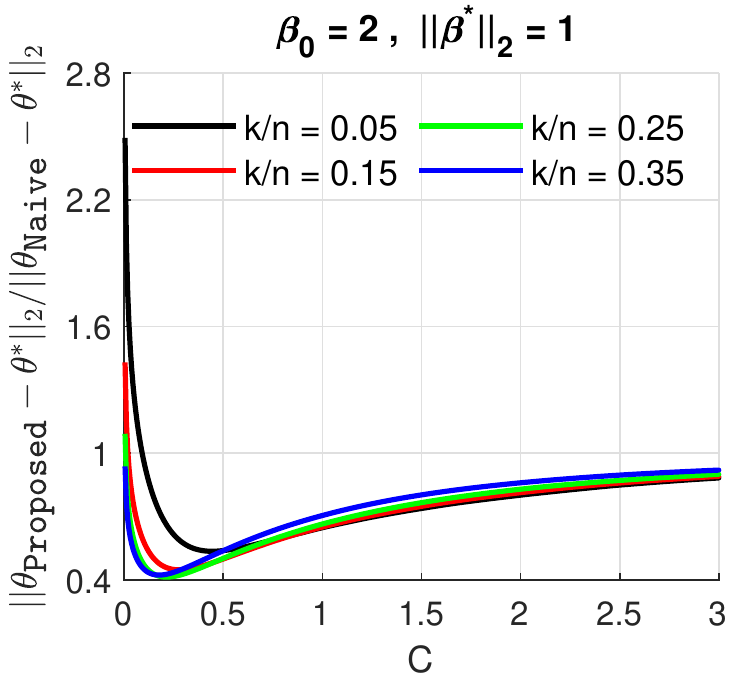}
    & \hspace*{-3.5ex} \includegraphics[width = 0.33\textwidth]{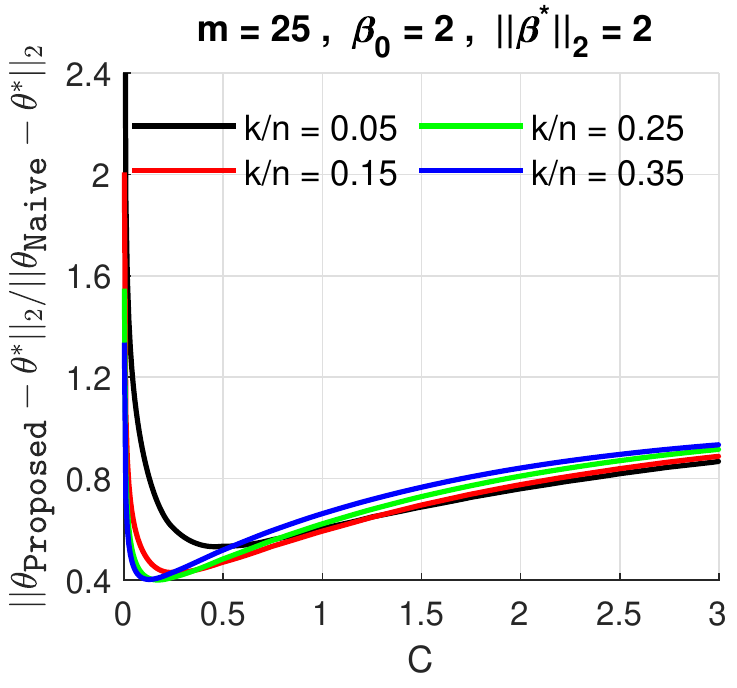} 
   & \hspace*{-3.5ex} \includegraphics[width = 0.33\textwidth]{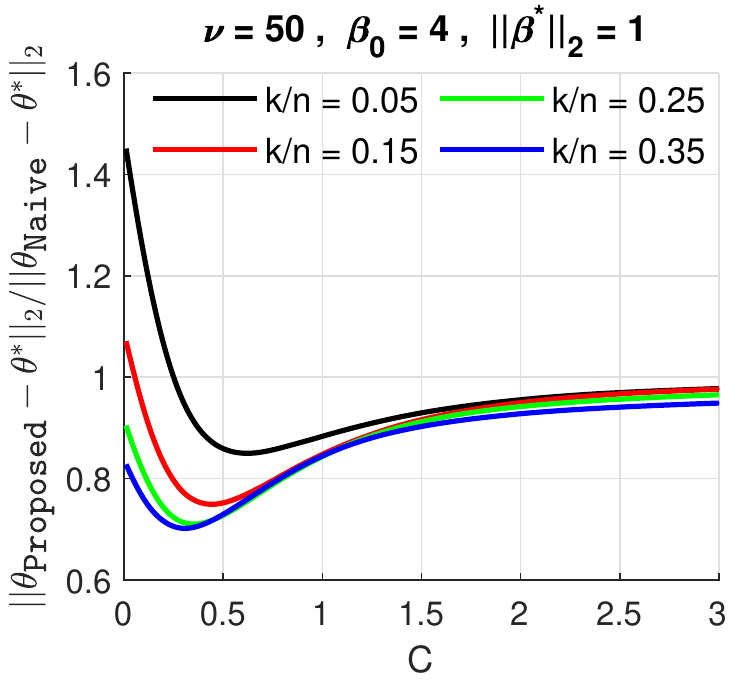}\\
    \hspace*{-3.5ex} \includegraphics[width = 0.33\textwidth]{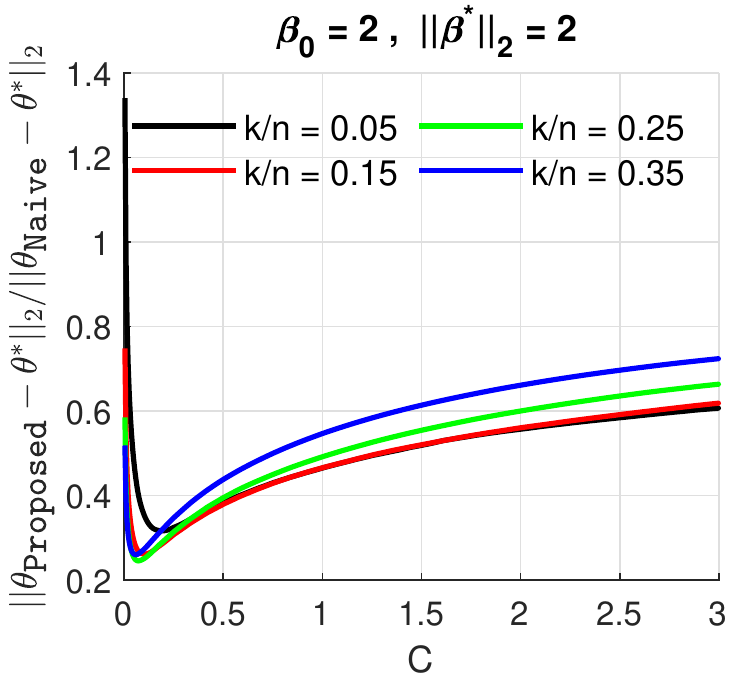}
    & \hspace*{-3.5ex} \includegraphics[width = 0.33\textwidth]{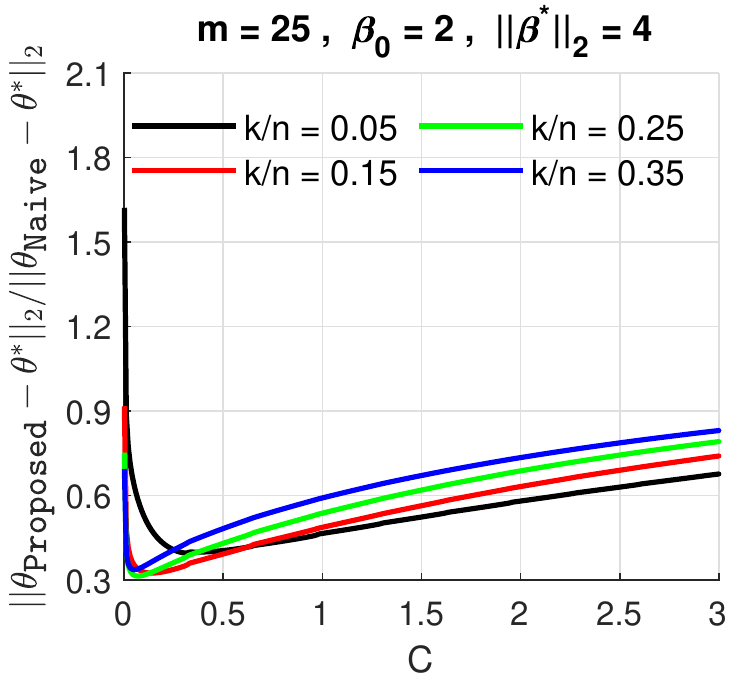} 
   & \hspace*{-3.5ex} \includegraphics[width = 0.33\textwidth]{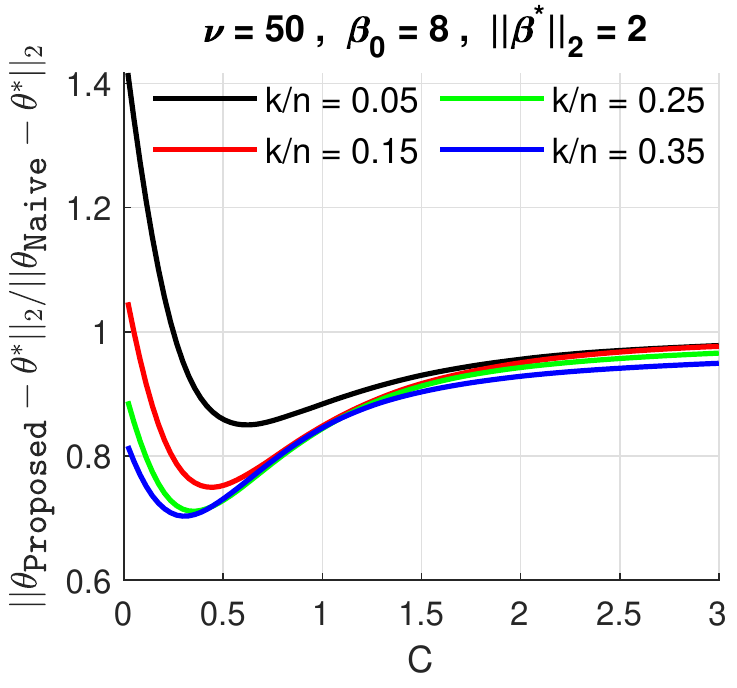}
\end{tabular}
\end{center}
\vspace*{-3ex}
\caption{Estimation error ratios $\nnorm{\wh{\theta} - \theta^*}_{2} / \nnorm{\wh{\theta}^{\text{naive}} - \theta^*}_{2}$ in dependence of the pre-factor $C$ appearing in the tuning parameter $\lambda$.} \label{fig:lam}
\end{figure}

In alignment with Theorem \ref{theo:estimation_error}, the regularization parameter $\lambda$ is chosen as $\lambda = C \cdot \sigma_{y} \cdot \sqrt{\frac{\log (n+d)}{n}}$, where $C$ is referred to as ``pre-factor" and $\sigma_y$ is a
calibration factor depending on the distribution of the response. The calibration factor is chosen as an approximation of the expected standard deviation of the response variables $\{ y_i \}_{i = 1}^n$, where the expectation is taken with respect to the random predictors\footnote{This expectation is evaluated using numerical integration, approximating the linear predictor by a $N(\beta_0^*, \nnorm{\beta^*}_2^2)$-random variable justified by the central limit theorem.}. The use of 
$\sigma_y$ is motivated by results on linear regression in \cite{SlawskiBenDavid2017} and an analysis of the $\ell_2$-estimation error for $\xi^*$ if $\beta^*$ were \emph{known} (omitted for space reasons). In practice, $\sigma_y$ can be approximated by taking the average of the variance function
of the corresponding GLM evaluated at the $\{ y_i \}_{i = 1}^n$; in order to enable Figure \ref{fig:lam} that is specifically dedicated to the selection of $\lambda$, the factor $\sigma_y$ is fixed so that it does not vary across different randomly generated data sets.  

For each triplet $(\beta_{0}^*,\nnorm{\beta^*}_{2},  k/n)$, 100 independent replications are considered. 
The following approaches are compared. \par
\noindent \textsf{``naive}". Plain GLM estimation based on $\{(\x_{i}, y_{i}\})^{n}_{i=1}$ without adjustment for mismatches. Note that the fitted values of this approach coincide with those
of the proposed approach (see below) if $\lambda \in (\lambda_{\max}, \infty)$, where $\lambda_{\max} = \nnorm{\nabla_{\xi} \ell(\wh{\theta}^{\text{naive}})}_{\infty}$ and $\wh{\theta}^{\text{naive}} = ((\wh{\beta}^{\text{naive}})^{\T} \; \M{0}^{\T}_{n})^{\T}$ with $\wh{\beta}^{\text{naive}}$ denoting
the plain GLM estimate; this is an immediate consequence of the KKT conditions of \eqref{eq:objective_theory}. 
\par
\noindent \textsf{oracle}. Plain GLM estimation based on the mismatch-free data $\{(\x_{i}, y^*_{i})\}^{n}_{i=1}$. \par
\begin{figure}
\begin{center}
\begin{tabular}{ccc}
    Poisson & Binomial & Gamma \\ 
     \hspace*{-3.5ex} \includegraphics[width = 0.33\textwidth]{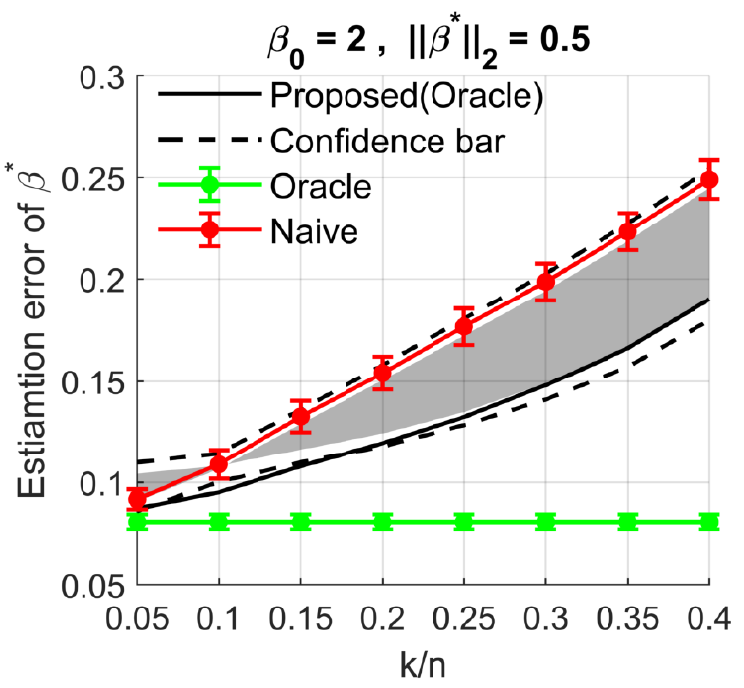}
    & \hspace*{-1.5ex} \includegraphics[width = 0.33\textwidth]{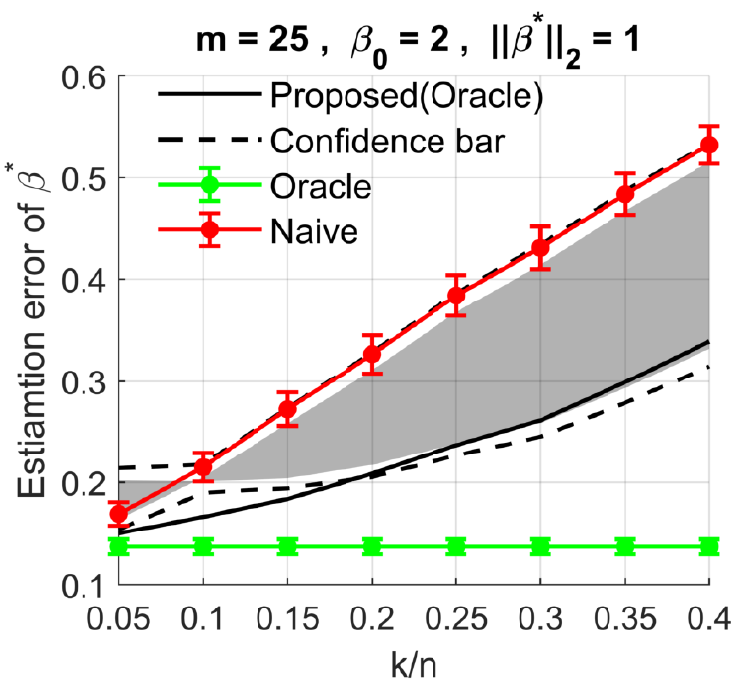} 
    & \hspace*{-1.5ex} \includegraphics[width = 0.33\textwidth]{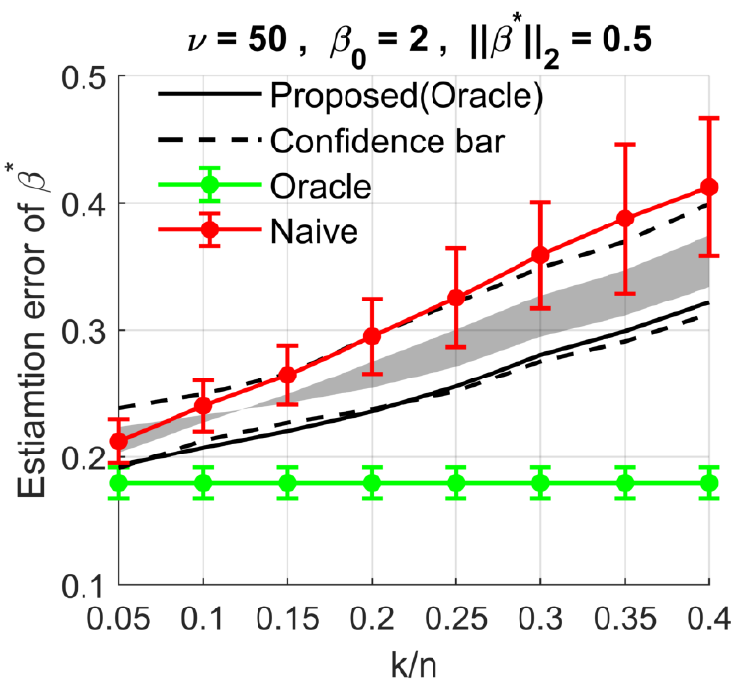}\\
    \hspace*{-3.5ex} \includegraphics[width = 0.33\textwidth]{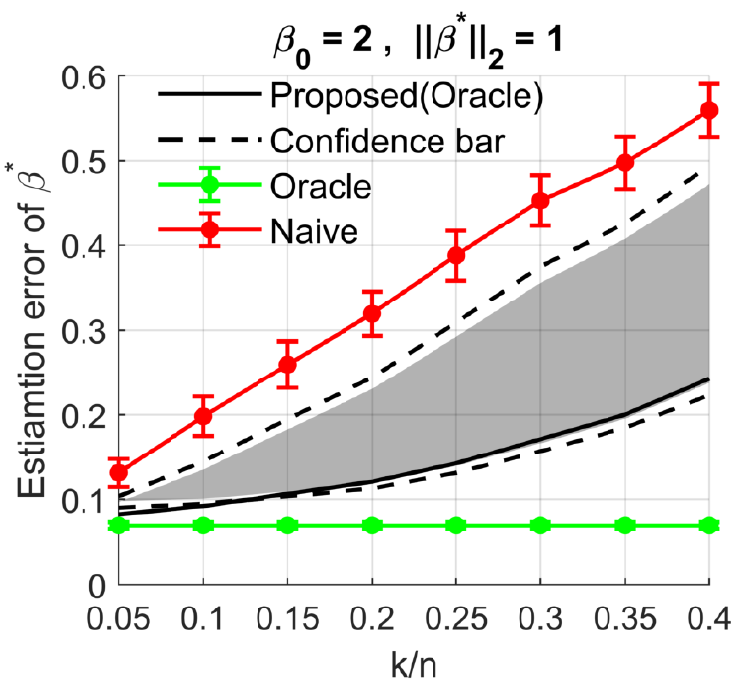}
    & \hspace*{-1.5ex} \includegraphics[width = 0.33\textwidth]{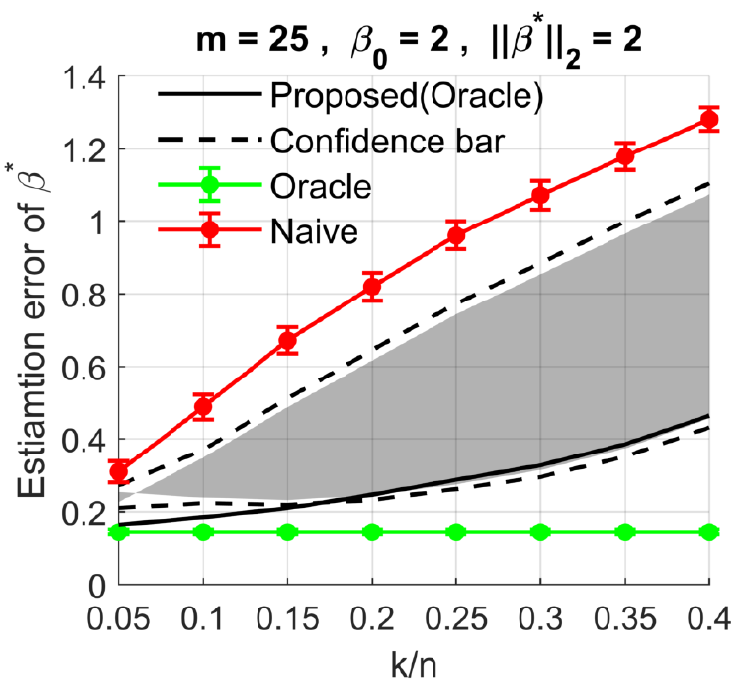}
    & \hspace*{-1.5ex} \includegraphics[width = 0.33\textwidth]{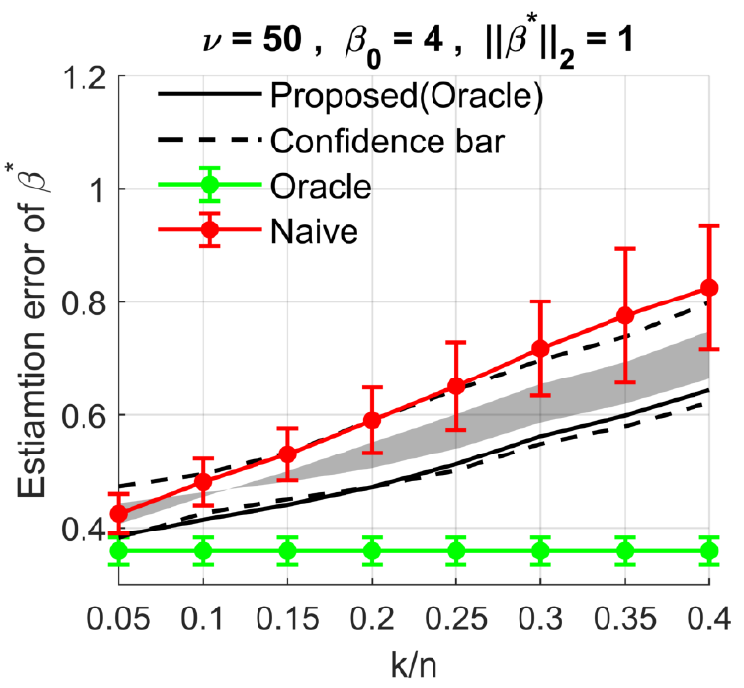}\\
    \hspace*{-3.5ex} \includegraphics[width = 0.33\textwidth]{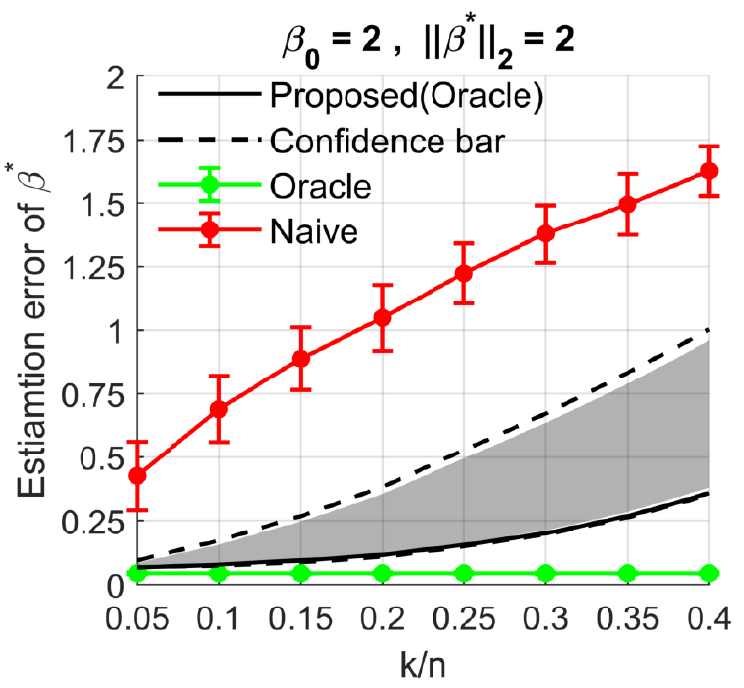} 
    & \hspace*{-1.5ex} \includegraphics[width = 0.33\textwidth]{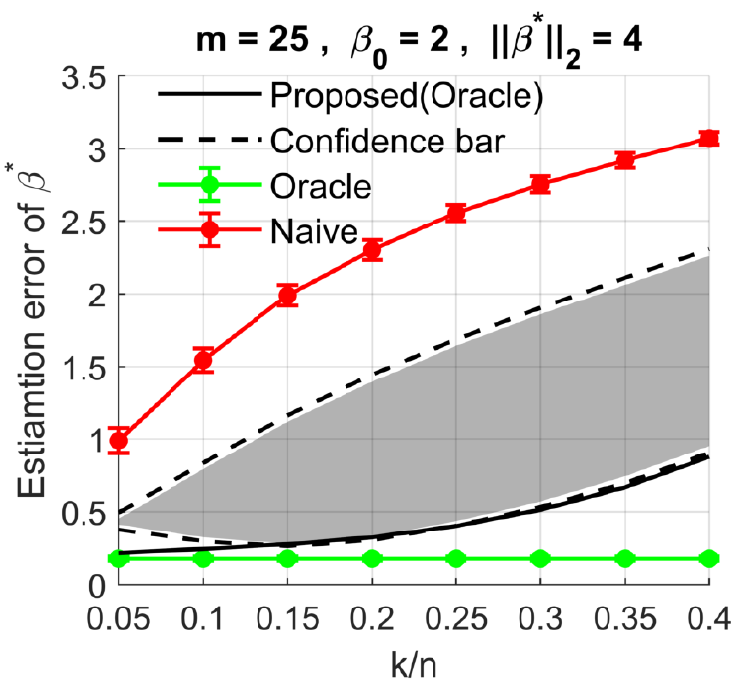}
    & \hspace*{-1.5ex} \includegraphics[width = 0.33\textwidth]{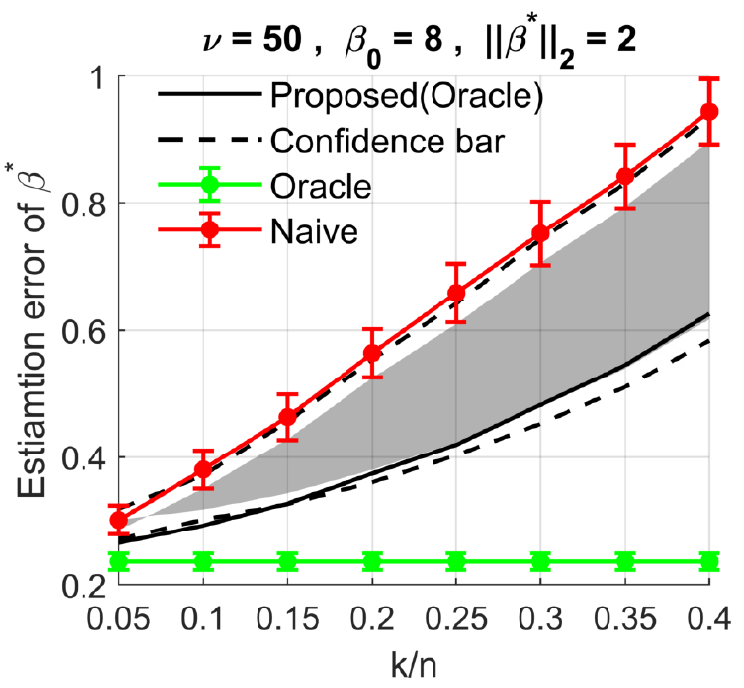}
\end{tabular}
\end{center}
\vspace*{-3ex}
\caption{Average estimation errors $\nnorm{\beta^{\text{est}} - \beta^*}_2$. The lower and upper boundary of the shaded area show the minimum and maximum error over all choices of the pre-factor $C \in [0.1, 2]$, and the corresponding dashed lines represent $\pm$5 $\times$ standard error.}\label{fig:MSE}
\end{figure}

\begin{figure}
\begin{center}
\begin{tabular}{ccc}
    Poisson & Binomial & Gamma \\ 
    \hspace*{-3.5ex} \includegraphics[width = 0.33\textwidth]{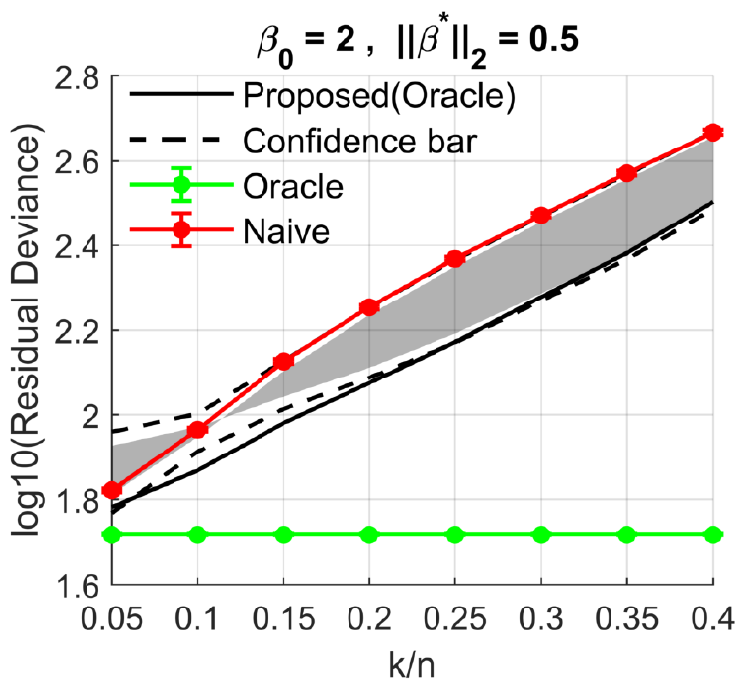}
    & \hspace*{-3.5ex} \includegraphics[width = 0.33\textwidth]{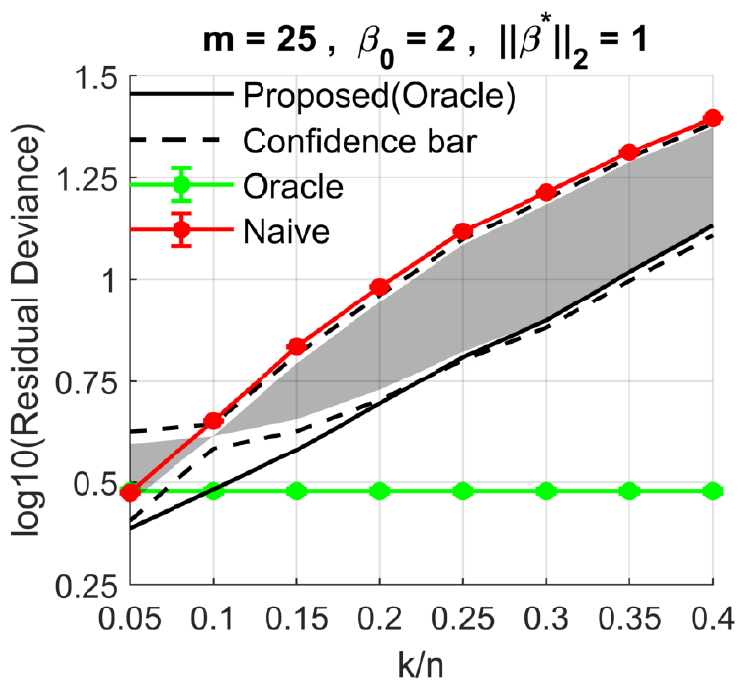}  
    & \hspace*{-3.5ex} \includegraphics[width = 0.33\textwidth]{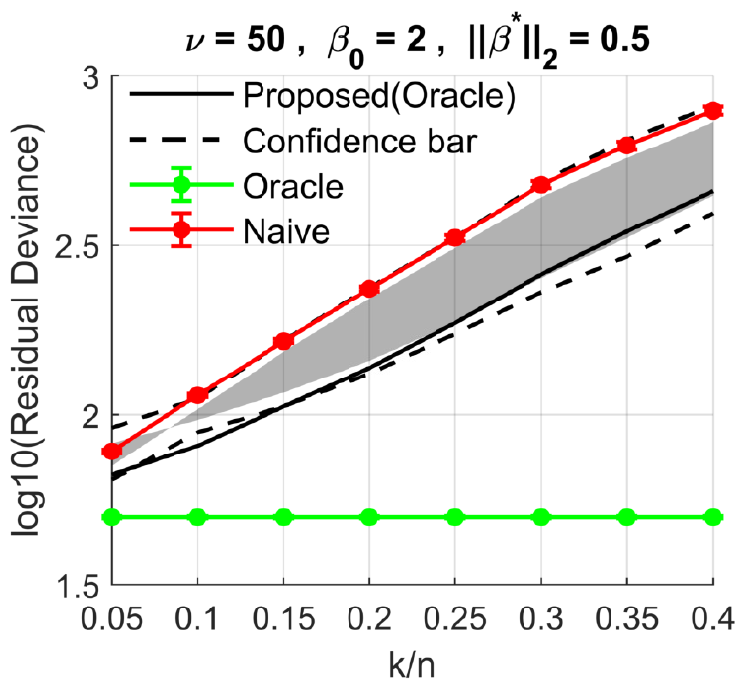} \\
    \hspace*{-3.5ex} \includegraphics[width = 0.33\textwidth]{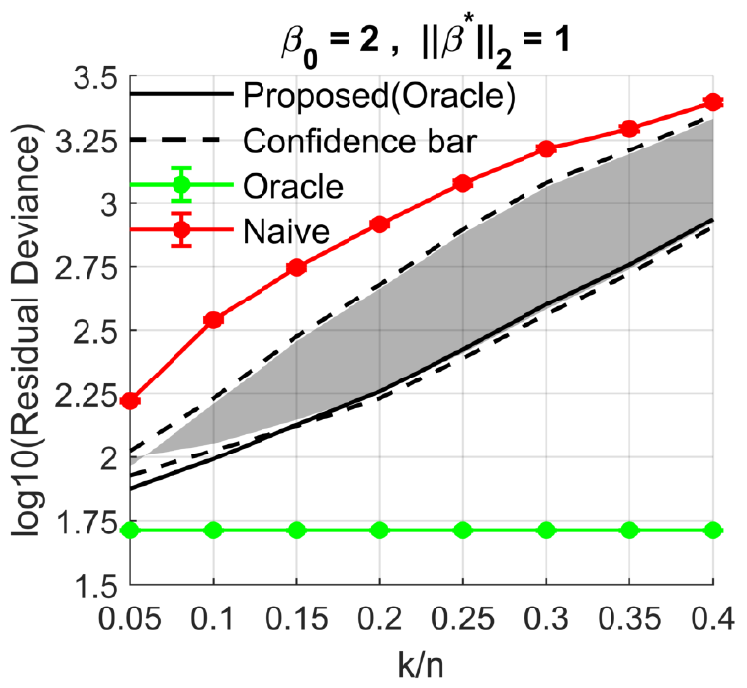}
    & \hspace*{-3.5ex} \includegraphics[width = 0.33\textwidth]{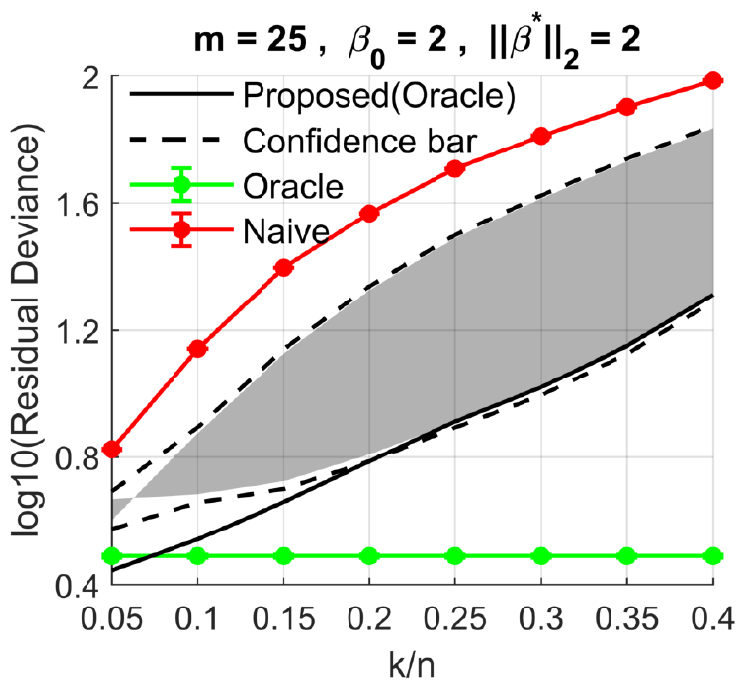}  
    & \hspace*{-3.5ex} \includegraphics[width = 0.33\textwidth]{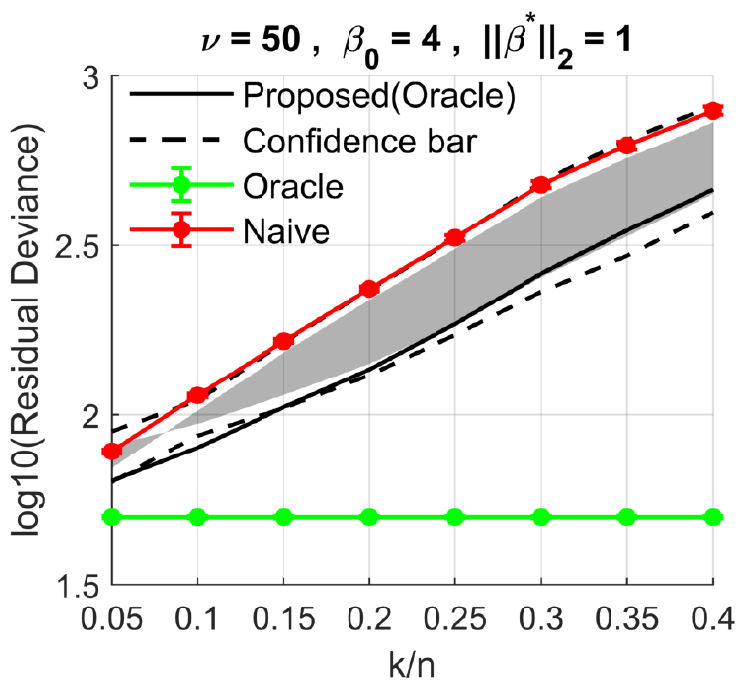}\\
    \hspace*{-3.5ex} \includegraphics[width = 0.33\textwidth]{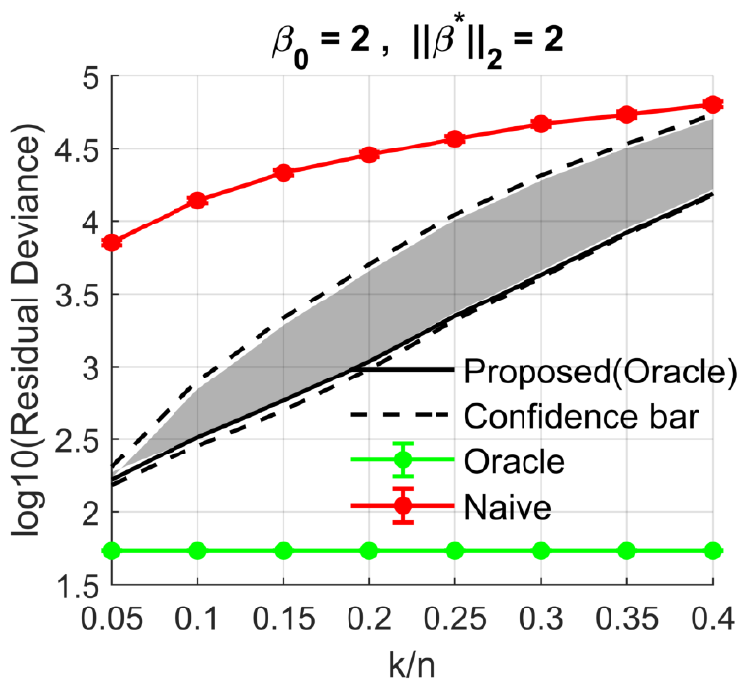}
    & \hspace*{-3.5ex} \includegraphics[width = 0.33\textwidth]{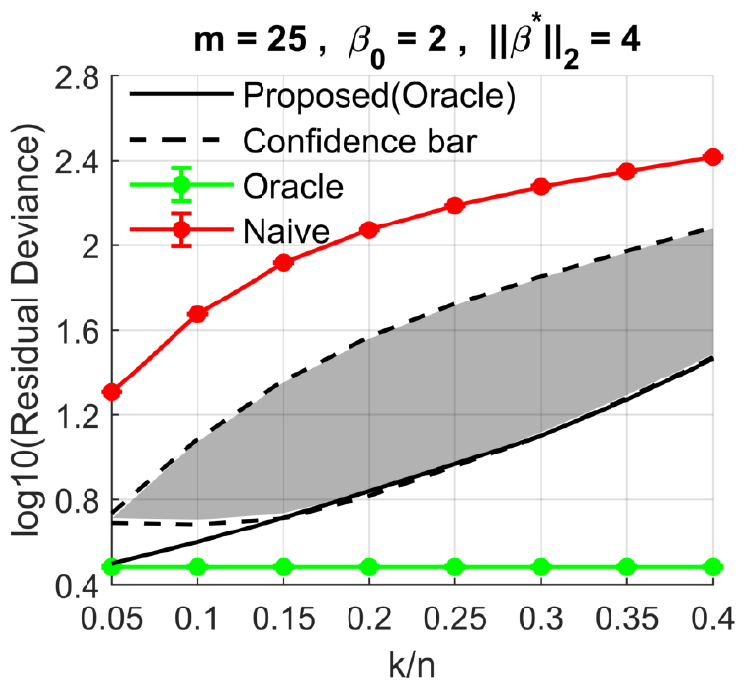}
    & \hspace*{-3.5ex} \includegraphics[width = 0.33\textwidth]{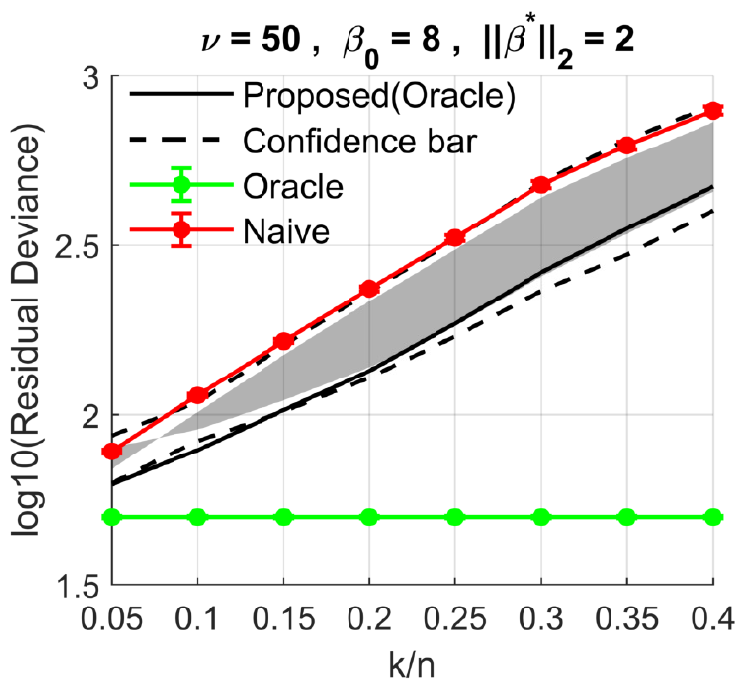}
\end{tabular}
\end{center}
\vspace*{-3ex}
\caption{Average deviances between $\mu^*$ and $\mu^{\text{est}}$. The lower and upper boundary of the shaded area show the minimum and maximum error over all choices of the pre-factor $C \in [0.1, 2]$, and the corresponding dashed lines represent $\pm$5 $\times$ standard error.}\label{fig:dev}
\end{figure}

 
\noindent ``\textsf{proposed}". $\wh{\theta}(\lambda) = (\wh{\beta}^{\T} \; \wh{\xi}^{\T})^{\T}$ is estimated according to Algorithm 1 with $\lambda$ chosen as explained above. The pre-factor $C$
is varied over a logarithmically spaced grid. When presenting the results in Figures \ref{fig:MSE} and \ref{fig:dev}, we display the oracle
selection of $\lambda = \lambda(C)$ minimizing $\nnorm{\wh{\theta}(\lambda) - \theta^*}_2$ as well as the range of a quantity of interest for $C \in [C_{\text{Lower}}, C_{\text{Upper}}]$, where 
$C_{\text{Lower}}$ and $C_{\text{Upper}}$ are fixed numbers. The resulting upper and lower bounds
over this range are complemented by confidence bars with height $5 \times$ standard error. Figure \ref{fig:lam} displays explicitly how the (normalized) estimation error $\nnorm{\wh{\theta} - \theta^*}_{2} / \nnorm{\wh{\theta}^{\text{naive}} - \theta^*}_{2}$ depends on the choice of $C$; the division by $\nnorm{\wh{\theta}^{\text{naive}} - \theta^*}_{2}$ with $\wh{\theta}^{\text{naive}}$ defined under ``\textsf{naive}" makes the results interpretable across different settings. 

The above three approaches are evaluated in terms of their $\ell_2$-estimation error for the
regression parameter, i.e., $\nnorm{\beta^{\text{est}} - \beta^*}_2$, and the deviance (Kullback-Leibler divergence) between $\mu^{*} = \E[\M{y}^* | \M{x}]$ and $\mu^{\text{est}} = \big(h(\M{x}_i^{\T} \beta^{\text{est}}) \big)_{i = 1}^n$, where we recall that $h(\cdot)$ denotes
the response function in GLMs. The notation $\beta^{\text{est}}$ represents a placeholder for any of the three estimators introduced above. The results shown in Figures \ref{fig:lam}, \ref{fig:MSE}, and \ref{fig:dev} are averages over the 100 replications obtained for each setting.  
%
%

%
Figure \ref{fig:lam} confirms that as the pre-factor $C$ increases, the error ratio approaches one as expected since $\wh{\theta} \rightarrow \wh{\theta}^{\text{naive}}$ as $C$ grows. Second, we note that the error ratio increases sharply beyond one as $C \rightarrow 0$; this corresponds to a regime of overfitting. Note that as $C \rightarrow 0$ the proposed approach effectively yields an over-parameterized model achieving perfect fit on a given data set. This in agreement with Theorem \ref{theo:estimation_error} which requires a lower bound on $\lambda$ for its results to hold. Figure \ref{fig:lam} indicates 
that $C \in [0.2, 1]$ typically yields satisfactory results independent of the specific setting
or the specific distribution of the response variable. 

Average estimation errors for the regression parameter are shown in Figure \ref{fig:MSE}. Shaded areas are used to represent the error range the for the proposed approach in dependence of the pre-factor $C$: the upper and lower margins of the shaded areas represent the maximum and minimum error over $C \in [0.1, 2]$, while the dashed lines outside the shaded areas indicate 
$\pm$5 $\times$ standard error. Overall, Figure \ref{fig:MSE} shows that the proposed estimator
can achieve substantial improvements over the naive estimator in a variety of settings. The extent
of the improvement generally increases with the fraction of mismatches and the signal level as measured by $\nnorm{\beta^*}_{2}$. For example, in the Poisson case the proposed approach with 
optimally calibrated $\lambda$ roughly achieves a three-fold reduction in average estimation error over the naive solution when $\nnorm{\beta^*}_{2} = 1$, whereas a five-fold to six-fold reduction is achieved when $\nnorm{\beta^*}_{2} = 2$. Despite the improvements that are obtained, the performance of the proposed estimator is still somewhat far from the oracle, specifically as $k/n$ increases, which indicates a potential for further reductions in error.  
 The results shown in Figure \ref{fig:dev} concerning the deviance between $\mu^{\ast}$ and $\mu^{\text{est}}$ as explained above agree with what is expected based on the results 
 for the estimation error in Figure \ref{fig:MSE}. Again, the results indicate a dependence
 of the improvement achieved by the proposed estimator on the signal level and the fraction of mismatches. 

\begin{figure}
\begin{center}
\begin{tabular}{ccc}
 & Bernoulli &  \\ 
    \hspace*{-3.5ex} \includegraphics[width = 0.33\textwidth]{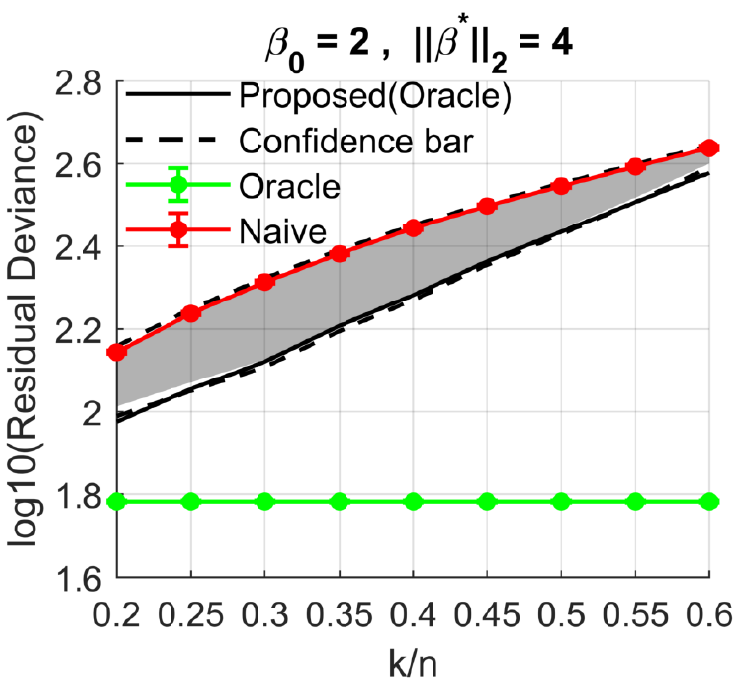}
    & \hspace*{-3.5ex} \includegraphics[width = 0.33\textwidth]{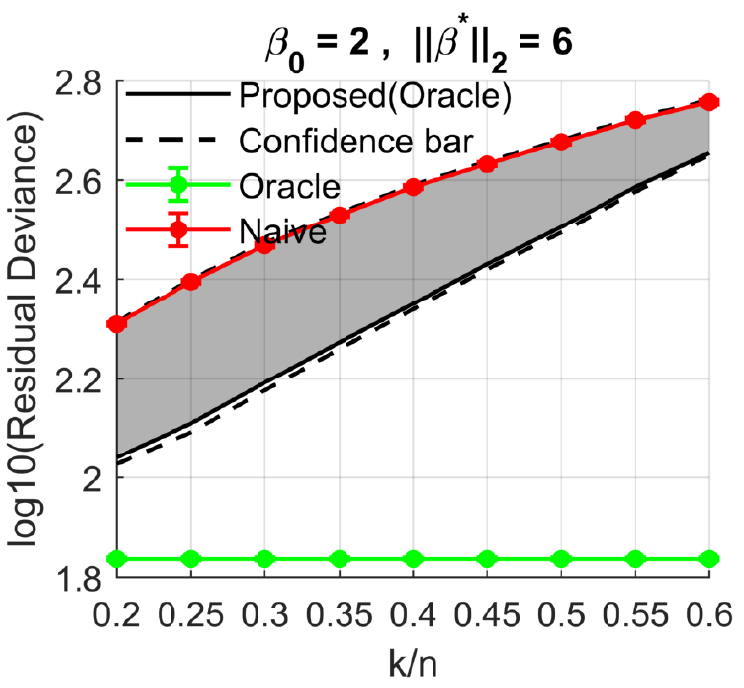}  
    & \hspace*{-3.5ex} \includegraphics[width = 0.33\textwidth]{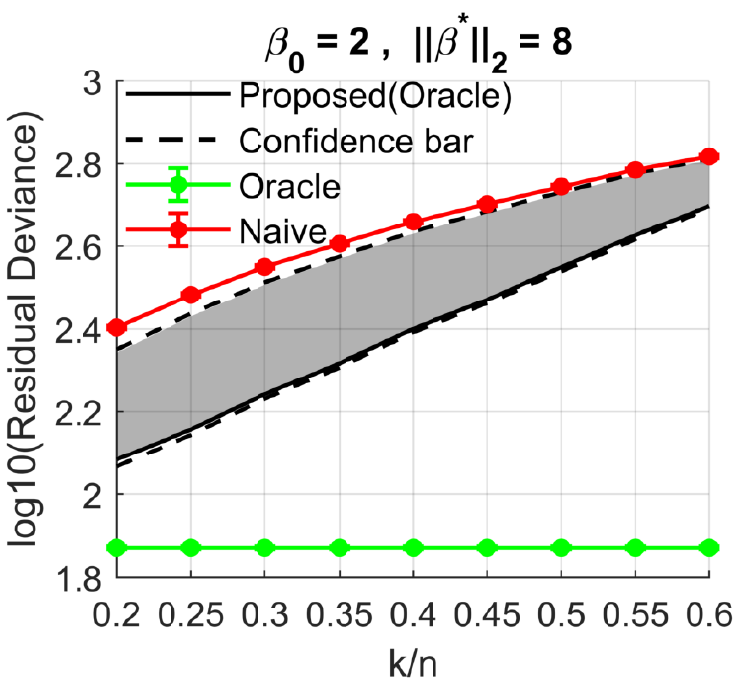} 
\end{tabular}
\end{center}
\vspace*{-3ex}
\caption{Average deviances between $\mu^*$ and $\mu^{\text{est}}$ for Bernoulli response. The lower and upper boundary of the shaded area show the minimum and maximum error over all choices of the pre-factor $C \in [0.5, 2]$, and the corresponding dashed lines represent $\pm$5 $\times$ standard error.}\label{fig:dev_log}
\end{figure}

\vskip1ex
\noindent {\bfseries Binary response}. We now address the case of binary logistic regression. Note that this case behaves differently
from the three distributions for the response considered above, including binomial response with a significant number of trials: it is clear that a massive number of samples is required in order to estimate $\beta^*$ accurately from binary response if $\nnorm{\beta^*}_2$ is large; at the same time, if $\nnorm{\beta^*}_2$ is small, the separation between the two classes corresponding to the two values
of the response variable is weak and thus the inherent noise is scarcely distinguishable from mismatch error in the response variable, which amounts to what has been extensively studied in the machine learning literature under the terms ``label noise" (e.g., \cite{Frenay2013}). For this reason, we adopt the simulation setup above with sufficiently strong signal, i.e., $\nnorm{\beta^*}_{2} \in \{4,6,8\}$ while $\beta_{0}^* = 2$, but the three competitors are evaluated only in terms of the deviance between $\mu^*$ and $\mu^{\text{est}}$.
Figure \ref{fig:dev_log} shows that the proposed approach achieves improvements over the naive estimator, but the improvements are less pronounced than for the three other settings. The observed moderate improvement is in alignment with what is reported in the paper 
\cite{tibshirani2013robust} that studies the empirical performance of the proposed estimator exclusively in the setting of binary response with noisy labels.    


\section{Permutation Recovery}\label{sec:PR}
In this section, we suppose throughout that $N = n$ and present sufficient conditions for specific GLMs under which the maximum likelihood (ML) estimator $\wh{\pi}$ of $\pi^*$ for known $\beta^*$ \eqref{eq:nloglik_pi} achieves perfect recovery in the sense of $\{ \wh{\pi} = \pi^* \}$. While the assumption of known $\beta^*$ may
appear limiting, the results of this section can still be useful from at least two considerations: first, it is not unreasonable to expect that they continue to be valid if $\beta^*$ is replaced
by an accurate estimator; second, they provide some insights into what is at best achievable in practice.  

The first result states that ML estimation of $\pi^*$ for known $\beta^*$ is computationally tractable as already indicated in the introduction of this paper. 
\vskip4ex
\begin{prop}\label{prop:sorting}
Consider ML estimation of $\pi^*$, i.e., optimization problem \eqref{eq:nloglik_pi} for 
$\beta = \beta^*$. We then have 
 \begin{equation}\label{eq:MLE2}
\min_{\pi  \in \mc{P}(n)} -\su \{ y_i \M{x}_{\pi(i)}^{\T} \beta^*  + \psi(\M{x}_{\pi(i)}^{\T} \beta^*) \} = \min_{\pi \in \mc{P}(n)} -\su y_i \M{x}_{\pi(i)}^{\T} \beta + c = -\su y_{(i)} (\M{x}^{\T} \beta^*)_{(i)} + c,
\end{equation}   
where $c = \su \psi(\M{x}_i^{\T} \beta^*)$ and the subscript $(i)$ refers to the $i$-th order statistic, i.e., for $v = (v_i)_{i = 1}^n$, $v_{(1)} \leq \ldots \leq v_{(n)}$. 
\end{prop}
Proposition \ref{prop:sorting} states that the MLE $\wh{\pi}$ of $\pi^*$ is given by the permutation 
that pairs the corresponding order statistics of $\{ \M{x}_i^{\T} \beta^* \}_{i = 1}^n$ and $\{ y_i \}_{i = 1}^n$. For later reference, it is worth noting that the conclusion of Proposition \ref{prop:sorting} with regard to the form of $\wh{\pi}$ continues to hold if the link function is not the canonical link. This is an immediate consequence of the fact that $\wh{\pi}$ is invariant under monotonically increasing transformations of $\{ \M{x}_i^{\T} \beta^* \}_{i = 1}^n$. 
The proof of Proposition \ref{prop:sorting} follows immediately from existing results in the literature
on linear assignment problems (cf., e.g., \cite{Burkard2009}), and is hence omitted. 

In the sequel, we refrain from presenting a unified analysis 
applicable to an entire class of GLMs for two reasons: first, sharper results can be obtained from case-specific analysis; second, permutation recovery turns out to be entirely or at least largely infeasible for a variety of GLMs, e.g., i) binomial response with a small number of trials due to excessive ties among the $\{ y_i \}_{i = 1}^n$, ii) exponential response with canonical (i.e., reciprocal) link since in this case recovery fails for a wide range of random designs (cf.~Theorem \ref{theo:sep_beta} below).    

\subsection{Recovery Results}
We start the presentation of our results by conditioning on the predictors
$\{ \M{x}_i \}_{i = 1}^n$, and hence for fixed conditional expectations of the responses. The extension to random predictors is considered subsequently.

In this subsection, it is appropriate to distinguish between random variables $\{ Y_i \}_{i = 1}^n$ and their realizations $\{ y_i \}_{i = 1}^n$. We let $\mu_i = \E[Y_i | \M{x}_{\pi^*(i)}] = h(\M{x}_{\pi^*(i)}^{\T} \beta^*)$, $1 \leq i \leq n$, where $h$ denotes the inverse link function of the underlying GLM. Unless stated otherwise, $h$ refers to the canonical link. 
\begin{theo}\label{theo:sep_mu}
Suppose without loss generality that $\mu_{1} \leq \mu_{2} \leq \cdots \leq \mu_{n}$, and consider the MLE $\wh{\pi}$ given by the minimizer of \eqref{eq:MLE2}. For any $\delta > 0$, we have $\p(\wh{\pi} \neq \pi^* | \M{X}) < \delta$ if
\begin{itemize}
\item[(a)] $Y_{i} \sim N(\mu_{i}, \sigma^{2}), \; 1\leq i \leq n \, \text{\emph{:}}\;\;
       \underset{{1 \leq i \leq  n-1}}{\min} (\mu_{i+1} - \mu_{i}) > 2\sigma \sqrt{\log \frac{n-1}{\delta}},
$

\item[(b)] $Y_{i} \sim \text{Poisson}(\mu_{i}) , \; 1\leq i \leq n \, \text{\emph{:}}\;\;
       \underset{{1 \leq i \leq  n-1}}{\min} (\sqrt{\mu_{i+1}} - \sqrt{\mu_{i}}) > \sqrt{\log \frac{n-1}{\delta}},
$
\item[(c)] $Y_{i} \sim \text{Gamma}(\nu,\mu_{i}) , \; 1\leq i \leq n \, \text{\emph{:}}\;\; \underset{1 \leq i \leq n-1}{\min} \frac{\mu_{i+1}}{\mu_{i}} > 4 \left (\frac{n-1}{\delta} \right )^{1/\nu}$. 
\end{itemize}
\end{theo}
\noindent Part (a) already appears in similar form in \cite{SlawskiBenDavid2017}. Part (b) can be linked to (a) by noting that the standard deviation of a Poisson random variable with mean $\mu$ equals $\sqrt{\mu}$. Substituting $\sigma$ in (a)
by $\sqrt{\mu}_i$ and dividing both sides by this quantity then approximately yields (b). Part (c) can be understood
according to a similar heuristic: observing that the standard deviation of $Y_i \sim \text{Gamma}(\nu,\mu_{i})$ is given by
$\mu_i / \sqrt{\nu}$, $1 \leq i \leq n$, substituting $\sigma$ in (a) by $\mu_i$ yields a requirement on the ratio 
$\mu_{i + 1} / \mu_i$. Note that for $\nu \asymp \log n$, the ratios need to exceed a constant factor $C > 1$, which 
still requires $\mu_n / \mu_1 = C^{n-1}$. For the exponential distribution, $\nu = 1$, and there is thus little hope that
(c) can be satisfied in practice even for small $n$.

Building on Theorem \ref{theo:sep_mu}, we next consider random predictors $\{ \M{x}_i \}_{i = 1}^n$, thereby providing specific examples in which the 
recovery conditions are satisfied with high probability. 


\begin{theo}\label{theo:sep_beta}
Consider the MLE  $\wh{\pi}$ given by the minimizer of \eqref{eq:MLE2}. Suppose that the $\{ \M{x}_i \}_{i = 1}^n$ are i.i.d.~random vectors with independent, unit variance entries whose densities are bounded by $K < \infty$ almost everywhere.  
For any $\delta > 0$, we have $\p(\wh{\pi} \neq \pi^*) < \delta$ if
\begin{itemize}
\item[(a)] $Y_{i} \sim N(\mu_{i}, \sigma^{2}), \; 1\leq i \leq n \, \text{\emph{:}}\;\; \nnorm{\beta^*}^{2}_{2} > \frac{8\sigma^{2}K^{2}n^{2}(n-1)^{2}}{\delta^{2}}\log\left(\frac{n(n-1)}{\delta}\right),$
\item[(b)] $Y_{i} \sim \text{Poisson}(\mu_{i}), \; 1\leq i \leq n \, \text{\emph{:}}\;\; \nnorm{\beta^*}^{2}_{2} > \frac{16K^{2}n^{2}(n-1)^{2}}{\delta^{2}}\log\left(\frac{n(n-1)}{\delta}\right),$ and $\beta_0^*/\nnorm{\beta^*}_2$ is such that $\sup_{u: \nnorm{u}_2 = 1} \p\left(\min_{1 \leq i \leq n} \scp{u}{\M{x}_i} \leq -\frac{\beta_0^*}{\nnorm{\beta^*}_2} \right) < \delta/2,$
\item[(c)] $Y_{i} \sim \text{Gamma}(\nu,\mu_{i}),\; \mu_i = \exp(\eta_i), \; 1\leq i \leq n \, \text{\emph{:}}\;\; \nnorm{\beta^*}^{2}_{2} > \frac{K^{2}n^{2}(n-1)^{2}}{2\nu^{2}\delta^{2}}\left (\log 4 \left (\frac{n(n-1)}{\delta} \right )^{1/\nu}\right)^{2}.$
\end{itemize}
\end{theo}
\noindent Part (a) appears in similar form for isotropic Gaussian $\{ \M{x}_i \}_{i = 1}^n$ in \cite{SlawskiBenDavid2017}. The statement here extends that result to a much broader class of designs, without imposing any condition on the tails of the distribution of the $\{ \M{x}_i \}_{i = 1}^n$. Part (b) for the Poisson distribution involves an extra condition compared to (a)
which in essence requires a lower bound on $\beta_0^* / \nnorm{\beta^*}_2$ to ensure that all of the $\{ \mu_i \}_{i = 1}^n$ are sufficiently bounded away from one with high probability. For example, if the entries of the $\{ \M{x}_i \}_{i = 1}^n$ are i.i.d.~ and symmetric around zero, and $\beta^* = (1,\ldots,1)^{\T}$, say, the resulting linear predictor will assume negative values with probability $1/2$ which then translate to expectations between zero and one via the inverse link function (exponential). According to Theorem 
\ref{theo:sep_mu}, we need the spacing between the $\{ \mu_i \}_{i = 1}^n$ to be at least proportional to the corresponding standard
deviations, which is violated in the range $[0,1]$ since the standard deviations are given by $\{ \sqrt{\mu_i} \}_{i = 1}^n$. Depending on the distribution of the $\{ \M{x}_i \}_{i = 1}^n$, the condition on $\beta_0^* / \nnorm{\beta^*}_2$ can be made explicit: 
in the simplest case with $\{ \M{x}_i \}_{i = 1}^n \overset{\text{i.i.d.}}{\sim} N(0, I_d)$, we have $\{ \scp{\M{x}_i}{u} \}_{i = 1}^n \overset{\text{i.i.d.}}{\sim} N(0,1)$ for any unit vector $u$, and it is then not hard to show that the extra condition 
in (b) holds if $\beta_0^* / \nnorm{\beta^*}_2 \geq \sqrt{2 \log n} + \sqrt{\log(2n/\delta)}$. Regarding part (c), let us emphasize that the result here concerns the log-link rather than the canonical (reciprocal) link. A recovery result for the latter appears
out of reach since the use of the reciprocal link would lead to a clustering of the $\{ \mu_i \}_{i = 1}^n$ in $[0,1]$ independent 
of $\beta_0^*$ for many common choices for the distribution of the $\{ \M{x}_i \}_{i = 1}^n$.
\vskip1ex
\noindent {\bfseries Incorporating blocking variables}. Following up on the discussion in $\S$\ref{subsec:blocking}, it is worth pointing out that permutation recovery based on \eqref{eq:MLE2} decouples across blocks, and hence can be performed in a block-by-block fashion. The recovery conditions in Theorems \ref{theo:sep_mu} and \ref{theo:sep_beta} can be applied for each block in that $n$ gets replaced by the number of elements belonging to the respective block, and are thus easier to satisfy. For example,
if $n_j = n/K$, $j = 1,\ldots,K$, and $\delta = \delta_0 / K$ for a given failure probability $\delta_0$, it is easy to check 
that all terms involving $n-1$ get replaced by $n / K - 1$. This yield substantial benefits particularly if $n / K = O(1)$, i.e., in the case of many small blocks.  

\vskip1ex
\noindent {\bfseries Mismatch recovery}. Note that \eqref{eq:MLE2} does not take advantage of a sparse mismatch regime if the 
latter is known to hold. Unfortunately, it turns out that replacing the minimum in \eqref{eq:MLE2} by the minimum over all 
permutations moving at most $k$ indices gives rise to a considerably harder optimization problem unlike the simple solution
via sorting obtained in the absence of such constraint. In spite of this, there is a natural workaround involving two steps: 1. identify the set of mismatches $S_* = \{i: \; \pi^*(i) \neq i \}$, 2. solve the minimization problem in \eqref{eq:MLE2} restricted
to the observations in $S_*$. With step 2. set and analyzed according to the preceding theorems, it remains to consider step 1., which we refer to as ``mismatch recovery". We suggest to address this task by assessing the fit
of each $y_i$ to its counterpart $\mu_i = h(\M{x}_i^{\T} \beta^*)$. Conditional on $\pi^*(i) = i$, the distribution of the
$y_i$ is known, and we may thus evaluate 
\begin{equation*}
p_i^* = \begin{cases} \p_{\pi^*(i) = i}(Y_i \geq y_i) \;\;& \text{if} \; y_i \geq \mu_i \\
                      \p_{\pi^*(i) = i}(Y_i \leq y_i) \;\;& \text{if} \; y_i \leq \mu_i,  \quad i=1,\ldots,n, 
        \end{cases}
\end{equation*}
where the probability is with respect to the underlying random variables $\{ Y_i \}_{i = 1}^n$ conditional on $\pi_i^*(i) = i$, $1 \leq i \leq n$. Note that similar to the notion of p-value, small $p_i^*$ can be considered as evidence against $\pi_i^*(i) = i$, $1 \leq i \leq n$. Accordingly, we may estimate $S_*$ as the set of indices corresponding to the $k$ smallest values among the $\{ p_i^* \}_{i=1}^n$; alternatively, if $k$ is unknown, we may estimate $S_*$ by $\{i:\; p_i^* \leq \tau \}$ for a threshold $\tau \in (0,1)$. 

As an illustration, let us consider linear regression with Gaussian errors. Observe that if $\pi^*(i) = i$, $1 \leq i \leq n$, we have for all $t > 0$
\begin{equation*}
\p \left(y_i - \M{x}_i^{\T} \beta^* > \sigma t \right) = \p \left(y_i - \M{x}_i^{\T} \beta^* < \sigma t \right) = 1 - \Phi(t), \quad 1 \leq i \leq n,
\end{equation*}
where $\Phi$ denotes the CDF of an $N(0,1)$-random variable, hence $p_i^* = 1 - \Phi(|y_i - \M{x}_i^{\T} \beta^*|/\sigma)$, $1 \leq i \leq n$. In order to fix $\tau$, a natural options is to require that $\su p_i^* \leq \delta$ for $\delta \in (0,1)$ and thus $p_i^* \leq \delta/n$, $1 \leq i \leq  n$. Using the standard Gaussian tail bound $1 - \Phi(t) \leq \exp(-t^2/2)$ for $t > 0$ yields that $p_i^* \leq \delta/n$ once $|y_i - \M{x}_i^{\T} \beta^*| > \sigma \sqrt{2 \log(n/\delta)}$, $1 \leq i \leq n$. Accordingly, in order for mismatches $i$ with $\pi^*(i) \neq i$ to be detectable, it is required that $|(\M{x}_{\pi^*(i)} - \M{x}_i)^{\T} \beta^*| = |\mu_{\pi^*(i)} - \mu_i|\geq 2 \sigma \sqrt{2 \log(n/\delta)}$, which is almost identical to the requirement in Theorem \ref{theo:sep_mu} (a). We conclude that mismatch recovery, i.e., the estimation of $S_*$, and permutation recovery obey similar regimes.   

\subsection{Simulation}
We complement Theorems \ref{theo:sep_mu} and \ref{theo:sep_beta} with simulation results. The entries of the 
design matrix are sampled i.i.d.~from three unit-variance distributions: (i) standard Normal, (ii) the uniform distribution
on $[-\sqrt{3}, \sqrt{3}]$, and (iii) (rescaled) $t$-distribution with five degrees of freedom, and responses are subsequently generated according to model \eqref{eq:expfamily} based on the Normal distribution with identity link , and the Poisson and Gamma distribution with log-link. The regression
parameter $\beta^*$ is drawn uniformly at random from spheres in dimension $d$ of varying radii $\nnorm{\beta^*}_2$. 
The intercept is taken as $\beta_0^* = c \cdot \nnorm{\beta^*}_2$ with $c \in \{0.5, 5\}$. The sample size is fixed as $n = 200$. For each configuration consisting of the distribution of the design matrix, the radius $\nnorm{\beta^*}_2$, and the value of $\beta_0^*$, 1000 replications are performed. In each replication, we evaluate the normalized Hamming distance 
$n^{-1} \su \mathbb{I}(i \neq \wh{\pi}(i))$, where $\wh{\pi}$ is the minimizer of \eqref{eq:MLE2}. 

\begin{figure}[!ht]
\begin{center}
\begin{tabular}{ccc}
   \hspace*{-3.5ex} \includegraphics[width = 0.33\textwidth]{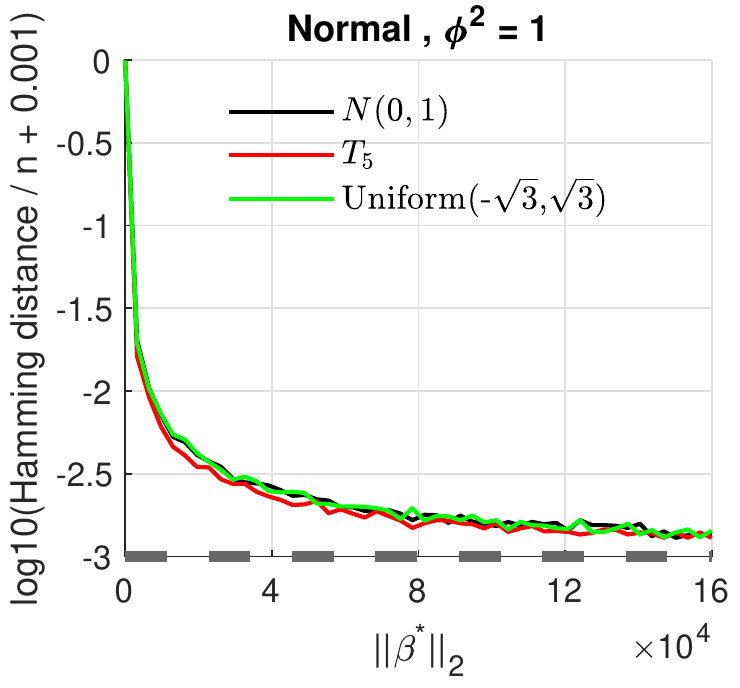}
   & \hspace*{-1.5ex} \includegraphics[width = 0.33\textwidth]{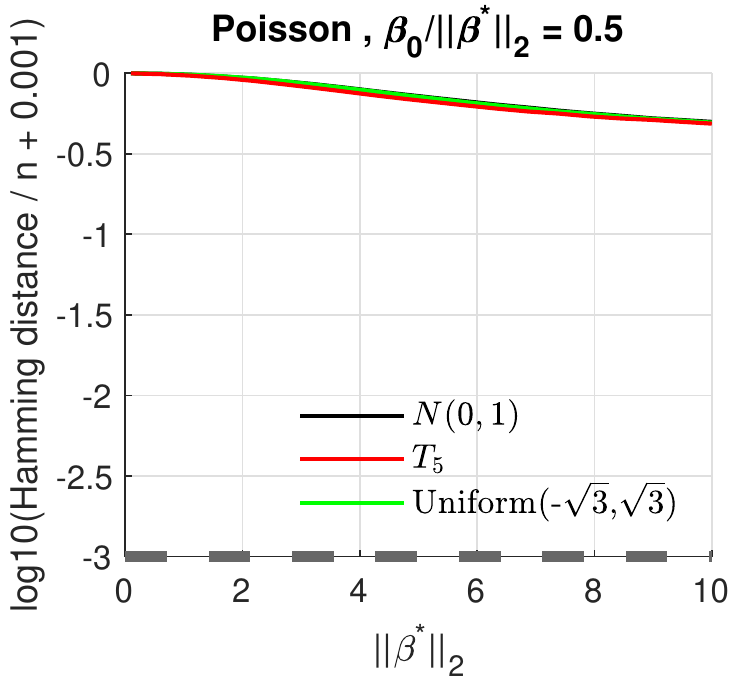}
   & \hspace*{-1.5ex} \includegraphics[width = 0.33\textwidth]{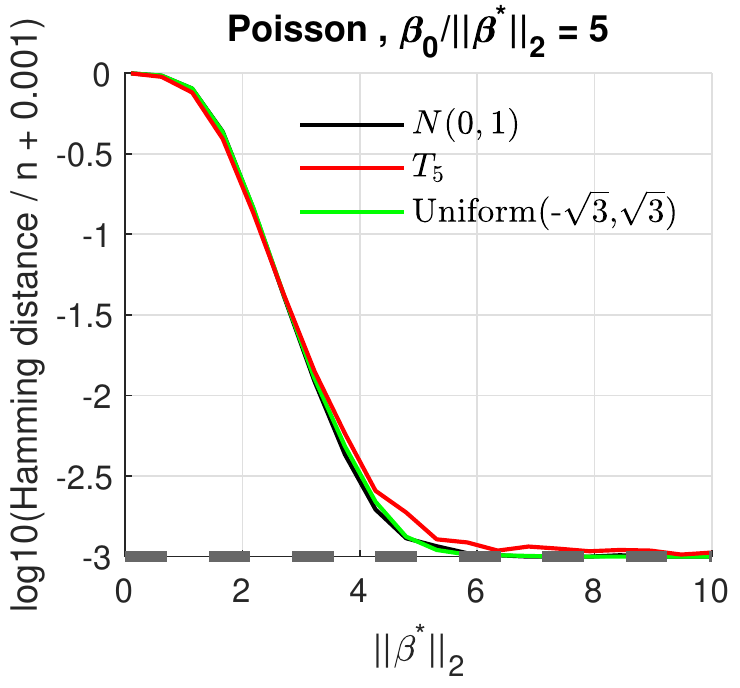} \\
   \hspace*{-3.5ex} \includegraphics[width = 0.33\textwidth]{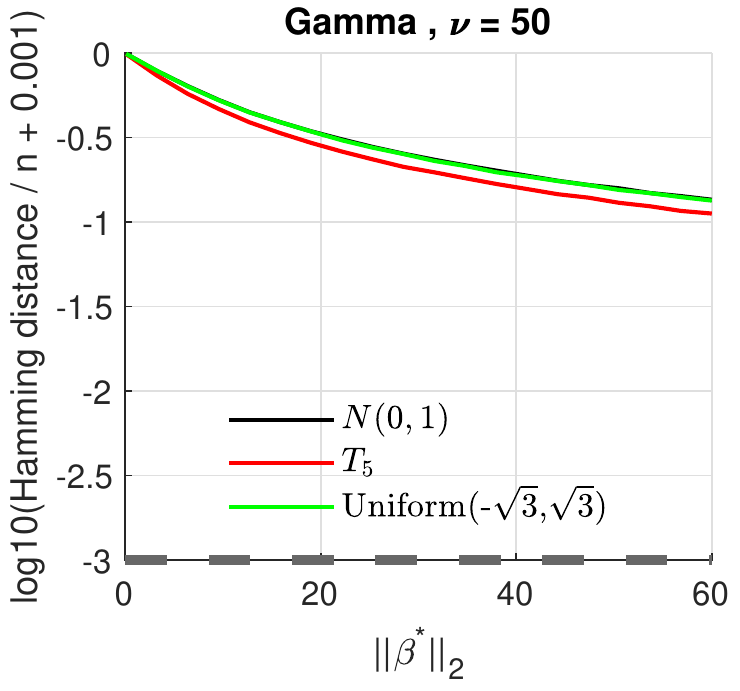}
   & \hspace*{-1.5ex} \includegraphics[width = 0.33\textwidth]{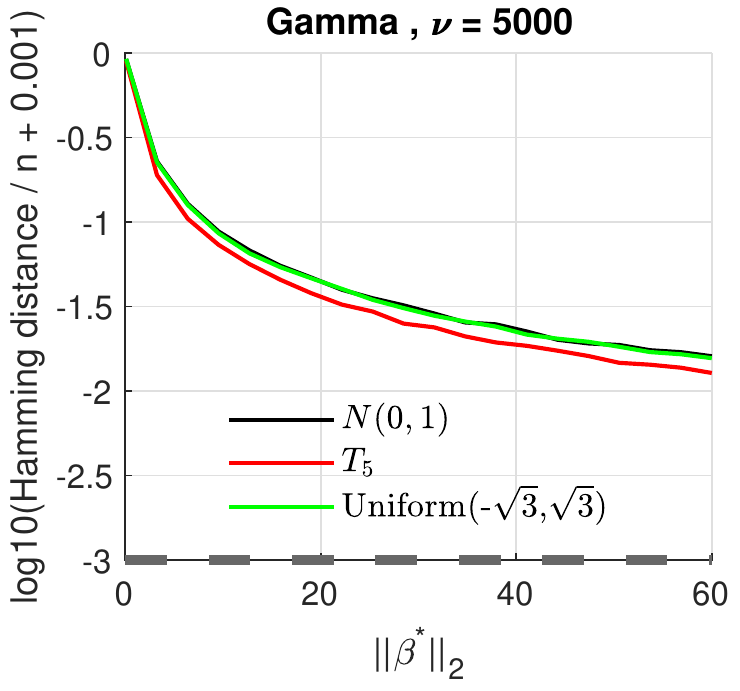}
   & \hspace*{-1.5ex} \includegraphics[width = 0.33\textwidth]{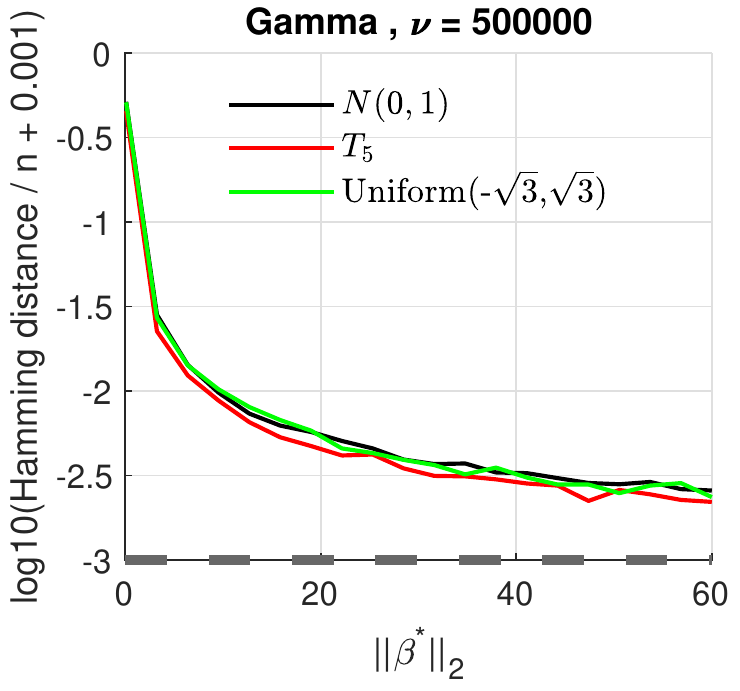}
     \end{tabular}
\end{center}
\caption{Hamming distance between $\wh{\pi}(\beta^*)$ and $\pi^*$ (on a $\log_{10}(\cdot + \, \epsilon)$ scale with $\epsilon = 0.001$), averaged over 1000 replications. Different curves correspond to different random design.}\label{fig:per_rec}
\end{figure}

In light of Theorem \ref{theo:sep_beta}, permutation recovery 
can be achieved if and only if $\nnorm{\beta^*}_{2}$ is large enough, and specifically for Poisson case, if in addition $\beta_{0}^*/\nnorm{\beta^*}_{2}$ exceeds a certain threshold. Figure \ref{fig:per_rec} confirms this qualitatively. In particular, we observe that the ratio $\beta_{0}^*/\nnorm{\beta^*}_{2}$ is crucial in the Poisson case unlike the other cases. Furthermore, we note that the distribution of the design does not have a significant impact on the results: all three random design can achieve (at least approximate) permutation recovery given sufficient signal as quantified by $\nnorm{\beta^*}_2$. Finally, note that for the Gamma distribution, recovery results improve as the shape parameter $\nu$ increases. 

\section{Comparison to the Lahiri-Larsen \& Chambers methods}\label{sec:Com_LL}
In this section, we aim to provide a short comparison of the proposed method and an established method whose prototype was proposed in \cite{Lahiri05} and later extended to a wider class of regression models including generalized linear models \cite{Han2019}, and will hence be referred to as the Lahiri-Larsen (LL) method. A closely related approach is due to Chambers \cite{Chambers2009, Chambers2019improved, KimChambers2012}. The former turns out to be somewhat easier to analyze in the block-structured permutation setting outlined in $\S$\ref{subsec:blocking} that will also be adopted in the sequel. The ultimate goal of our discussion is to delineate scenarios in which the proposed estimator tends to be superior and inferior, respectively, relative to the LL method. Both methods differ noticeably with regard to the assumptions on $\pi^*$ and the required amount of knowledge about the linkage process, and the regimes of interest differ accordingly. 

In a nutshell, the LL method assumes that $\M{y} = \Pi^* \M{y}^*$ with $\M{y}_j^* | \M{x}_j$, $1 \leq j \leq N$, following a GLM as specified in $\S$\ref{sec:PS} and $\Pi^*$ being a generalized random permutation matrix associated with the map $\pi^*: \{1,\ldots,N\} \rightarrow \{1,\ldots,n\}$ whose
$(i,j)$-th entry equals one if $\pi^*(i) = j$ and zero otherwise, $1 \leq i \leq n, \, 1 \leq j \leq N$. The LL method further assumes that $\Pi^*$ is conditionally independent of $\M{y}^*$ given $\{ \M{x}_j \}_{j = 1}^N$ and that the corresponding conditional expectation of $\Pi^*$ is given by $\M{Q} \in \R^{n \times N}$. Let $\M{X}_N$ denote the design matrix associated with the full set of covariates $\{\M{x}_i \}_{i = 1}^N$\footnote{Relevant to the sample-to-register linkage setting only, cf.~$\S$\ref{sec:PS}. Note that $\M{X}_N = \M{X}$ if $N = n$.}.   
Equipped with the above assumptions, it is readily shown that 
\begin{equation}\label{eq:estimating_equation}
\M{X}_N^{\T} \M{Q}^{\T} (\M{y} - \M{Q} \bm{\mu}^*(\beta)) = \M{0} \; \, \Leftrightarrow \; \M{X}_N^{\T} \M{Q}^{\T} (\Pi^* \M{y}^* - \M{Q} \bm{\mu}^*(\beta)) = \M{0}, \qquad \bm{\mu}^*(\beta) \coloneq \big( \psi'(\M{x}_j^{\T} \beta) \big )_{j=1}^N
\end{equation}
is an unbiased estimating equation in the sense that 
\begin{equation*}
\E_{\Pi^*, \M{y^*}}[\M{X}_N^{\T} \M{Q}^{\T} (\Pi^* \M{y}^* - \M{Q} \bm{\mu}^*(\beta^*))] = \E_{\M{y}^*}[\M{X}_N^{\T} \M{Q}^{\T} \M{Q} (\M{y}^* - \bm{\mu}(\beta^*))] = \M{0}. 
\end{equation*}
The estimator is particularly easy to understand in the setting in which $\pi^*$ is a block-structured permutation ($N = n$) as discussed in $\S$\ref{subsec:blocking} and the additional assumption that for each
block $G_j$, the corresponding permutation is chosen uniformly at random. Without loss of generality, let $\Pi^* = \text{bdiag}(\Pi_1^*, \ldots, \Pi_K^*)$ be the block diagonal matrix associated with $\pi^*$. It then follows 
that 
\begin{equation}\label{eq:averaging_operator}
\M{Q} = \text{bdiag}(\mathbbm{1}_{n_1}, \ldots, \mathbbm{1}_{n_K}), 
\end{equation}
where for any integer $m$, the symbol $\mathbbm{1}_m$ denotes an $m$-by-$m$ matrix of ones, multiplied by $1/m$. Since each matrix block matrix
is a projection (averaging operator), $\M{Q} = \M{Q}^{\T} = \M{Q}^2$ is a projection as well. Note that even
if the block-wise permutations are not chosen uniformly at random, we may still use \eqref{eq:averaging_operator} for $\M{Q}$ in \eqref{eq:estimating_equation} to obtain an unbiased estimating equation since $\M{Q}^{\T} \Pi^* = \M{Q}$ for \emph{any} permutation matrix $\Pi^*$ block-structured as 
$\M{Q}$. Letting $\wh{\beta}^{\text{LL}}$ denote a solution of this estimating equation, asymptotic theory implies that $\wh{\beta}^{\text{LL}}$ is a consistent estimator of $\beta^*$ with asymptotic covariance matrix 
\begin{equation}\label{eq:sandwich}
\E_{\M{y}}[J_{\Gamma}(\beta^*)]^{-1} \, \cov_{\M{y}}(\Gamma(\beta^*)) \, \E_{\M{y}}[J_{\Gamma}(\beta^*)]^{-1},
\end{equation}
where $\Gamma: \R^{d} \rightarrow \R^d$ denotes the function defining the estimating equation, and 
$J_{\Gamma}$ denotes the Jacobian of $\Gamma$. Straightforward calculations yield that 
\begin{equation}\label{eq:sandwich_parts}
\E_{\M{y}}[J_{\Gamma}(\beta^*)]  = \M{X}^{\T} \M{Q} \M{V}(\beta^*) \M{X}, \qquad \cov_{\M{y}}(\Gamma(\beta^*)) = \M{X}^{\T} \M{Q} \M{V}(\beta^*) \M{Q} \M{X}, 
\end{equation}
where $\M{V}(\beta^*)$ is a diagonal matrix whose diagonal entries are given by the variances of $y_i^*$ given by $\psi''(\M{x}_i^{\T} \beta^*), \; 1 \leq i \leq n$.  In the sequel, we shall argue that for a wide range of
random designs, the entries of \eqref{eq:sandwich} are of the order $O_{\p}(K^{-1})$, i.e., the estimator
$\wh{\beta}^{\text{LL}}$ converges asymptotically at the same rate, and that rate will be recovered exactly
for linear regression with Gaussian errors. 

For this purpose, observe first that the rows of $\M{Q} \M{X}$ are given by $n_j$ replications of $\overline{\M{x}}_j = \frac{1}{n_j} \sum_{i \in G_j} \M{x}_i$, $1 \leq j \leq K$. Accordingly, we have
\begin{equation*}
\cov_{\M{y}}(\Gamma(\beta^*)) = \sum_{j = 1}^K \sum_{i \in G_j} \psi''(\M{x}_i^{\T} \beta^*) \, \overline{\M{x}}_j \overline{\M{x}}_j^{\T}, 
\end{equation*}
which scales as $O_{\p}(K)$ for a wide range of random designs that satisfy (i) $\overline{\M{x}}_j = O_{\p}(n_j^{-1/2})$, $1 \leq j \leq K$,  and (ii) $\psi''(\M{x}_i^{\T} \beta^*) = O_{\p}(1)$, $1 \leq i \leq n$. With a similar reasoning, one also obtains $\E_{\M{y}}[J_{\Gamma}(\beta^*)] = O_{\p}(K)$ and thus in combination $O_{\p}(K^{-1})$ for the asymptotic covariance \eqref{eq:sandwich}.

For linear regression with Gaussian errors, the same order additionally holds in a non-asymptotic fashion. Note that in this case, the underlying estimation equation has the closed form solution $\wh{\beta}^{\text{LL}} = (\M{X}^{\T} \M{Q}\M{X})^{-1} \M{X}^{\T} \M{Q} \M{y}$ and hence $\cov(\wh{\beta}^{\text{LL}}) = \phi^2 (\M{X}^{\T} \M{Q} \M{X})^{-1} = \phi^2 \left( \sum_{j = 1}^K n_j \overline{\M{x}}_j \overline{\M{x}}_j^{\T} \right)^{-1}$. The same expression is obtained from \eqref{eq:sandwich} and \eqref{eq:sandwich_parts} by using $\M{V}(\beta^*) = \phi^2 I_d$ and the fact that $\M{Q} = \M{Q}^2$. 

While the above discussion does not provide a comprehensive analysis of the estimator $\wh{\beta}^{\text{LL}}$ given the focus on a specific choice of $\M{Q}$ in the setting of a block-structured permutation, we feel that this very choice is presumably among 
the most relevant in practice. In their landmark paper \cite{Lahiri05}, Lahiri and 
Larsen consider the entries of $\M{Q}$ being taken as the probability that 
observation $i$ in file $F_{\M{x}}$ and observation $j$ in File $F_{\M{y}}$ are a match, $1 \leq i,j \leq n$,
computed from the Fellegi-Sunter model \cite{Fellegi69} given a set of comparison variables. However, it is generally not guaranteed that this choice is misspecified. In addition, the LL method only asserts unbiasedness when averaging over random $\Pi^*$, whereas in practice, the data analyst has to deal with
a merged data set arising from a single realization of $\Pi^*$. Choosing $\M{Q}$ according to \eqref{eq:averaging_operator} based on a uniform-at-random model within blocks avoids these
issues, producing an unbiased estimator from the comparison of blocking variables only.

In light of the above findings and discussions, we summarize advantages and disadvantages
of the estimator $\wh{\beta}^{\text{LL}}$ in relation to the estimator \eqref{eq:objective_theory} proposed herein.
\vskip1ex
\emph{Advantages}.
\begin{itemize}
\item The estimator $\wh{\beta}^{\text{LL}}$ does not rely on a sparsely occuring mismatches. In fact,
      $\wh{\beta}^{\text{LL}}$ can tolerate a fraction of mismatches close to one, and
      is still guaranteed to be unbiased. 
\item Asymptotic confidence intervals can be obtained from the expression for the asymptotic
      covariance \eqref{eq:sandwich}.
\item The approach is free of tuning parameters.      
\end{itemize}
\emph{Disadvantages}.
\begin{itemize}
\item The approach can suffer from a high variance, and is generally not guaranteed to be consistent
      as the sample size $n$ grows. Instead, consistency requires that the number of blocks
      $K \rightarrow \infty$. Even if the latter holds true, the asymptotic rate of convergence
      $O_{\p}(K^{-1})$ is suboptimal unless $n/K = O(1)$.
\item In view of the previous bullet, $\wh{\beta}^{\text{LL}}$ hinges on the availability of additional information that gives rise to a sufficiently fine block partitioning (i.e., consisting of a good number of blocks).
\item The approach does not account for potential errors in variables used to generate the block partitioning (cf., e.g., \cite{Dalzell2018}). 
\item The estimation equation \eqref{eq:estimating_equation} is not guaranteed to have a unique
      root, and thus practical algorithms for its solution such as Newton's method may deliver
      roots that are not consistent. 
\end{itemize}

\noindent \emph{Chambers' method}. For the sake of completeness, we present a brief account of the approach
due to Chambers in relation to the LL method. Chambers' method is based on the estimating equation
\begin{equation}\label{eq:estimation_equation_Chambers}
\M{X}^{\T} (\M{y} - \M{Q} \bm{\mu}^*(\beta)) = \M{0},  
\end{equation}
which differs from the LL estimation equation \eqref{eq:estimating_equation} only in that $\M{Q} \M{X}$ is 
replaced by $\M{X}$. Estimation equation \eqref{eq:estimation_equation_Chambers} is unbiased, but the resulting
estimator tends to converge at a slower rate than the LL estimator in the block setting discussed above. Using a similar arguments as above, it can be shown that the asymptotic covariance of the Chambers estimator is given by 
\begin{equation}\label{eq:sandwich_Chambers}
(\M{X}^{\T} \M{Q}^{\T} \M{V}(\beta^*) \M{X})^{-1} (\M{X}^{\T} \Pi^* \M{V}(\beta^*) \Pi^{*\T} \M{X} + \bm{\Xi}) (\M{X}^{\T} \M{Q}^{\T} \M{V}(\beta^*) \M{X})^{-1},
\end{equation}
for some positive semidefinite matrix $\bm{\Xi}$, and all other quantities are as above. The slower rate of convergence results from the middle matrix. We have 
\begin{equation*}
(\M{X}^{\T} \Pi^* \M{V}(\beta^*) \Pi^{*\T} \M{X}) = \sum_{i = 1}^n \M{x}_{\pi^*(i)} \psi''(\M{x}_i^{\T} \beta^*) \M{x}_{\pi^*(i)}^{\T} = O_{\p}(n)
\end{equation*}
for typical random designs. The two outer matrices in \eqref{eq:sandwich_Chambers} are the same as in the expression
for the asymptotic covariance of the LL method as provided in \eqref{eq:sandwich} and \eqref{eq:sandwich_parts}. Using the results for the LL method above then yields that altogether \eqref{eq:sandwich_Chambers} scales as $O_{\p}(n/K^2)$, which is generally slower than the rate $O_{\p}(K^{-1})$ of the LL method. 


\section{Case Study}\label{sec:realdata}
We consider the bike sharing data available on the UCI machine learning repository \cite{fanaee2014event}. The data set contains seasonal and weather information along with daily counts of rental bikes used between 2011 and 2012 in Washington, DC. The objective is to predict the number of ride-sharing bikes used during any given day (variable \texttt{count}) based on the categorical predictor variables \texttt{season} (spring, summer, fall, winter), \texttt{year} (2011 or 2012), \texttt{weathersit} (describing the weather situation on a given day in four categories from good to highly inclement weather), \texttt{weekday} (Monday through Sunday, numbered 0 to 6) and 
\texttt{workingday} (binary variable indicating whether a given day is a working day as opposed
to a holiday or Saturday/Sunday), as well as the three continuous predictor variables \texttt{atemp} (feeled temperature), \texttt{hum} (a humidity index) and \texttt{wind} (windspeed). 

Overall, the data set consists of 731 instances (days). We apply a square root transformation to the response variable \texttt{count}. The transformed response variable is treated as if it followed
a Poisson GLM with log-link, which can formally be regarded as a quasi-likelihood approach. The use of the transformation yields substantial improvement in terms of model fit compared to a Poisson model in which the raw counts are used as the response variable.  

In order to further improve the fit of the model, we delete observations satisfying one of the following criteria : (1) days affected by an extreme weather condition (e.g., blizzard, hurricane or windstorm), (2) national holidays (including Thanksgiving and Christmas), (3) especially hot days with temperatures exceeding 31.8 degrees Celsius. The resulting thinned data set includes 692 instances that are used to fit the following (Poisson-like) regression model:
\begin{align}\label{eq:model_bike}
 \log(\sqrt{\texttt{count}}) & = \beta_0^* + \beta_{\texttt{s2}}^* \cdot \mathbb{I}(\texttt{season} = 2) + \beta_{\texttt{s3}}^* \cdot \mathbb{I}(\texttt{season} = 3) + \beta_{\texttt{s4}}^* \cdot \mathbb{I}(\texttt{season} = 4) + \beta_{\texttt{y}}^* \cdot \texttt{year}   \notag\\ & +  \beta_{\texttt{wor}}^* \cdot \texttt{workingday} + \beta_{\texttt{a}}^* \cdot \texttt{atemp} + \beta_{\texttt{hum}}^* \cdot \texttt{hum} +  \beta_{\texttt{wi}}^* \cdot \texttt{wind} + \beta_{\texttt{y*a}}^* \cdot \texttt{yr}*\texttt{atemp} \notag\\
 &  +  \beta_{\texttt{we}}^* \cdot \mathbb{I}(\texttt{weekday} \in \{4,5,6\}) 
 + \beta_{\texttt{wea2}}^* \cdot \mathbb{I}(\texttt{weathersit} = 2) + \beta_{\texttt{wea3}}^* \cdot \mathbb{I}(\texttt{weathersit} = 3) \notag\\ 
 & + \beta_{\texttt{wea4}}^* \cdot \mathbb{I}(\texttt{weathersit} = 4) + \beta_{\texttt{s2*a}}^* \cdot \mathbb{I}(\texttt{season} = 2)*\texttt{atemp} \notag\\
 & + \beta_{\texttt{s3*a}}^* \cdot \mathbb{I}(\texttt{season} = 3)*\texttt{atemp} + \beta_{\texttt{s4*a}}^* \cdot \mathbb{I}(\texttt{season} = 4)*\texttt{atemp}
\end{align}
In total, the linear predictor consists of 17 terms apart from the intercept including interaction term between \texttt{season} and \texttt{atemp} and between \texttt{year} and \texttt{atemp}.  

\begin{figure}[!ht]
\begin{center}
\begin{tabular}{cc}
     \includegraphics[width = 0.49\textwidth]{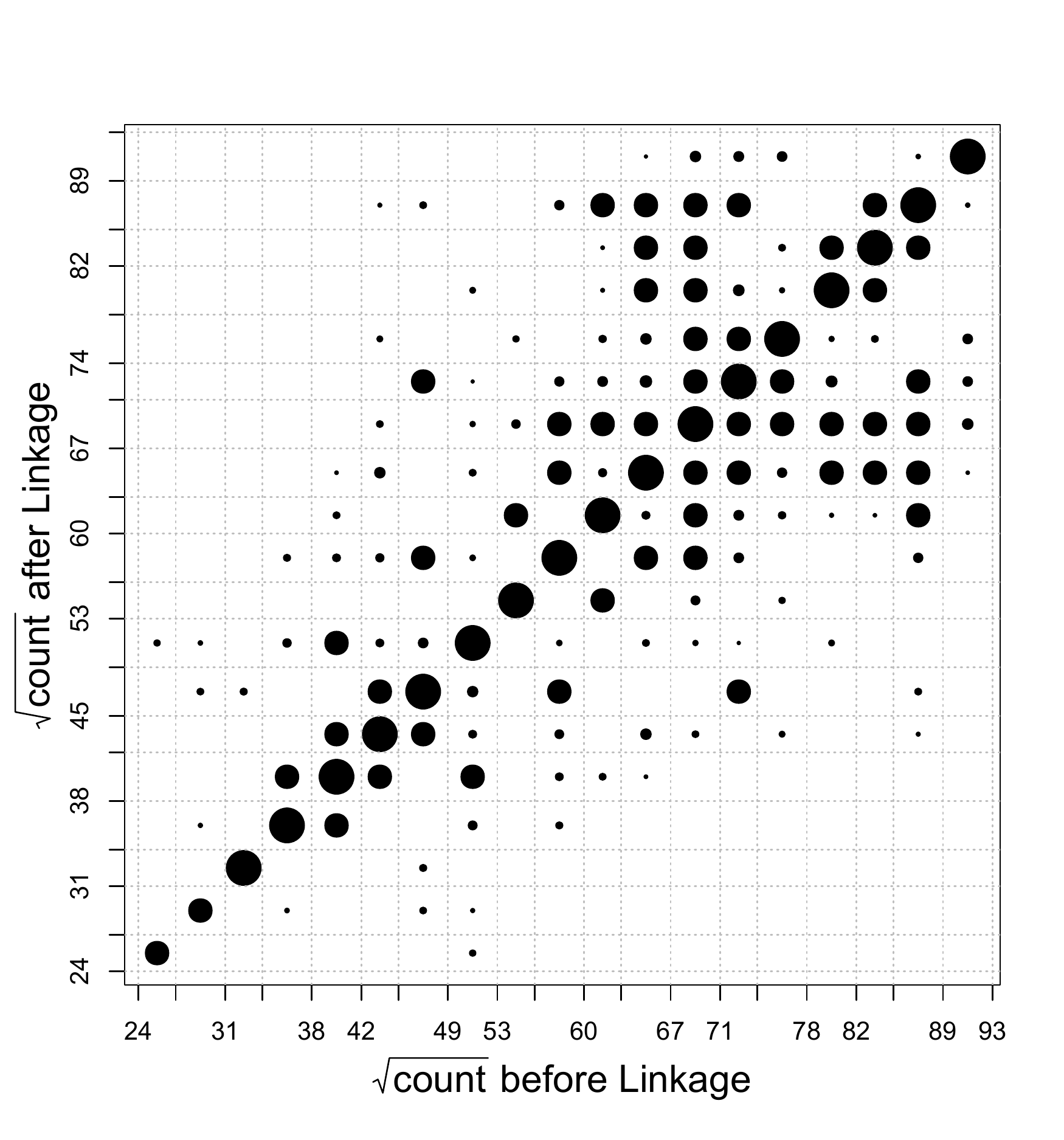}
   & \hspace*{-4.5ex}  \includegraphics[width = 0.5\textwidth]{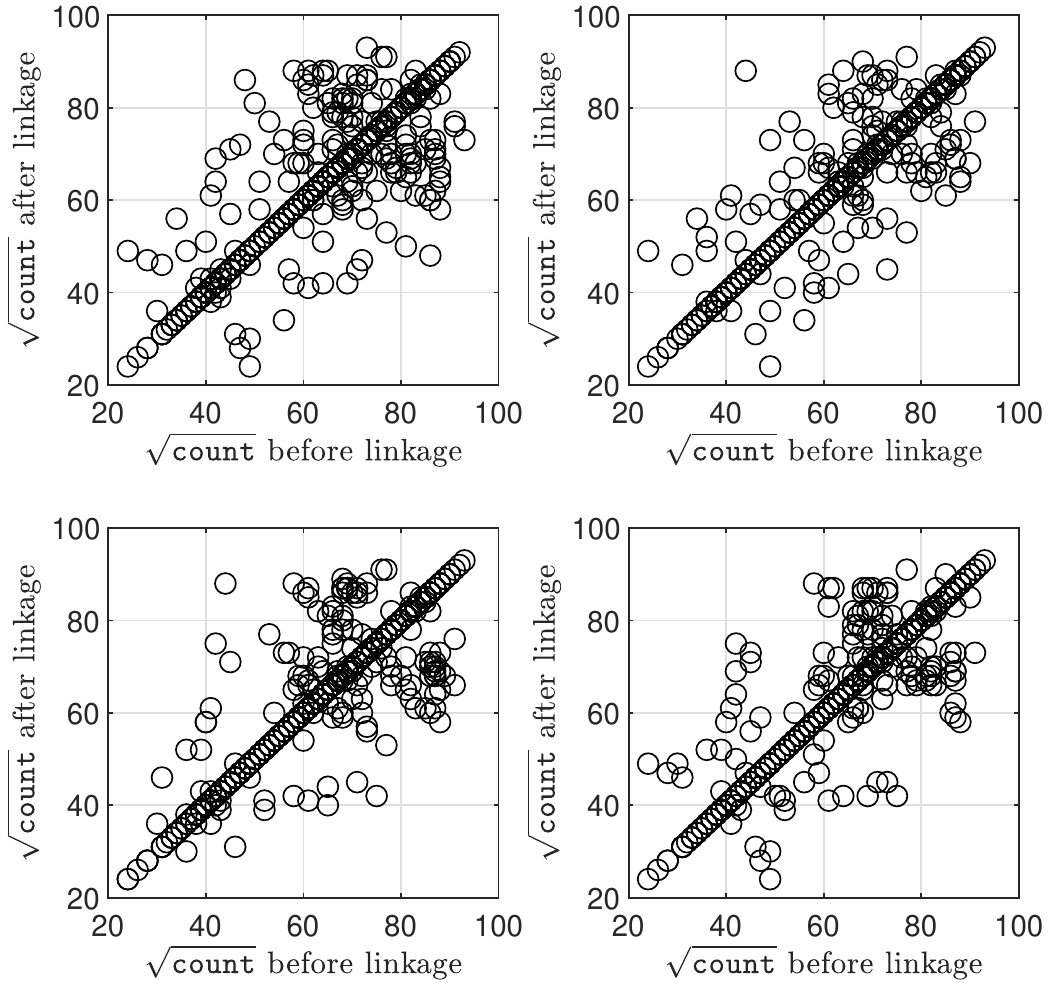}\\
\end{tabular}
\end{center}
\caption{Left: 2-D histogram of the response variable before and after linkage resulting based on 100 random permutations. Right: Scatterplots of the response variable before and after linkage for four selected permutations.}\label{fig:real_reponse}
\end{figure}

In order to mimic mismatch error introduced by record linkage, the response variable (\texttt{count}) is put into a separate file that additionally contains varying combinations of variables that are used for record linkage (see below for a list of those combinations). In order to enforce one-to-one linkage, ties between potentially matching records given the variables used for matching are broken uniformly at random; in order to account for that randomness, we consider 100 independent replications, and the results reported subsequently represent averages over those replications unless noted otherwise. Put differently, the underlying random permutations $\Pi^*$ are of the block form $\Pi^* = \text{bdiag}(\Pi^*_1, \ldots, \Pi^*_K)$, where $K$ denotes the number of unique combinations of values assumed for the matching variables. The average fraction of mismatches $k/n$ obtained in this way approximately equals $0.23$. Figure \ref{fig:real_reponse} illustrates the discrepancy between actual response and response after linkage based on the following combination of matching variables: \texttt{month}, \texttt{holiday}, \texttt{weekday}, \texttt{workingday}, $\texttt{temp}$\footnote{\texttt{temp} denotes the temperature in degree Celsius on a given day (integer-valued).}. In practice, the combination of variables used for matching may not be (fully) disclosed to the data analyst that operates on the merged data. Therefore, in addition to the case in which the matching variables are fully known, we also consider cases in which only one or two of the matching variables out of five overall are known (cf.~Table \ref{table:real}).  

\noindent \textit{Results.} In order to assess the performance of the proposed approach and
the baseline competitors discussed in $\S$\ref{sec:Com_LL}, estimation of the regression 
parameters of model \eqref{eq:model_bike} is performed based on the merged file contaminated 
by mismatch error. The resulting parameter estimates are then used to evaluate the deviance on the file $\{ (\M{x}_i, y_i^*) \}_{i = 1}^n$ containing predictors and response in their correct correspondence, i.e., 
\begin{equation}\label{eq:deviance_real}
2 \su \left[y_i^* \log \left(\frac{y_i^*}{\mu_i^{\text{est}}}\right) - (y_i^* - \mu_i^{\text{est}}) \right], \quad \text{where} \; \mu_i^{\text{test}} = \exp(\beta_0^{\text{est}} + \M{x}_i^{\T} \beta^{\text{est}}), 
\end{equation}
where $\beta^{\text{est}}$ is a placeholder for the estimates based on the merged file $\{(\M{x}_i, y_i) \}_{i = 1}^n$ delivered by i) the proposed method,
ii) the methods of Lahiri-Larsen (LL) and Chambers, and (iii) the naive estimator without any adjustment for mismatches. Both i) and ii) are evaluated for full and only partial knowledge of the matching variables. The ultimate reference for the quantity \eqref{eq:deviance_real} is obtained by substituting $\beta^{\text{est}} = \wh{\beta}^{\text{oracle}}$, where the oracle refers to the estimator based on the file  $\{ (\M{x}_i, y_i^*) \}_{i = 1}^n$ without any mismatched pairs. The quantity \eqref{eq:deviance_real} is hence intended to measure the drop in model fit that is induced by parameter estimates in the presence of mismatch error.  

Specific figures regarding \eqref{eq:deviance_real} are presented in Table \ref{table:real}. Note that in case that all matching variables are known to the data analyst, the performance of the LL estimator is rather close to the oracle and better than that of the proposed estimator \eqref{eq:objective_constrained} in its constrained form (265.46 vs.~269.92 with a standard error of 0.21). However, as less information about matching variables is available, the proposed estimator
performs on par (2nd column) respectively dramatically better (3rd column) than the LL estimator; in the latter case, the performance of the LL estimator is even considerably worse than that of the naive estimator. This is somewhat in alignment with what is predicted in $\S$\ref{sec:Com_LL}: as the number of blocks $K$ drops, so does the performance of the LL estimator. By contrast, the proposed estimator achieves a solid improvement over the naive estimator even in the complete absence of information about the matching variables used. The Chambers estimator performs weaker than the proposed and the LL estimator if the full set of matching variables is provided; in the two other cases with less information, we experience numerical difficulties (hence the value \textsf{NA}): Newton iterations used to obtain a root of the estimating equations \eqref{eq:estimation_equation_Chambers} converge to far suboptimal points leading to a deviance exceeding that of the intercept-only model. 

\begin{figure}[!ht]
\begin{center}
\begin{tabular}{ccc}
    $\Pi_{\min}$ & $\Pi_{\max}$\\
     \hspace*{-3.5ex} \includegraphics[width = 0.45\textwidth]{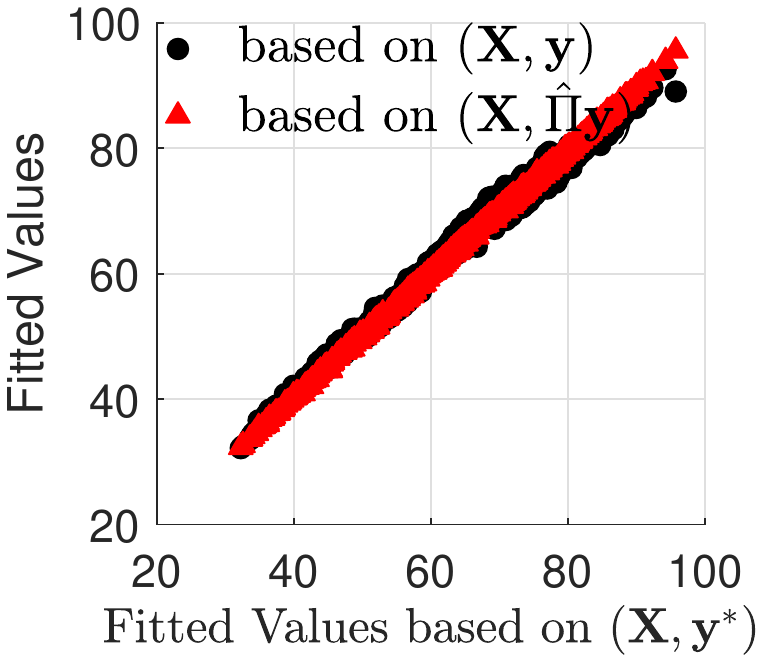}
   & \hspace*{-1.5ex} \includegraphics[width = 0.45\textwidth]{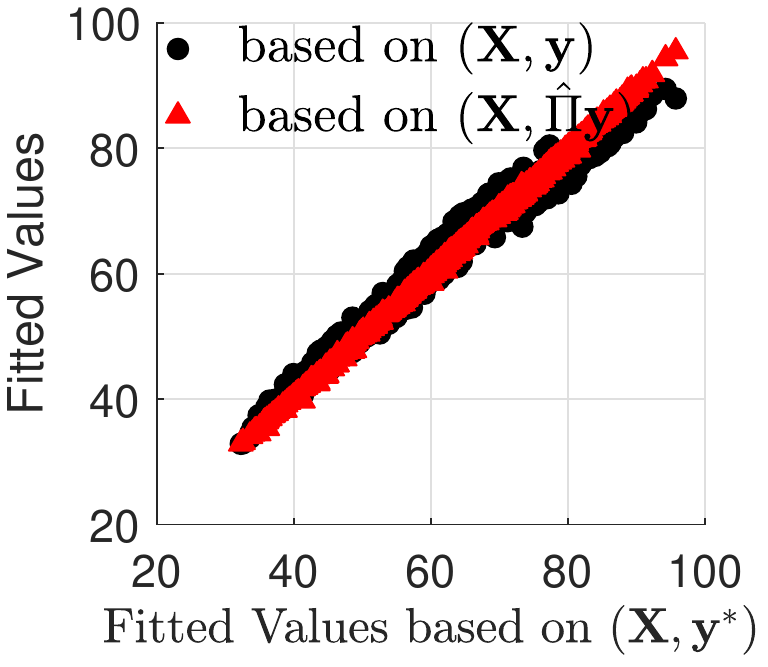}\\
        \hspace*{-3.5ex} \includegraphics[width = 0.45\textwidth]{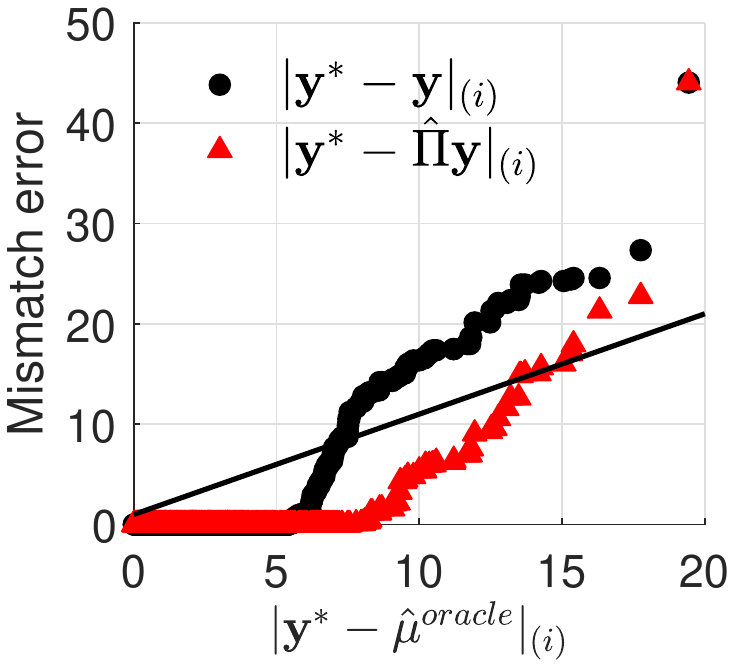}
   & \hspace*{-1.5ex} \includegraphics[width = 0.45\textwidth]{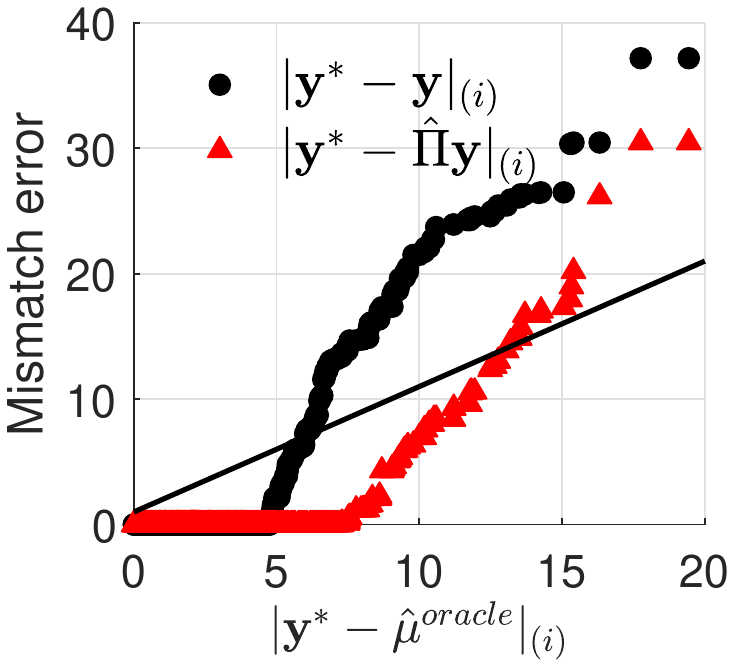}\\
     \end{tabular}
\end{center}
\caption{Top: Fitted values based on mismatch-free data $(\M{X},\M{y}^*)$ vs.~fitted values based on the merged data $(\M{X},\M{y})$ and corrected data $(\M{X},\wh{\Pi}\M{y})$ with $\wh{\Pi}$ denoting the minimizer of \eqref{eq:permutationrecovery_real}; ``fitted values" here refer to ordinary
(Quasi-) GLM estimation based on the data given in parentheses. Bottom: Q-Q plots of the absolute differences between the true responses and their fitted values based on the oracle estimator vs.~the absolute mismatch errors in the merged file $\{|y_i^* - y_i| \}_{i = 1}^n$ (dots) and their counterparts after correction based on \eqref{eq:permutationrecovery_real} (triangles).} \label{fig:real_per_rec}
\end{figure}

Regarding the selection of the tuning parameter $\lambda$ for the proposed approach, we create a separate validation set free of mismatches whose size is 20\% of the total number of samples, and select $\lambda$ so as to minimize the counterpart to \eqref{eq:deviance_real} on the validation set. Note that the existence of such a validation set is reasonably realistic, at least if sufficient information about matching variables is provided and at least some of the resulting combinations are unique, i.e., they yield singleton blocks for which mismatches can be ruled out. The corresponding results based on this data-driven selection of $\lambda$ are labelled ``Proposed($\lambda$)" in Table \ref{table:real}, to be distinguished from ``Proposed(oracle)" in which $\lambda$ is optimized to minimize the performance measure \eqref{eq:deviance_real} directly. We note that the performance of 
``Proposed($\lambda$)" is only slightly inferior to that of ``Proposed(oracle)". 
\begin{table}[!ht]
    \centering
    \begin{tabular}{|l|l|l|l|l|l|}
    \hline
    matching variables known & $K$ & \eqref{eq:objective_constrained} & \eqref{eq:averaging_operator}\\
    \hline
 \makecell{\texttt{month}, \texttt{holiday}, \texttt{weekday}, \texttt{workingday}, \texttt{temp}}  & 535 &  $\M{C}^1$ & $\M{Q}^1$\\
    \hline
  \texttt{month}, \texttt{temp} & 167 & $\M{C}^2$ & $\M{Q}^2$ \\
    \hline
    \texttt{temp} & 34 & $\M{C}^3$ & $\M{Q}^3$ \\
    \hline
    \end{tabular}
    
    \vspace{3mm}
    
    \begin{tabular}{|l|l|l|l|}        
    \hline
     & ($\M{Q}^1, \M{C}^1$) & ($\M{Q}^2 , \M{C}^2$) & ($\M{Q}^3, \M{C}^3$) \\ 
    \hline
   {\small Oracle} & \multicolumn{3}{c}{263.40} \vline\\    
    \hline
    {\small Lahiri-Larsen}  & 265.46 & 274.88 & 436.68\\  [0.5ex]
    \hline
    & & & \\[-3ex]
    {\small Chambers} & $\sss{272.70}{0.40}$ & \textsf{NA} & \textsf{NA} \\
    \hline
    & \multicolumn{3}{l}{} \vline  \\[-3ex]
    {\small Proposed(Oracle)}  & \multicolumn{3}{c}{$\sss{282.63}{0.39}$} \vline  \\
    \hline 
    & & &  \\[-3ex]
    $\small \text{Proposed}^{\M{C}}(\text{Oracle})$ & $\sss{269.92}{0.21}$ & $\sss{275.00}{0.29}$ & $\sss{278.34}{0.32}$ \\
    \hline
    & \multicolumn{3}{l}{} \vline  \\[-3ex]
    {\small Proposed($\lambda$)}  & \multicolumn{3}{c}{$\sss{285.04}{0.48}$} \vline  \\
    \hline 
    & & & \\[-3ex]
    $\small \text{Proposed}^{\M{C}}(\lambda)$ & $\sss{270.98}{0.36}$ & $\sss{276.29}{0.36}$ & $\sss{281.74}{0.52}$ \\
    \hline
    &  \multicolumn{3}{l}{} \vline \\[-3ex]
    {\small Naive} & \multicolumn{3}{c}{$\sss{316.86}{1.07}$} \vline \\
    \hline
    {\small Intercept only} & \multicolumn{3}{c}{2540.44} \vline \\
    \hline
\end{tabular}
\caption[Caption for LOF]{Top table: list of matching variables provided to the data analyst (the combination of five being the full list), the resulting number of blocks $K$ and the associated constraint matrices $\M{C}$ used for the proposed estimator ``Proposed$^{\M{C}}$" in its constrained form \eqref{eq:objective_constrained} as well as the associated matrices $\M{Q}$ used for the methods of Lahiri-Larsen and Chambers. Bottom table: deviances \eqref{eq:deviance_real} for several competitors as described in the text, averaged over 100 random block-structured permutations (the corresponding standard errors are given in parentheses\protect\footnotemark{The oracle estimators and the methods by Lahiri-Larsen and Chambers do not differ across permutations by construction, hence no standard error is reported.}
). ``Proposed(oracle)" and ``Proposed(oracle)$^{\M{C}}$" refer to the choice of $\lambda$ that directly minimizes \eqref{eq:deviance_real}, while ``Proposed($\lambda$)" and ``Proposed($\lambda$)$^{\M{C}}$" refer to the choice of $\lambda$ based on a validation set.}\label{table:real}
\end{table}
\noindent \emph{Permutation Recovery}. In this paragraph, we describe how the proposed approach can be leveraged to reduce mismatch error in the response variable contained in the merged file. We henceforth suppose that the data analyst is equipped with full knowledge of the matching variables \texttt{month}, \texttt{holiday}, \texttt{weekday}, \texttt{workingday}, \texttt{temp} used during the creation of the merged file. For ease of presentation, we here confine ourselves to two specific 
random permutations $\Pi_{\min}$ and $\Pi_{\max}$ out of the ensemble $\{\Pi^{(1)},\ldots,\Pi^{(100)} \}$ that were generated, defined by 
\begin{equation*}
\Pi_{\min} = \min_{1 \leq i \leq N} \nnorm{\Pi^{(i)} \M{y}^* - \M{y}^*}_2, \qquad 
\Pi_{\max} = \max_{1 \leq i \leq N} \nnorm{\Pi^{(i)} \M{y}^* - \M{y}^*}_2,
\end{equation*}
representing a best and a worst case scenario, respectively. Given $(\M{X}, \Pi_{\min} \M{y}^*)$ and
$(\M{X}, \Pi_{\max} \M{y}^*)$, we compute the proposed estimator with constraint matrix $\M{C} = \M{C}^1$ (cf.~Table \ref{table:real} top) and use the resulting solution $\wh{\beta}$ in place of $\beta^*$ in the optimization problem
defining the maximum likelihood estimator for the unknown permutation \eqref{eq:MLE2}. The resulting optimization problem is given by  
\begin{equation}\label{eq:permutationrecovery_real}
\min_{\Pi \in \mc{P}(\mc{G})}  -\nscp{\Pi \M{y}}{\M{X} \wh{\beta}},
\end{equation}
where $\M{y} = \Pi_{\min} \M{y}^*$ and $\M{y} = \Pi_{\max} \M{y}^*$, respectively, and $\mc{P}(\mc{G})$ denotes the set of all block-wise permutation matrices induced by the resulting index subsets $\mc{G} = \{ G_j \}_{j = 1}^K$, $K = 535$, corresponding to identical values for the matching variables. Note that the minimizer of \eqref{eq:permutationrecovery_real} can be obtained by pairing the order statistics of the linear predictor and the response within each of the sets $\{ G_j \}_{j = 1}^K$. While permutation recovery, i.e., $\{\wh{\Pi} = \Pi_{\min} \}$ and $\{\wh{\Pi} = \Pi_{\max} \}$, respectively, where $\wh{\Pi}$ denotes the minimizer of \eqref{eq:permutationrecovery_real} turns out to be out of reach here, a substantial reduction 
in mismatch error is achieved, i.e., $\wh{\Pi} \M{y}$ is visibly closer to $\M{y}^*$ than
$\M{y}$ as shown in Figure \ref{fig:real_per_rec}. The corrected response  $\wh{\Pi} \M{y}$ can
be used to refit the regression model. Figure \ref{fig:real_per_rec} indicates that the resulting
fitted values exhibit a much better agreement with the fitted values obtained from a mismatch-free
data set.
\footnotetext{The oracle estimators and the Lahiri-Larsen method do not differ across permutations by construction, hence no standard error is reported.} 

\section{Conclusion}\label{sec:Con}
In this paper, we have presented a method based on $\ell_1$-penalization to account for mismatch
error in the response in linked files, and have highlighted its benefits compared to established methods for this scenario. We have also explored how to directly reduce mismatch error by estimating the underlying permutation associated with the true correspondence between predictor-response pairs. 
The proposed approach is computationally appealing, supported by theoretical guarantees, and bears considerable potential regarding the adjustment for mismatch error in post-linkage analysis. At the same time, the approach presented herein prompts several directions of future research. Concerning the estimation of the regression parameter, it is worthwhile exploring the use of observation-specific penalization factors $\{ \lambda_i \}_{i = 1}^n$ instead of a single global 
value $\lambda$ that could prove particularly beneficial in Poisson and Gamma regression in light
of heteroscedasticity. Moreover, it is of great practical relevance to be able to conduct statistical inference for the regression parameter (confidence intervals and tests of linear hypotheses). A promising approach with regard to this aspect is the use of techniques developed for constructing confidence intervals in lasso regression, e.g., \cite{ZhangZhang2014, Javanmard2014,Buhlmann2014}.

Concerning estimation of the permutation, we have focused on exact permutation recovery and on approximate recovery with small Hamming distance. The results are coupled to stringent conditions, which are often not met in practice. Nevertheless, the results presented in the case study of the previous section indicate that approximate permutation recovery can be achieved with respect to 
alternative metrics like the $\ell_2$-distance between the true response $\M{y}^*$ and the estimator
$\wh{\Pi} \M{y}$, and elaborating the corresponding theory constitutes a further promising direction.   


\bibliographystyle{IEEEtran}
\bibliography{refs.bib}

\appendix

\section{Proof of Theorem \ref{theo:estimation_error}}
The proof presented herein builds on techniques developed in the long version of the article \cite{negahban2012unified}, the gist of which is also presented in the monograph \cite{wainwright_2019} (cf.~chapter 9 therein). In the following, we present the main thread of the proof, and defer supporting lemmas to the end of this section.   
\vskip1ex
\noindent A Taylor expansion of $\ell$ around $\theta^*$ yields
\begin{equation*}
\ell(\wh{\theta}) = \ell(\theta^*) + \nscp{\nabla \ell(\theta^*)}{\wh{\theta} - \theta^*} + \frac{1}{2} (\wh{\theta} - \theta^*)^{\T} \nabla^2 \ell(\wt{\theta}) (\wh{\theta} - \theta^*)
\end{equation*}
for some $\wt{\theta}$ in the line segment between $\theta^*$ and $\wh{\theta}$. Next, since $\wh{\theta}$ is a minimizer of $\ell_{\text{pen}}$, we have $\ell(\wh{\theta}) + \lambda \nnorm{\wh{\xi}}_1 \leq \ell(\theta^*) + \lambda \nnorm{\xi^*}_1$. 
Combining this with the above Taylor expansion, we obtain 
\begin{equation}\label{eq:basic_inequality}
  \nscp{\nabla \ell(\theta^*)}{\wh{\theta} - \theta^*} + \lambda \nnorm{\wh{\xi}}_1
  \leq \lambda \nnorm{\xi^*}_1. 
\end{equation}
Denote $\wh{\delta} = \wh{\theta} - \theta^*$, $\wh{\delta}^{\beta} = \wh{\beta} - \beta^*$, and
$\wh{\delta}^{\xi} = \wh{\xi} - \xi^*$. Furthermore, according to condition ({\bfseries C1}), $\nnorm{\nabla \ell(\theta^*)}_{\infty} \leq \nu_n$ with probability tending to one. Conditional on this event, inequality \eqref{eq:basic_inequality} and the use of H\"older's inequality $\nscp{\nabla \ell(\theta^*)}{\wh{\theta} - \theta^*} \leq  \nnorm{\nabla \ell(\theta^*)}_{\infty}\nnorm{\wh{\theta} - \theta^*}_1 \leq \nu_n \nnorm{\wh{\delta}}_1$ yield
\begin{align*}
\lambda \nnorm{\wh{\xi}}_1 \leq \nu_n (\nnorm{\wh{\delta}^{\beta}}_1 + \nnorm{\wh{\delta}^{\xi}}_1) +
   \lambda \nnorm{\xi^*}_1. 
\end{align*}
Let $S \subset \{1,\ldots,d + n \}$ denote the support of $\theta^*$, and let $T \subset \{1,\ldots,n \}$ denote the support of $\xi^*$. We have $|T| = k$ and $|S| = d + k$. Using the previous inequality and the triangle inequality, we obtain that 
\begin{align*}
\lambda \nnorm{\wh{\xi}_{T^c}}_1 \leq \nu_n (\nnorm{\wh{\delta}^{\beta}}_1 + \nnorm{\wh{\delta}^{\xi}}_1) +
   \lambda (\nnorm{\xi^*}_1 - \nnorm{\wh{\xi}_{T}}_1) \leq \nu_n (\nnorm{\wh{\delta}^{\beta}}_1 + \nnorm{\wh{\delta}^{\xi}}_1) + \lambda \nnorm{\wh{\delta}_{T}^{\xi}}_1. 
\end{align*}
Noting that $\wh{\xi}_{T^c} = \wh{\delta}_{T^c}^{\xi}$, the above inequality can be written as 
\begin{equation*}
\lambda \nnorm{\wh{\delta}_{T^c}^{\xi}}_1 \leq \nu_n (\nnorm{\wh{\delta}^{\beta}}_1 + \nnorm{\wh{\delta}_T^{\xi}}_1 + \nnorm{\wh{\delta}_{T^c}^{\xi}}_1) + \lambda \nnorm{\wh{\delta}_{T}^{\xi}}_1.
\end{equation*}
Re-arranging yields 
\begin{equation*}
(\lambda - \nu_n) \nnorm{\wh{\delta}_{T^c}^{\xi}}_1 \leq \nu_n (\nnorm{\wh{\delta}^{\beta}}_1 + \nnorm{\wh{\delta}_T^{\xi}}_1) + \lambda \nnorm{\wh{\delta}_{T}^{\xi}}_1.
\end{equation*}
Since $\lambda > \nu_n$ by assumption, dividing both sides by $(\lambda - \nu_n)$ yields
\begin{equation*}
 \nnorm{\wh{\delta}_{T^c}^{\xi}}_1 \leq \frac{\lambda + \nu_n}{\lambda - \nu_n} \nnorm{\wh{\delta}_T^{\xi}}_1 + \frac{\nu_n}{\lambda - \nu_n} \nnorm{\wh{\delta}^{\beta}}_1. 
\end{equation*}
Observing that $\wh{\delta}_{T^c}^{\xi} = \wh{\delta}_{S^c}$ and that $\nnorm{\wh{\delta}_T^{\xi}}_1 +  \nnorm{\wh{\delta}^{\beta}}_1 = \nnorm{\wh{\delta}_S}_1$, we arrive at the inequality
\begin{equation}\label{eq:cone_condition}
\nnorm{\wh{\delta}_{S^c}}_1 \leq  \frac{\lambda + \nu_n}{\lambda - \nu_n} \nnorm{\wh{\delta}_S}_1. 
\end{equation}
For $\rho > 0 $, we denote the set
\begin{equation}\label{eq:K_S}
\mc{K}_S(\rho) = \{ \delta \in \mc{C}_S:\; \nnorm{\delta}_2 = \rho \}, \qquad \mc{C}_S = \left\{\delta \in \R^{d + n}:\; \nnorm{\delta_{S^c}}_1 \leq  \frac{\lambda + \nu_n}{\lambda - \nu_n} \nnorm{\delta_{S}}_1 \right\}.
\end{equation}
We note that for any $\rho > 0$, $\mc{K}_S(\rho)$ is non-empty since $\mc{C}_S$ is a non-empty cone. In the sequel, the proof will be based on the following two steps. 
\begin{description}
\item[\hspace*{3ex}{\bfseries Step 1}:] We assume for a moment that $\nnorm{\wh{\delta}}_2 \leq R$, where $R$ is the radius 
                           defined in condition {\bfseries (C2)}. Given $\nnorm{\wh{\delta}}_2 \leq R$ and {\bfseries (C2)}, we use assumption {\bfseries (A1)}, random matrix techniques, and the conditions on $n$ and $\lambda$ specified in Theorem \ref{theo:estimation_error} to establish a so-called \emph{restricted strong convexity} (RSC) condition \cite{negahban2012unified} of the form
                           \begin{equation}\label{eq:RSC}
                           \ell(\theta^* + \delta) - \ell(\theta^*) - \nscp{\nabla \ell(\theta^*)}{\delta} \geq \varphi \nnorm{\delta}_2^2  \qquad \text{for all} \; \delta \in \bigcup_{0 < \rho \leq R} \mc{K}_S(\rho),
                           \end{equation}
                           which will hold with probability tending to one. In the above condition, $\varphi \geq \frac{1}{32} \sigma_{\min} \lambda_R$,
                           for a constant $c > 0$ and $\lambda_R$ as in {\bfseries (C2)}. 
\item[\hspace*{3ex}{\bfseries Step 2}:] Following the proof technique reviewed in \cite{wainwright_2019}, we show that the 
                                        RSC condition \eqref{eq:RSC} implies the $\ell_2$-estimation error bound in Theorem \ref{theo:estimation_error}. 
\end{description}
\emph{Step 2}: We first elaborate Step 2. since Step 1. is somewhat more involved. We start with the following result,
which is an application of Lemma 9.21 in \cite{wainwright_2019}. 
\begin{lemmaA} Consider the centered objective function $\mc{F}(\delta) = \ell(\theta^* + \delta) - \ell(\theta)^* + \lambda (\nnorm{\xi + \delta^{\xi}}_1 - \nnorm{\xi^*}_1)$ with $\delta = ([\delta^{\beta}]^{\T} \; [\delta^{\xi}]^{\T})^{\T}$ partitioned in the same fashion as $\theta$. For any $\rho > 0$, if $\mc{F}(\delta) > 0$ for all $\delta \in \mc{K}_S(\rho)$, then 
$\nnorm{\wh{\delta}}_2 \leq \rho$, where $\wh{\delta} = \wh{\theta} - \theta^*$ as above, and with $\mc{K}_S(\rho)$ as defined 
in \eqref{eq:K_S}.
\end{lemmaA}
In light of the above lemma, the goal is to find the smallest possible $\rho \leq R$ with $R$ defined according to the RSC condition \eqref{eq:RSC} such that $\mc{F}(\delta) > 0$
for all $\delta \in \mc{K}_S(\rho)$. We have 
\begin{alignat*}{2}
\mc{F}(\delta) &= \ell(\theta^* + \delta) - \ell(\theta^*) + \lambda (\nnorm{\delta + \theta^*}_1 - \nnorm{\theta^*}_1) \\
                 &\geq \nscp{\nabla \ell(\theta^*)}{\delta} + \varphi \nnorm{\delta}_2^2  + \lambda (\nnorm{\delta + \theta^*}_1 - \nnorm{\theta^*}_1) \qquad \qquad \qquad&&\text{[using \eqref{eq:RSC}]}\\
                 &\geq -\nu_n \nnorm{\delta}_1 + \varphi \nnorm{\delta}_2^2 + \lambda \left( \nnorm{\delta_{S^c}}_1 + \nnorm{\delta_S + \theta_S^*}_1 - \nnorm{\theta_S^*}_1 \right) &&\text{[using ($\mathbf{C1}$)]} \\
                 &\geq -\nu_n (\nnorm{\delta_S}_1  + \nnorm{\delta_{S^c}}_1) + \varphi \nnorm{\delta}_2^2 + \lambda \left( \nnorm{\delta_{S^c}}_1 - \nnorm{\delta_S}_1 \right) \\             &= \varphi \nnorm{\delta}_2^2 + (\lambda - \nu_n) \nnorm{\delta_{S^c}}_1 - (\lambda + \nu_n) \nnorm{\delta_S}_1 \\
                 &\geq \varphi \nnorm{\delta}_2^2 - (\lambda + \nu_n) \nnorm{\delta_S}_1 &&\text{[using that $\lambda \geq \nu_n$]} \\
                 &\geq \varphi \nnorm{\delta}_2^2 - (\lambda + \nu_n) \sqrt{d+k} \nnorm{\delta}_2 \\
      &\geq \varphi \nnorm{\delta}_2 (\nnorm{\delta}_2 - \varphi^{-1} (\lambda + \nu_n) \sqrt{d+k}),                
\end{alignat*}
which is positive for all $\delta \in \mc{K}_S(\rho)$ with $\varphi^{-1} (\lambda + \nu_n) \sqrt{d + k} < \rho \leq R$. Taking the infimum over this set of $\rho$'s yields the error bound in Theorem \ref{theo:estimation_error}. 
\vskip1ex
\noindent \emph{Step 1}: We now proceed with the technically more involved portion of the proof which entails establishing the RSC condition \eqref{eq:RSC}. Fix $\delta \in \mc{K}_S(\rho)$ for some $\rho \leq R$. Note that there exists 
$\theta_{\delta}$ contained in the line segment between $\theta^*$ and $\theta + \delta$ such that 
$\ell(\theta^* + \delta) - \ell(\theta^*) - \nscp{\nabla \ell(\theta^*)}{\delta} = \frac{1}{2} \delta^{\T} \nabla^2 \ell(\theta_{\delta}) \delta$. Let $\bm{\ddot{\Psi}}^{\delta} = \text{diag}(\psi''_1(\theta_{\delta}), \ldots, \psi_n''(\theta_{\delta}))$, and partition $\delta$ into $\delta^{\beta}$ and $\delta^{\xi}$. We thus have
\begin{align}
\frac{1}{2} \delta^{\T} \nabla^2 \ell(\theta_{\delta}) \delta &= \frac{1}{2}   \begin{pmatrix}
       \delta^{\beta} \\
       \delta^{\xi}
     \end{pmatrix}^{\T}  \frac{1}{n} \begin{pmatrix} X^{\T} \\[1ex]
                        \sqrt{n} I
       \end{pmatrix}  \ddot{\bm{\Psi}}^{\delta}    \begin{pmatrix} X \; \sqrt{n} I 
       \end{pmatrix} \begin{pmatrix}
       \delta^{\beta} \\
       \delta^{\xi}
     \end{pmatrix} \notag \\
     &= \frac{1}{2} (\delta^{\beta})^{\T} \frac{1}{n} X^{\T}  \ddot{\bm{\Psi}}^{\delta}  X \delta^{\beta} + \frac{1}{2} \nnorm{\ddot{\bm{\Psi}^{\delta}}^{1/2} \delta^{\xi}}_2^2 + 
  (\delta^{\beta})^{\T} \frac{X^{\T}}{\sqrt{n}} \ddot{\bm{\Psi}}^{\delta} \delta^{\xi} \label{eq:decomposition_quadratic}. 
\end{align}
We split
$\delta_{T^c}^{\xi}$ into subvectors of cardinality $d+k$ such that
$\delta_{T_1}^{\xi}$ contains the $d+k$-largest entries of $\delta_{T^c}^{\xi}$ in absolute value, $\delta_{T_2}^{\xi}$ contains the next $d+k$ largest entries of $\delta_{T^c}^{\xi}$ in absolute value, and so forth. Accordingly, we bound
\begin{align}
\left| (\delta^{\beta})^{\T} \frac{X^{\T}}{\sqrt{n}} \ddot{\bm{\Psi}}^{\delta} \delta^{\xi} \right|   &\leq  \nnorm{\delta^{\beta}}_2  \frac{1}{n^{1/2}}\left( \nnorm{(X_{T,:})^{\T} \ddot{\bm{\Psi}}_{T, T}^{\delta} \, \delta_T^{\xi}}_2 + \sum_{j \geq 1} \nnorm{{(X_{T_j,:})^{\T} \ddot{\bm{\Psi}}_{T_j, \, T_j}^{\delta} \, \delta_{T_j}^{\xi}}}_2 \right) \notag \\
&\leq \Lambda_R  \nnorm{\delta^{\beta}}_2 \left[ \left( \frac{1}{n^{1/2}} \max \left\{\nnorm{X_{T,:}}_2,  \max_{j \geq 1} \nnorm{X_{T_j,:}}_2   \right \} \right) \left( \nnorm{\delta_{T}^{\xi}}_2 + \sum_{j \geq 1} \nnorm{\delta_{T_j}^{\xi}}_2 \right) \right] \notag \\
&\leq \Lambda_{R} \nnorm{\wh{\delta}^{\beta}}_2 \frac{1}{n^{1/2}} \max\left\{\nnorm{X_{T,:}}_2 , \max_{j \geq 1} \nnorm{X_{T_j,:}}_2 \right \} \left( \nnorm{\delta_{T}^{\xi}}_2 + \nnorm{\delta_{T_1}^{\xi}}_2  + \frac{1}{\sqrt{d + k}} \nnorm{\delta_{T^c}^{\xi}}_1\right) \notag \\
&\leq \Lambda_{R} \nnorm{\wh{\delta}^{\beta}}_2 \frac{1}{n^{1/2}} \max\left\{\nnorm{X_{T,:}}_2 , \max_{j \geq 1} \nnorm{X_{T_j,:}}_2 \right \} \left( 2 \nnorm{\delta^{\xi}}_2 + \frac{1}{\sqrt{d + k}} \nnorm{\delta_{T^c}^{\xi}}_1\right), \label{eq:bound_cross}
\end{align}
where the third inequality follows from \cite[p.~8]{Candes2006}. Since $\delta$ is contained in the cone $\mc{C}_S$ \eqref{eq:K_S}, we have 
\begin{align}\label{eq:bound_cross_inner}
  \frac{1}{\sqrt{d+k}} \nnorm{\delta_{T^c}^{\xi}}_1 \leq  \frac{1}{\sqrt{d+k}} \frac{\lambda + \nu_n}{\lambda - \nu_n} \left( 
                                                         \nnorm{\delta_T^{\xi}}_1 +  \nnorm{\delta^{\beta}}_1 \right) \notag\\
  \leq \frac{\lambda + \nu_n}{\lambda - \nu_n} \left( \sqrt{\frac{k}{d + k}}\nnorm{\delta^{\xi}}_2 + \sqrt{\frac{d}{d+k}} \nnorm{\delta^{\beta}}_2  \right)
\end{align}
The maximum spectral norm over submatrices $\max \{\nnorm{X_{T,:}}_2 , \, \max_{j \geq 1} \nnorm{X_{T_j,:}}_2 \}$ can be controlled according to the following Lemma whose proof is delegated to Section \ref{sec:technical} of this appendix. 
\begin{lemmaA}\label{lem:submatrices} Under assumption \emph{({\bfseries A})}, there exists a universal constant $C > 0$ such that 
\begin{equation}\label{eq:maximumTj}
\p\left(\max \left\{ \nnorm{X_{T,:}}_2 , \, \max_{j \geq 1} \nnorm{X_{T_j,:}}_2 \right \} > C \sqrt{\sigma_{\max} \{ \log\left(\textstyle\frac{n}{d+k} \right) (d +k) \vee \log n\}} \right) \leq 1/n. 
\end{equation}
\end{lemmaA}
\noindent Combining \eqref{eq:bound_cross}, \eqref{eq:bound_cross_inner}, and \eqref{eq:maximumTj}, we obtain that
\begin{align}
\left| (\delta^{\beta})^{\T} \frac{X^{\T}}{\sqrt{n}} \ddot{\bm{\Psi}}^{\delta} \delta^{\xi} \right|   
&\leq \Lambda_R \left(2 + \frac{\lambda + \nu_n}{\lambda - \nu_n} \right) C \sqrt{\frac{\sigma_{\max}\{ (d + k) \log\left(\textstyle\frac{n}{d+k} \right) \vee \log n\}}{n}} \nnorm{\delta^{\beta}}_2 \nnorm{\delta^{\xi}}_2  + \notag \\
&\quad+\Lambda_R \frac{\lambda + \nu_n}{\lambda - \nu_n} C \sqrt{\frac{\sigma_{\max}\{ d \log\left(\textstyle\frac{n}{d+k} \right) \vee \textstyle \log n \}}{n}} \nnorm{\delta^{\beta}}_2^2, \label{eq:bound_cross_final}
\end{align}
which concludes the control of the cross-term in \eqref{eq:decomposition_quadratic}. Using {\bfseries (C2)}, the quadratic terms in 
\eqref{eq:decomposition_quadratic} are bounded as follows: with probability at least $1 - \epsilon_n'$, we have
\begin{align}\label{eq:lowerbound_quadratic}
\frac{1}{2} (\delta^{\beta})^{\T} \frac{1}{n} X^{\T}  \ddot{\bm{\Psi}}^{\delta}  X \delta^{\beta} + \frac{1}{2} \nnorm{\ddot{\bm{\Psi}^{\delta}}^{1/2} \delta^{\xi}}_2^2 \geq \underbrace{\frac{1}{8}  \sigma_{\min} \lambda_R}_{\invcoloneq \overline{\varphi}} (\nnorm{\delta^{\beta}}_2^2 + \nnorm{\delta^{\xi}}_2^2),
\end{align}
conditional on the event $\{ s_{\min}(X/\sqrt{n}) \geq 1/2 \}$, which holds with probability at least $1-1/n$ as long as $d \vee \log n \lesssim n$ according to Theorem A.\ref{theo:vershynin_subgaussian}. Now suppose that with $\overline{\varphi}$ as defined in \eqref{eq:lowerbound_quadratic} 
\begin{align}
&\Lambda_R \left(2 + \frac{\lambda + \nu_n}{\lambda - \nu_n} \right) C \sqrt{\frac{\sigma_{\max}\{ (d + k) \log\left(\textstyle\frac{n}{d+k} \right) \vee \log n\}}{n}} \leq \overline{\varphi}/2   \label{eq:term_preceding} \\
&\Leftrightarrow n \geq C'\,\frac{\sigma_{\max}}{\sigma_{\min}} \cdot \left(\frac{\Lambda_R}{\lambda_R} \right)^2 \big\{ (d+k) \log\left(\textstyle\frac{n}{d+k} \right) \vee \log n \big\},  \label{eq:samplecomplexity_RSC}
\end{align}
where we have used that $\frac{\lambda + \nu_n}{\lambda - \nu_n} \leq 3$. Since the term preceding $\nnorm{\delta^{\beta}}_2^2$ in \eqref{eq:bound_cross_final} is upper bounded by the left hand side of \eqref{eq:term_preceding}, Eq.~\eqref{eq:lowerbound_quadratic}, condition \eqref{eq:samplecomplexity_RSC}, and the elementary inequality $|x y| \leq \frac{1}{2}(x^2 + y^2), \, x,y \in \R$ imply that for all $\delta \in \bigcup_{0 < \rho \leq R} \mc{K}_S(\rho)$
\begin{equation*}
\frac{1}{2} \delta^{\T} \nabla^2 \ell(\theta_{\delta}) \delta =   \frac{1}{2} (\delta^{\beta})^{\T} \frac{1}{n} X^{\T}  \ddot{\bm{\Psi}}^{\delta}  X \delta^{\beta} + \frac{1}{2} \nnorm{\ddot{\bm{\Psi}^{\delta}}^{1/2} \delta^{\xi}}_2^2 + 
  (\delta^{\beta})^{\T} \frac{X^{\T}}{\sqrt{n}} \ddot{\bm{\Psi}}^{\delta} \delta^{\xi} \geq \frac{\overline{\varphi}}{4} \nnorm{\delta}_2^2, 
\end{equation*} 
which demonstrates that under \eqref{eq:samplecomplexity_RSC}, the RSC condition \eqref{eq:RSC} with $\varphi = \overline{\varphi}/4$ holds with probability at least $1 - \epsilon_n - \epsilon_n' - 2/n$.

\section{Proof of Theorem \ref{theo:sep_mu}}

Without loss of generality, we assume that $\pi^*$ is the identity, i.e., $\pi^*(i) = i$, $1 \leq i \leq n$. In virtue of Proposition \ref{prop:sorting}, it is clear that $\wh{\pi} \neq \pi^*$ whenever there exists an index $i$ such that $(y_i - y_{i+1})(\x_{i}^{\T}\beta^* - \x_{i+1}^{\T}\beta^*) < 0$. Since by assumption, $\mu_1 \leq \ldots \leq \mu_n$, and the link function is monotonically increasing, we also have that $\x_{i}^{\T}\beta^* - \x_{i+1}^{\T}\beta^* < 0$ for $1 \leq i \leq n - 1$. By conditioning on the random design matrix $\M{X}$, we have that 
\begin{align}
        \p(\hat{\pi} \neq \pi^* | \M{X}) & = 1 - \p(\hat{\pi} = \pi^* | \M{X})  = 1 - \p(Y_{1} \leq Y_{2} \leq \cdots \leq Y_{n}) \notag \\
        & = 1 -  \p\left ( \bigcap_{i = 1}^{n-1}  Y_{i} \leq Y_{i+1} \right )
        = \p\left (\bigcup_{i=1}^{n-1}Y_{i} > Y_{i+1} \right ) \leq \sum_{i=1}^{n-1} \p(Y_{i} > Y_{i+1}) \label{eq:thm2proof_template}
\end{align}
The main efforts now goes in finding bounds on $\{ \p(Y_{i} > Y_{i+1}) \}_{i = 1}^{n-1}$ in terms of the separation of the $\{ \mu_i \}_{i = 1}^n$. This done below in case-by-case fashion for different distributions in the GLM family. 

\subsection{Proof of Theorem \ref{theo:sep_mu} (a)}

 Since $Y_i-Y_{i+1} \mid \{ \M{x}_{i}, \M{x}_{i + 1} \} \sim N(\mu_{i} - \mu_{i + 1}, 2\sigma^2)$, using the usual tail bound for the Gaussian distribution for each term in the above sum, we obtain
 \begin{equation*}
\p(Y_{i} > Y_{i+1}) \le \exp\left( -\dfrac{\left(\mu_{i} - \mu_{i + 1}\right)^2}{4\sigma^2} \right), \quad 1 \leq i \leq n-1. 
\end{equation*}
Inserting this result into \eqref{eq:thm2proof_template}, we obtain that  \begin{align*}
        \p(\hat{\pi} \neq \pi^* | \M{X}) & \leq \sum_{i=1}^{n-1} \exp\left( -\dfrac{\left(\mu_{i} - \mu_{i + 1}\right)^2}{4\sigma^2} \right) \leq (n-1) \underset{{1\leq i \leq n-1}}{\max}  \exp\left( -\dfrac{\left(\mu_{i} - \mu_{i + 1}\right)^2}{4\sigma^2} \right) \\
        & \leq (n-1) \exp\left( -\underset{{1\leq i \leq n-1}}{\min}\dfrac{\left(\mu_{i} - \mu_{i + 1}\right)^2}{4\sigma^2} \right)
\end{align*}
Hence, for any $\delta>0$, $\p(\wh{\pi}\ne \pi^* \mid \M{X}) <\delta$ if
\begin{equation*}
\min_{1\leq i \leq n-1}(\mu_{i+1} - \mu_{i}) > 2\sigma\sqrt{\log \left(\dfrac{n-1}{\delta} \right)}.
\end{equation*}

\subsection{Proof of Theorem \ref{theo:sep_mu} (b)}
Since $Y_{i} | \M{x}_{i} \sim \text{Poisson}\big(\exp(\M{x}_i^{\T} \beta^*) \big)$, we have that
\begin{align*}
        \p(Y_{i} > Y_{i+1}) \notag 
        &= \p(t(Y_{i} - Y_{i+1}) > 0)  \text{ for } t> 0  \notag \\ 
        &= \p(e^ {t(Y_{i} - Y_{i+1})} > 1) \notag \leq \E[e^{t(Y_{i} - Y_{i+1})}]  \quad\text{(Markov's inequality)} \notag \\ 
        & = \E[e^{tY_{i}}] \E[e^{-tY_{i+1}}] \notag = \exp(\mu_{i} (e^{t} - 1) + \mu_{i+1} (e^{-t} - 1)) \notag \\
        &= \exp(- \mu_{i} - \mu_{i+1} + \mu_{i} \cdot e^{t} + \mu_{i+1}\cdot e^{-t}) .  
\end{align*}
 Since the above inequality is true for every $t>0$, we minimize the right hand side with respect to $t$. We have
    \begin{align}
     \p(Y_{i} > Y_{i+1}) & \leq \inf_{ t > 0 } \; \exp(- \mu_{i} - \mu_{i+1} + \mu_{i} \cdot e^{t} + \mu_{i+1}\cdot e^{-t}) \notag\\
     & \leq \exp\left(-\mu_{i} - \mu_{i+1}\right) \cdot \exp\left(\inf_{ t > 0 } \; \mu_{i} \cdot e^{t} + \mu_{i+1}  \cdot e^{-t}\right) \label{eq:Markov_end}
    \end{align}
    Differentiating the convex map $t \mapsto \mu_{i} \cdot e^{t} + \mu_{i+1} \cdot e^{-t}$ with respect to $t$ and setting the result equal to zero yields the minimizer $t_{0} = \frac{\log(\mu_{i+1}) - \log(\mu_{i})}{2}$.
Backsubstitution into \eqref{eq:Markov_end} yields
\begin{align*}
        \p(Y_{i} > Y_{i+1}) & \leq \exp\left(-\mu_{i} - \mu_{i+1}\right) \cdot \exp\left(\inf_{ t > 0 } \; \mu_{i} \cdot e^{t} + \mu_{i+1}  \cdot e^{-t}\right) \\
        & \leq \exp(-\mu_{i} - \mu_{i+1} + 2\sqrt{\mu_{i}\mu_{i+1}}) \\ 
        & \leq \exp(-(\sqrt{\mu_{i}} - \sqrt{\mu_{i+1}})^{2}), \quad 1 \leq i \leq n-1. 
    \end{align*} 
Using this result in \eqref{eq:thm2proof_template}, we obtain that \begin{align*}
        \p(\hat{\pi} \neq \pi^* | \M{X}) & \leq \sum_{i=1}^{n-1} \exp(-(\sqrt{\mu_{i}} - \sqrt{\mu_{i+1}})^{2}) \\
        & \leq (n-1) \underset{{1\leq i \leq n-1}}{\max}  \exp(-(\sqrt{\mu_{i}} - \sqrt{\mu_{i+1}})^{2}) \\
        & \leq (n-1)  \exp(- \underset{{1\leq i \leq n-1}}{\min} (\sqrt{\mu_{i}} - \sqrt{\mu_{i+1}})^{2}). 
    \end{align*}
Hence, for any $\delta > 0$, $\p(\hat{\pi} \neq \pi^* | \M{X}) < \delta$ if \begin{equation*}
        \underset{{1 \leq i \leq n-1}}{\min}  \left (\sqrt{\mu_{i}} - \sqrt{\mu_{i+1}} \right )  > \sqrt{\log \left( \frac{n-1}{\delta} \right)} . 
\end{equation*} 

\subsection{Proof of Theorem \ref{theo:sep_mu} (c)}
Since $Y_{i} | \M{x}_{i} \sim \text{Gamma}(\nu, \mu_{i})$, we obtain parallel to the proof of Theorem 
\ref{theo:sep_mu} (b) that
\begin{align*}
\p(Y_{i} > Y_{i+1}) &\leq \E \left[ e^{tY_{i}} \right] \E \left[ e^{-tY_{i+1}} \right] \; \text{for $t > 0$}\\
    &= \left(1 - \frac{\mu_{i}t}{\nu} \right)^{-\nu}\left(1 + \frac{\mu_{i+1}t}{\nu} \right)^{-\nu} \text{ for } \; 0 < t < \frac{\nu}{\mu_{i}} \\
    & = (-\mu_{i}\mu_{i+1}t^{2} + \mu_{i+1}t\nu - \mu_{i}t\nu + \nu^{2})^{-\nu}\nu^{2\nu}.  
\end{align*}
Since the above inequality is true all $0 < t < \frac{\nu}{\mu_{i}}$, we obtain that 
\begin{align*}
        \p(Y_{i} > Y_{i+1}) & \leq \underset{0 < t < \frac{\nu}{\mu_{i}}}{\inf} (-\mu_{i}\mu_{i+1}t^{2} + \mu_{i+1}t\nu - \mu_{i}t\nu + \nu^{2})^{-\nu}\nu^{2\nu} \\
        & \leq \underset{0 < t < \frac{\nu}{\mu_{i}}}{\inf} \left ( (-\mu_{i}\mu_{i+1})\left(t^{2} - (\frac{1}{\mu_{i}} - \frac{1}{\mu_{i+1}})t\nu\right)  + \nu^{2} \right ) ^{-\nu}\nu^{2\nu} \\ 
        & \leq \left ( \frac{\mu_{i}\mu_{i+1}}{4}(\frac{1}{\mu_{i}} - \frac{1}{\mu_{i+1}})^{2}\nu^{2} + \nu^{2} \right)^{-\nu}\nu^{2\nu},
    \end{align*}
    where we have used that the above infimum is attained at $t = \frac{\nu}{2\mu_{i}} - \frac{\nu}{2\mu_{i+1}}$. Further simplifying the previous term, we obtain that 
      \begin{align*}
        \p(Y_{i} > Y_{i+1}) & \leq \left ( \frac{\mu_{i}\mu_{i+1}}{4}(\frac{1}{\mu_{i}} - \frac{1}{\mu_{i+1}})^{2}\nu^{2} + \nu^{2} \right)^{-\nu}\nu^{2\nu}  \leq \left ( \frac{\mu_{i}\mu_{i+1}}{4}(\frac{1}{\mu_{i}} - \frac{1}{\mu_{i+1}})^{2} + 1 \right)^{-\nu}\\
        & \leq \left ( \frac{\mu_{i+1}}{4\mu_{i}}  +  \frac{\mu_{i}}{4\mu_{i+1}} + \frac{1}{2} \right)^{-\nu}, \quad 1 \leq i \leq n-1. 
    \end{align*}
Inserting the previous bound into \eqref{eq:thm2proof_template}, we obtain  
\begin{align*}
    \p(\hat{\pi} \neq \pi^* | \M{X}) & \leq \sum_{i=1}^{n-1}\p(Y_{i} > Y_{i+1})  \leq \sum_{i=1}^{n-1}\left ( \frac{\mu_{i+1}}{4\mu_{i}}  +  \frac{\mu_{i}}{4\mu_{i+1}} + \frac{1}{2} \right)^{-\nu} \\
    & \leq (n-1) \underset{1 \leq i \leq n-1}{\max} \left ( \frac{\mu_{i+1}}{4\mu_{i}}  +  \frac{\mu_{i}}{4\mu_{i+1}} + \frac{1}{2} \right)^{-\nu} \\
     & \leq (n-1) \underset{1 \leq i \leq n-1}{\max} \left ( \frac{z_{i}}{4} + \frac{1}{4z_{i}} + \frac{1}{2} \right)^{-\nu} \; \text{ where } z_{i} = \frac{\mu_{i+1}}{\mu_{i}}\\
     & \leq \frac{n-1}{\underset{1 \leq i \leq n-1}{\min} \left ( \frac{z_{i}}{4} + \frac{1}{4z_{i}} + \frac{1}{2} \right)^{\nu}}.
\end{align*}
Requiring $\p(\hat{\pi} \neq \pi^* | \M{X}) < \delta$ and using that $z_{i} \geq 1$, $1 \leq i \leq n-1$ by assumption,  we have that \begin{align*}
    & \underset{1 \leq i \leq n-1}{\min} \left ( \frac{z_{i}}{4} + \frac{1}{4z_{i}} + \frac{1}{2} \right)^{\nu} > \frac{n-1}{\delta} \\
    \Leftrightarrow & \underset{1 \leq i \leq n-1}{\min} \frac{\sqrt{z_{i}}}{2} + \frac{1}{2\sqrt{z_{i}}} > \left (\frac{n-1}{\delta} \right )^{1/2\nu}\\
    \Leftrightarrow & \underset{1 \leq i \leq n-1}{\min} z_{i} - 2\left (\frac{n-1}{\delta} \right )^{1/2\nu}\sqrt{z_{i}} > -1\\
    \Leftrightarrow & \underset{1 \leq i \leq n-1}{\min} \left (\sqrt{z_{i}} - \left (\frac{n-1}{\delta} \right )^{1/2\nu}\right )^{2} > \left (\frac{n-1}{\delta} \right )^{1/\nu}-1\\
    \Leftrightarrow & \underset{1 \leq i \leq n-1}{\min} \sqrt{z_{i}} > \sqrt{\left (\frac{n-1}{\delta} \right )^{1/\nu}-1} + \left (\frac{n-1}{\delta} \right )^{1/2\nu} 
\end{align*}
Note that $\sqrt{\left (\frac{n-1}{\delta} \right )^{1/\nu}-1} + \left (\frac{n-1}{\delta} \right )^{1/2\nu} < 2 \left (\frac{n-1}{\delta} \right )^{1/2\nu}$, so we have that $\p(\hat{\pi} \neq \pi^* | \M{X}) < \delta$ if 
    \begin{equation*}
        \underset{1 \leq i \leq n-1}{\min} \frac{\mu_{i+1}}{\mu_{i}} > 4 \left (\frac{n-1}{\delta} \right )^{1/\nu}. 
    \end{equation*}
\section{Proof of Theorem \ref{theo:sep_beta}}
Since the $\{ \M{x}_i \}_{i = 1}^n$ are now considered random, we may no longer assume without loss of generality that $\mu_1 \leq \ldots \leq \mu_n$. Instead, the proof now involves a union bound over $\binom{n}{2}$ pairs:  
\begin{align*}
        \p(\hat{\pi} \neq \pi^* | \M{X}) & = 2 \p\left ( \bigcup_{i < j} \left \{ Y_{i} > Y_{j} \mid \x^{\T}_{i}\beta^* < \x^{\T}_{j}\beta^* \right \} \right ) \\
        & \leq 2\sum_{i < j} \p(Y_{i} > Y_{j} \mid \x^{\T}_{i}\beta^* < \x^{\T}_{j}\beta^*)
\end{align*}
Accordingly, we state the following Lemma B.\ref{lemma:sep_mu1} which arises as a modification of Theorem \ref{theo:sep_mu}. The proof of the latter can be carried over easily since the only the union bound changes while the individual terms inside the sum can be estimated in the same way as before; therefore, the proof of the subsequent lemma is omitted. 
\begin{lemmaB}\label{lemma:sep_mu1}
Consider the MLE $\wh{\pi}$ given by the minimizer of \eqref{eq:MLE2}. 
For any $\delta > 0$, we have $\p(\wh{\pi} \neq \pi^* | \M{X}) < \delta$ if
\begin{itemize}
\item[(a)] $Y_{i} \sim N(\mu_{i}, \sigma^{2}), \; 1\leq i < j \leq n \, \text{\emph{:}}\;\;
       \underset{{i<j}}{\min} \; |\mu_{i} - \mu_{j}| > 2\sigma \sqrt{\log \frac{n(n-1)}{\delta}},
$
\item[(b)] $Y_{i} \sim \text{Poisson}(\mu_{i}) , \; 1\leq i < j \leq n \, \text{\emph{:}}\;\;
       \underset{{i<j}}{\min}\; |\sqrt{\mu_{i}} - \sqrt{\mu_{j}}| > \sqrt{\log \frac{n(n-1)}{\delta}},
$
\item[(c)] $Y_{i} \sim \text{Gamma}(\nu,\mu_{i}) , \; 1 \leq  i < j \leq n \, \text{\emph{:}}\;\; \underset{i < j}{\min} \; \frac{\mu_{j}}{\mu_{i}} > 4 \left (\frac{n(n-1)}{\delta} \right )^{1/\nu}$. 
\end{itemize}
\end{lemmaB}
Let $T_{i} = \frac{\x^{\T}_{i}\beta^*}{\nnorm{\beta^*}_{2}}$, $1 \leq i \leq n$. Note that by assumption, the 
$\{ T_i \}_{i = 1}^n$ are i.i.d.~random variables with a density, which we here denote by $f_{T}$. Since the density of the $\{ \x_{i} \}_{i = 1}^n$ is bounded by a constant $K < \infty$ almost everywhere, application of Theorem D. \ref{lemma:bouden} in Appendix \ref{sec:technical} yields that $f_{T}$ is bounded by $\sqrt{2}K$ almost everywhere. This property will be used repeatedly below.
\begin{lemmaB}\label{lemma:SBP} \hfill \\
Let $X_{1}, X_{2}, \ldots, X_{n}$ be real - valued independent random variables whose densities are bounded by $K$ almost everywhere. Let $a_{1}, ..., a_{n}$ be real numbers with $\sum_{i=1}^{n}a^{2}_{i} = 1$, then $\p \left (\abs{\su a_{i}X_{i}} \leq \epsilon \right ) \leq 2\sqrt{2}K\epsilon, \; \forall \epsilon > 0$. 
\end{lemmaB}
Furthermore, note that the $\{ \mu_i \}_{i = 1}^n$ are transformations of the
$\{ T_i \}_{i = 1}^n$ and are i.i.d.~random variables as well. The corresponding density is denoted by $f_{\mu}$. The main ingredient of the proof is the derivation of suitable lower bound on $\nnorm{\beta^*}^{2}_{2}$ such that the recovery conditions stated in terms of the $\{ \mu_i \}$ in Lemma B.\ref{lemma:sep_mu1} are satisfied. As before, we assume without loss of generality that $\pi^*(i) = i$, $1 \leq i \leq n$. 
\subsection{Proof of Theorem \ref{theo:sep_beta} (a)}
According to Lemma B. \ref{lemma:sep_mu1} (a), when $Y_{i}|\M{x}_i \sim N(\mu_{i},\sigma^{2})$, $1 \leq i \leq n$, we have that $\p(\hat{\pi} \neq \pi^* \mid \M{X}) < \delta$ is implied by the event $\left \{\underset{i<j}{\min}\; |\mu_{i+1} - \mu_{i}| > 2\sigma\sqrt{\log \frac{n(n-1)}{\delta}} \right \}$.
Consider the probability 
\begin{align*}
    \p \left ( \underset{i<j}{\min}\; | \mu_{i} - \mu_{j}| < \epsilon \right)& \leq \p\left ( \bigcup_{i < j} \{| \mu_{i} - \mu_{j}| < \epsilon\}\right)\\
    & \leq \sum_{i<j}\p \left ( |\mu_{i} - \mu_{j}| < \epsilon \right) \\
    & \leq \frac{n(n-1)}{2}\p \left ( |\mu_{i} - \mu_{j}| < \epsilon \right)
\end{align*}
Fix an arbitrary $i \in \{1,\ldots,n\}$. Observe that $\mu_{i} = \beta_{0} + T_{i} \cdot \nnorm{\beta^*}_{2}$ with $T_i$ defined above, and hence $\mu_{i} = m(T_{i})$ where the map $m$ is defined by $t \mapsto m(t) \coloneq \beta_0 + t \cdot \nnorm{\beta^*}_2$. Accordingly, its inverse $m^{-1}$ is given by $z \mapsto m^{-1}(z) = \frac{z - \beta_0^*}{\nnorm{\beta^*}_2}$. By the transformation formula, we obtain that the density 
of $\mu_i$ is given by $f_{\mu}(\cdot) = \frac{f_T(m^{-1}(\cdot))}{\nnorm{\beta^*}_2}$. Since
$f_T$ is bounded by $\sqrt{2} K$ almost everywhere (a.e.), it follows that $f_{\mu}$ is bounded by 
 $\frac{\sqrt{2} K}{\nnorm{\beta^*}_2}$ a.e. Consequently, invoking Lemma B.\ref{lemma:SBP} yields that for any $\epsilon > 0$
\begin{align*}
    \p \left ( |\mu_{i} - \mu_{j}| < \epsilon \right) & = \p \left ( \left|\frac{ \mu_{i}}{\sqrt{2}} - \frac{\mu_{j}}{\sqrt{2}} \right| < \frac{\epsilon}{\sqrt{2}} \right) \leq 2\sqrt{2} \cdot \frac{\sqrt{2}K}{\nnorm{\beta^*}_{2}} \cdot \frac{\epsilon}{\sqrt{2}} = \frac{2\sqrt{2}K\epsilon}{\nnorm{\beta^*}_{2}}. 
\end{align*}
By requiring that $\p \left ( \underset{i<j}{\min}\; | \mu_{i} - \mu_{j}| < \epsilon \right) < \delta$ for $\delta > 0$, resolving for $\nnorm{\beta^*}^{2}_{2}$,
we have that $\p(\hat{\pi} \neq \pi^*) < \delta$ if 
\begin{equation*}
    \nnorm{\beta^*}^{2}_{2} > \frac{2K^{2}\epsilon^{2}}{\delta^{2}}n^{2}(n-1)^{2}. 
\end{equation*}
Setting $\epsilon$ by the requirement on $\underset{i < j}{\min}\; |\mu_i - \mu_j|$ given in Lemma B.\ref{lemma:sep_mu1} (a), we obtain the condition
\begin{equation*}
    \nnorm{\beta^*}^{2}_{2} > \frac{8\sigma^{2}K^{2}n^{2}(n-1)^{2}}{\delta^{2}}\log\left(\frac{n(n-1)}{\delta}\right). 
\end{equation*}

\subsection{Proof of Theorem 3 (b)}
The first part of the proof parallels the preceding proof. Fix an arbitrary $i \in \{1,\ldots,n\}$. We have $Y_{i}|\M{x}_i \sim 
\text{Poisson}(\mu_{i})$ with $\mu_{i} = \exp(\beta_{0} + T_{i} \cdot \nnorm{\beta^*}_{2})$. Denote $Z_{i} = \sqrt{\mu_{i}} \invcoloneq  m(T_{i})$, where the inverse of $m$ is given by $z \mapsto m^{-1}(z) = \frac{2\log(z) - \beta_{0}^*}{\nnorm{\beta^*}_{2}}$. Letting $f_Z$ denote the density of the $\{ Z_i\}_{i = 1}^n$, an application of the transformation formula yields that 
\begin{align*}
\begin{split}
    f_{Z}(z) & = f_{T}(m^{-1}(z)) \cdot \frac{2}{z \nnorm{\beta^*}_{2}}, \quad z > 0. 
\end{split}
\end{align*}
Note that for any $z \geq 1$, we have $f_Z(z) \leq \frac{2 \sqrt{2}K}{\nnorm{\beta^*}_2}$. In order to apply
this bound, we note that 
\begin{equation*}
Z_i \geq 1 \; \, \Leftrightarrow  \; \, \beta_{0}^* + T_{i} \cdot \nnorm{\beta^*}_{2} \geq 0 \;\, \Leftrightarrow
\; \, \nscp{\beta^* / \nnorm{\beta^*}_{2}}{\M{x}_i} \geq - \beta_{0} / \nnorm{\beta^*}_{2}. 
\end{equation*}
Let $u^* = \frac{\beta^*}{\nnorm{\beta^*}_{2}}$. Note that
\begin{equation*}
    \left\{ f_Z(Z_i) \leq \frac{2\sqrt{2}K}{\nnorm{\beta^*}_{2}} \right \} \supseteq \left\{\scp{u^*}{\M{x}_i} \geq  -\frac{\beta_{0}}{\nnorm{\beta^*}_{2}} \right\}.  
\end{equation*}
Therefore, by applying Lemma B.\ref{lemma:SBP}, we obtain that for any pair $i < j$ and any $\epsilon > 0$
\begin{align}
    \p \left ( |\sqrt{\mu_{i}} - \sqrt{\mu_{j}}| < \epsilon \big| \scp{u^*}{\M{x}_{i}} \wedge \scp{u^*}{\M{x}_{j}} \geq - \frac{\beta_{0}}{\nnorm{\beta^*}_{2}} \right) & = \p \left ( \left |\frac{Z_{i}}{\sqrt{2}} - \frac{Z_{j}}{\sqrt{2}} \right | < \frac{\epsilon}{\sqrt{2}} \right) \notag \\
    &\leq 2 \sqrt{2} \cdot \frac{2\sqrt{2}K}{\nnorm{\beta^*}_{2}} \cdot \frac{\epsilon}{\sqrt{2}} = \frac{4\sqrt{2}K\epsilon}{\nnorm{\beta^*}_{2}}. \label{eq:smallgap_conditional} 
\end{align}
For an arbitrary fixed unit vector $u$ in $\R^d$, define the events $\mc{A}_u = \{\min_{1 \leq i \leq n} \scp{u}{\M{x}_i} \geq -\frac{\beta_0}{\nnorm{\beta^*}_2}\}$ and $\mc{B} = \bigcup_{i < j} \left \{| \sqrt{\mu_{i}} - \sqrt{\mu_{j}}| < \epsilon \right \}$. We then have 
\begin{align*}
    \p \left ( \underset{i<j}{\min}\; | \sqrt{\mu_{i}} - \sqrt{\mu_{j}}| < \epsilon \right)& \leq \p\left ( \mc{B} \mid \mc{A}_{u^*} \right)\p (\mc{A}_{u^*}) + \p\left ( \mc{B} \mid \mc{A}^{c}\right)\p (\mc{A}_{u^*}^{c})
    \leq \p\left ( \mc{B} \mid \mc{A}_{u^*}\right) + \p (\mc{A}_{u^*}^{c}). 
\end{align*}
Now observe that in view of \eqref{eq:smallgap_conditional} 
\begin{align*}
    \p\left ( \mc{B} \mid \mc{A}_{u^*}\right)  & \leq \sum_{i<j}\p \left ( |\sqrt{\mu_{i}} - \sqrt{\mu_{j}}| < \epsilon \mid \scp{u^*}{\M{x}_{i}} \wedge \scp{u^*}{\M{x}_{j}} \geq - \frac{\beta_{0}}{\nnorm{\beta^*}_{2}} \right)  \\
    & \leq \frac{n(n-1)}{2} \cdot \frac{4\sqrt{2}K\epsilon}{\nnorm{\beta^*}_{2}}. 
\end{align*}
By requiring that both $\p\left ( \mc{B} \mid \mc{A}_{u^*}\right) \leq \delta/2$  and $\sup_{u: \nnorm{u}_2 = 1}\p(\mc{A}_u^{c}) < \delta/2$, resolving for $\nnorm{\beta^*}^{2}_{2}$ in the previous display yields the condition
\begin{equation*}
    \nnorm{\beta^*}^{2}_{2} > \frac{16K^{2}\epsilon^{2}}{\delta^{2}}n^{2}(n-1)^{2}. 
\end{equation*}
By substituting $\epsilon$ by the lower bound imposed on $\underset{i < j}{\min}\; |\mu_i - \mu_j|$  in Lemma B. \ref{lemma:sep_mu1} (b), we have that $\p(\widehat{\pi} \neq \pi^*) < \delta$ if both 
\begin{equation*}
\nnorm{\beta^*}^{2}_{2} > \frac{16K^{2}n^{2}(n-1)^{2}}{\delta^{2}}\log\left (\frac{n(n-1)}{\delta} \right )
\end{equation*} and $\sup_{u: \nnorm{u}_2 = 1}\p (\min_{1 \leq i \leq n} \scp{u}{\M{x}_i} < -\frac{\beta_0}{\nnorm{\beta^*}_2}) < \delta/2$ hold.

\subsection{Proof of Theorem 3 (c)}
For Gamma regression with log link, we have that $\mu_{i} = \exp(\beta_{0} + \nu \cdot T_{i} \cdot \nnorm{\beta^*}_{2})$, $1 \leq i \leq n$. We therefore have for any pairs $i < j$
\begin{align*}
    \p \left ( \frac{\mu_{j}}{\mu_{i}} < \epsilon \right) & = \p \left (\exp\left \{\beta_{0} + \nu \cdot T_{j} \cdot \nnorm{\beta^*}_{2} - \beta_{0} - \nu \cdot T_{i} \cdot \nnorm{\beta^*}_{2}\right \} < \epsilon \right) \\
    & = \p \left ( T_{j} - T_{i} < \frac{\log(\epsilon)}{\nu\nnorm{\beta^*}_{2}} \right)  = \p \left ( \frac{T_{j}}{\sqrt{2}} - \frac{T_{i}}{\sqrt{2}} < \frac{\log(\epsilon)}{\sqrt{2}\nu\nnorm{\beta^*}_{2}} \right)
\end{align*}
Since the density of the $\{ T_{i} \}_{i = 1}^n$ is bounded by $\sqrt{2}K$ almost everywhere, by applying Lemma B.\ref{lemma:SBP}, we have that 
\begin{align*}
    \p \left ( \frac{\mu_{j}}{\mu_{i}} < \epsilon \right) & =  \p \left ( \frac{T_{j}}{\sqrt{2}} - \frac{T_{i}}{\sqrt{2}} < \frac{\log(\epsilon)}{\sqrt{2}\nu\nnorm{\beta^*}_{2}} \right)  =  \frac{1}{2} \p \left ( \left| \frac{T_{j}}{\sqrt{2}} - \frac{T_{i}}{\sqrt{2}} \right | < \frac{\log(\epsilon)}{\sqrt{2}\nu\nnorm{\beta^*}_{2}} \right)\\
    & \leq \frac{1}{2} \cdot 2\sqrt{2} \cdot \sqrt{2}K \cdot \frac{\log(\epsilon)}{\sqrt{2}\nu\nnorm{\beta^*}_{2}}  \leq \frac{\sqrt{2}\log(\epsilon)K}{\nu\nnorm{\beta^*}_{2}}.
\end{align*}
By requiring that $\p \left ( \underset{i < j}{\min}\; \; \frac{\mu_{j}}{\mu_{i}} < \epsilon \right) < \delta$ and resolving for $\nnorm{\beta^*}^{2}_{2}$ yields the condition 
\begin{equation*}
    \nnorm{\beta^*}^{2}_{2} > \frac{K^{2}(\log(\epsilon))^{2}}{2\nu^{2}\delta^{2}} n^{2}(n-1)^{2}. 
\end{equation*}
By substituting the $\epsilon$ by the lower bound on $\underset{i < j}{\min}\; \frac{\mu_j}{\mu_i}$ given in Lemma \ref{lemma:sep_mu1} (c), we have that $\p(\widehat{\pi} \neq \pi^*) < \delta$ if
\begin{equation*}
    \nnorm{\beta^*}^{2}_{2} > \frac{K^{2}n^{2}(n-1)^{2}}{2\nu^{2}\delta^{2}}\left (\log 4 \left (\frac{n(n-1)}{\delta} \right )^{1/\nu}\right)^{2}. 
\end{equation*}    

\section{Proofs of technical lemmas}\label{sec:technical}
\emph{Proof of Lemma A.\ref{lem:submatrices}}: In view of assumption ({\bfseries A}), the $d+k$ rows of $X_{T_j,:} \Sigma^{-1/2}$ are i.i.d.~sub-Gaussian for all $j$. As a result, for any fixed row subset $T_j$, Theorem D.\ref{theo:vershynin_subgaussian} below yields \begin{equation}\label{eq:fixedTj}
\p(\nnorm{X_{T_j,:}}_2 \leq \sigma_{\max}^{1/2}(\sqrt{d + k} + C' \sqrt{d} + t)) \leq \exp(-c' t^2), \quad t\geq 0,
\end{equation} 
with $\sigma_{\max} = \nnorm{\Sigma}_2$,  and $C' = C_K'$, $c' = c_K'$ only depending on the sub-Gaussian norm $K$ of the rows of 
$X_{T_j,:} \Sigma^{-1/2}$. Now observe that there are $\binom{n}{d+k} \leq \left( \frac{n e}{d+k} \right)^{d+k}$ possible subsets
$T_j$. Applying the union bound and invoking \eqref{eq:fixedTj} with the choice $t = C \sqrt{\log(\frac{n}{d+k})\cdot(d+k) \vee \log n}$, we obtain \eqref{eq:maximumTj}. Finally, note that $\nnorm{X_{T,:}}_2$ is stochastically smaller than $\max_{j \geq 1} \nnorm{X_{T_j,:}}_2$.  

\begin{theoremD}\label{theo:vershynin_subgaussian}\emph{(Theorem 5.39 in \cite{Vershynin2010})} Let $A$ be an $N \times M$ matrix whose
rows are independent sub-Gaussian isotropic random vectors in $\R^M$. Then for every $t \geq 0$, with probability
at least $1 - 2 \exp(-ct^2)$, one has 
\begin{equation*}
\sqrt{N} - C\sqrt{M} - t \leq \text{s}_{\min}(A) \leq s_{\max}(A) \leq \sqrt{N} + C \sqrt{M} + t,
\end{equation*}    
where $C, c > 0$ only depend on the maximum of the sub-Gaussian norms of the rows of $A$. 
\end{theoremD}

\noindent \emph{Proof of Lemma B.\ref{lemma:SBP}} 
Denote the probability density function of $\su a_{i}X_{i}$ as $f_{\su a_{i}X_{i}}$. We then have 
\begin{align*}
    \p \left (\abs{\su a_{i}X_{i}} \leq \epsilon \right ) & = \p \left (- \epsilon\leq \su a_{i}X_{i} \leq \epsilon \right ) \\
    & = \int_{-\epsilon}^{\epsilon}f_{\su a_{i}X_{i}}(x) \, dx \\
    & \leq \int_{-\epsilon}^{\epsilon}\sqrt{2} \, dx \quad \text{by applying Theorem D.\ref{lemma:bouden} below}\\
    & \leq 2\sqrt{2}K\epsilon.
\end{align*}

\begin{theoremD} (from Theorem 1.2 in \cite{rudelson2014small}) \;\label{lemma:bouden} \\
Let $X_{1}, X_{2}, \ldots, X_{n}$ be real - valued independent random variables whose densities are bounded by K almost everywhere. Let $a_{1}, \ldots, a_{n}$ be real numbers with $\sum_{i=1}^{n}a^{2}_{i} = 1$,Then the density of $\su a_{i}X_{i}$ is bounded by $\sqrt{2}K$ almost everywhere.
\end{theoremD}

%


\end{document}